\documentclass[12pt,a4paper]{article}
\usepackage{ifthen}
\newboolean{pdflatex}
\setboolean{pdflatex}{true}

\newboolean{articletitles}
\setboolean{articletitles}{true}

\newboolean{uprightparticles}
\setboolean{uprightparticles}{false}

\usepackage{booktabs}

\newif\ifEnableSectionTOCLinks
\EnableSectionTOCLinksfalse 

\def\paperauthors{Sebastian Schmitt$^1$, Amr Elmarassy$^1$, Michele Atzeni$^1$, Eluned Smith$^1$\\} 
\def\paperasciititle{Prospects for Measuring $CP$-Violation in Bs -> phi mu^+mu^- via Time-Dependent Angular Analysis}
\def\papertitle{Prospects for Measuring $CP$-Violation in \decay{\Bs}{\Pphi \mumu} via Time-Dependent Angular Analysis}

\def\paperkeywords{{High Energy Physics}} 
\def\papercopyright{\the\year\ } 
\def\paperlicence{CC BY 4.0 licence}
\def\paperlicenceurl{https://creativecommons.org/licenses/by/4.0/}

\usepackage[top=1in, bottom=1.25in, left=1in, right=1in]{geometry}

\usepackage{microtype}
\usepackage{lineno} 
\usepackage{xspace} 
\usepackage{caption}

\usepackage{graphicx}
\usepackage{color}
\usepackage{colortbl}
\graphicspath{{./figs/}}

\usepackage{amsmath}
\usepackage{amssymb}
\usepackage{amsfonts}
\usepackage{upgreek}

\newcommand*\patchAmsMathEnvironmentForLineno[1]{%
\expandafter\let\csname old#1\expandafter\endcsname\csname #1\endcsname
\expandafter\let\csname oldend#1\expandafter\endcsname\csname
end#1\endcsname
 \renewenvironment{#1}%
   {\linenomath\csname old#1\endcsname}%
   {\csname oldend#1\endcsname\endlinenomath}%
}
\newcommand*\patchBothAmsMathEnvironmentsForLineno[1]{%
  \patchAmsMathEnvironmentForLineno{#1}%
  \patchAmsMathEnvironmentForLineno{#1*}%
}
\AtBeginDocument{%
\patchBothAmsMathEnvironmentsForLineno{equation}%
\patchBothAmsMathEnvironmentsForLineno{align}%
\patchBothAmsMathEnvironmentsForLineno{flalign}%
\patchBothAmsMathEnvironmentsForLineno{alignat}%
\patchBothAmsMathEnvironmentsForLineno{gather}%
\patchBothAmsMathEnvironmentsForLineno{multline}%
\patchBothAmsMathEnvironmentsForLineno{eqnarray}%
}
\usepackage[pdftex,
            pdfauthor={\paperauthors},
            pdftitle={\paperasciititle},
            pdfkeywords={\paperkeywords}]{hyperref}
\usepackage{hyperxmp}
\hypersetup{
    pdfcopyright={Copyright (C) \papercopyright},
    pdflicenseurl={\paperlicenceurl}
}

\usepackage[colorinlistoftodos,textsize=scriptsize]{todonotes}

\usepackage[bottom,flushmargin,hang,multiple]{footmisc}

\usepackage[all]{hypcap}

\usepackage{xspace}
\usepackage{upgreek}

\def\btosll {\ensuremath{b \rightarrow s \ell\ell}\xspace}
\def\tk     {\ensuremath{\vartheta_K}\xspace}
\def\tl     {\ensuremath{\vartheta_\ell}\xspace}
\def\ctk    {\ensuremath{\cos\tk}\xspace}
\def\ctl    {\ensuremath{\cos\tl}\xspace}
\newcommand\varpm{\mathbin{\vcenter{\hbox{%
  \oalign{\hfil$\scriptstyle+$\hfil\cr
          \noalign{\kern-.3ex}
          $\scriptscriptstyle({-})$\cr}%
}}}}
\newcommand\varmp{\mathbin{\vcenter{\hbox{%
  \oalign{\hfil$\scriptstyle-$\hfil\cr
          \noalign{\kern-.3ex}
          $\scriptscriptstyle({+})$\cr}%
}}}}
\def\eip      {\ensuremath{e^{i \phi}}\xspace}
\def\emip     {\ensuremath{e^{-i \phi}}\xspace}
\def\Im       {\ensuremath{\mathcal{I}}\xspace}
\def\Re       {\ensuremath{\mathcal{R}}\xspace}

\def\lhcb   {\mbox{LHCb}\xspace}

\ifthenelse{\boolean{uprightparticles}}%
{

 \def\Pmu         {\ensuremath{\upmu}\xspace}

 \def\Ppi         {\ensuremath{\uppi}\xspace}

 \def\Pphi        {\ensuremath{\upphi}\xspace}

 \def\Ppsi        {\ensuremath{\uppsi}\xspace}

 \def\PDelta      {\ensuremath{\Delta}\xspace}
 \def\PXi         {\ensuremath{\Xi}\xspace}
 \def\PLambda     {\ensuremath{\Lambda}\xspace}
 \def\PSigma      {\ensuremath{\Sigma}\xspace}
 \def\POmega      {\ensuremath{\Omega}\xspace}
 \def\PUpsilon    {\ensuremath{\Upsilon}\xspace}
 \let\oldPi\Pi
 \def\PPi         {\ensuremath{\oldPi}\xspace}

 \def\PB      {\ensuremath{\mathrm{B}}\xspace}
 \def\PD      {\ensuremath{\mathrm{D}}\xspace}
 
 \def\PJ      {\ensuremath{\mathrm{J}}\xspace}
 \def\PK      {\ensuremath{\mathrm{K}}\xspace}

 \def\Pb      {\ensuremath{\mathrm{b}}\xspace}

 \def\Ps      {\ensuremath{\mathrm{s}}\xspace}

 \def\thebaroffset{0.0em}
}
{

 \def\Pmu         {\ensuremath{\mu}\xspace}

 \def\Ppi         {\ensuremath{\pi}\xspace}

 \def\Pphi        {\ensuremath{\phi}\xspace}

 \def\Ppsi        {\ensuremath{\psi}\xspace}
 
 \mathchardef\PDelta="7101
 \mathchardef\PXi="7104
 \mathchardef\PLambda="7103
 \mathchardef\PSigma="7106
 \mathchardef\POmega="710A
 \mathchardef\PUpsilon="7107
 \mathchardef\PPi="7105
 \def\PB      {\ensuremath{B}\xspace}
 \def\PD      {\ensuremath{D}\xspace}
 
 \def\PJ      {\ensuremath{J}\xspace}
 \def\PK      {\ensuremath{K}\xspace}

 \def\Pb      {\ensuremath{b}\xspace}

 \def\Ps      {\ensuremath{s}\xspace}

 \def\thebaroffset{0.18em}
}
\newcommand{\offsetoverline}[2][\thebaroffset]{\kern #1\overline{\kern -#1 #2}}%

\makeatletter
\ifcase \@ptsize \relax%
  \newcommand{\miniscule}{\@setfontsize\miniscule{4}{5}}%
\or%
  \newcommand{\miniscule}{\@setfontsize\miniscule{5}{6}}%
\or%
  \newcommand{\miniscule}{\@setfontsize\miniscule{5}{6}}%
\fi
\makeatother

\DeclareRobustCommand{\optbar}[1]{\shortstack{{\miniscule (\rule[.5ex]{1.25em}{.18mm})}
  \\ [-.7ex] $#1$}}
\DeclareRobustCommand{\opttilde}[1]{\shortstack{{\miniscule (\ensuremath{\thicksim})}
  \\ [-.3ex] $#1$}}

\def\mup        {{\ensuremath{\Pmu^+}}\xspace}
\def\mun        {{\ensuremath{\Pmu^-}}\xspace} %
\def\mumu       {{\ensuremath{\Pmu^+\Pmu^-}}\xspace}

\def\ellm       {{\ensuremath{\ell^-}}\xspace}
\def\ellp       {{\ensuremath{\ell^+}}\xspace}
\def\ellell     {\ensuremath{\ellp \ellm}\xspace}

\def\squark    {{\ensuremath{\Ps}}\xspace}

\def\bquark    {{\ensuremath{\Pb}}\xspace}

\def\pion   {{\ensuremath{\Ppi}}\xspace}
\def\pip    {{\ensuremath{\pion^+}}\xspace}
\def\pim    {{\ensuremath{\pion^-}}\xspace}

\def\kaon    {{\ensuremath{\PK}}\xspace}

\def\KorKbar {\kern \thebaroffset\optbar{\kern -\thebaroffset \PK}{}\xspace}

\def\Kp      {{\ensuremath{\kaon^+}}\xspace}
\def\Km      {{\ensuremath{\kaon^-}}\xspace}
\def\KS      {{\ensuremath{\kaon^0_{\mathrm{S}}}}\xspace}
\def\Kstarz  {{\ensuremath{\kaon^{*0}}}\xspace}

\def\D       {{\ensuremath{\PD}}\xspace}

\def\Dsm     {{\ensuremath{\D^-_\squark}}\xspace}

\def\B       {{\ensuremath{\PB}}\xspace}
\def\Bbar    {{\ensuremath{\offsetoverline{\PB}}}\xspace}

\def\Bz      {{\ensuremath{\B^0}}\xspace}
\def\Bd      {{\ensuremath{\B^0}}\xspace}

\def\Bu      {{\ensuremath{\B^+}}\xspace}

\def\Bp      {{\ensuremath{\Bu}}\xspace}

\def\Bs      {{\ensuremath{\B^0_\squark}}\xspace}
\def\Bsb     {{\ensuremath{\Bbar{}^0_\squark}}\xspace}
\def\BsorBsbar {\kern \thebaroffset\optbar{\kern -\thebaroffset \Bs}\xspace}

\def\jpsi     {{\ensuremath{{\PJ\mskip -3mu/\mskip -2mu\Ppsi}}}\xspace}

\def\Lz          {{\ensuremath{\PLambda}}\xspace}

\def\Lb           {{\ensuremath{\Lz^0_\bquark}}\xspace}

\newcommand{\decay}[2]{\mbox{\ensuremath{#1\!\to #2}}\xspace}

\def\to                 {\ensuremath{\rightarrow}\xspace}

\def\qsq       {{\ensuremath{q^2}}\xspace}

\def\CP                {{\ensuremath{C\!P}}\xspace}

\newcommand{\aunit}[1]{\ensuremath{\text{\,#1}}}

\newcommand{\tev}{\aunit{Te\kern -0.1em V}\xspace}
\newcommand{\gev}{\aunit{Ge\kern -0.1em V}\xspace}
\newcommand{\mev}{\aunit{Me\kern -0.1em V}\xspace}
\newcommand{\kev}{\aunit{ke\kern -0.1em V}\xspace}
\newcommand{\ev}{\aunit{e\kern -0.1em V}\xspace}

\newcommand{\mevc}{\ensuremath{\aunit{Me\kern -0.1em V\!/}c}\xspace}
\newcommand{\gevc}{\ensuremath{\aunit{Ge\kern -0.1em V\!/}c}\xspace}
\newcommand{\mevcc}{\ensuremath{\aunit{Me\kern -0.1em V\!/}c^2}\xspace}
\newcommand{\gevcc}{\ensuremath{\aunit{Ge\kern -0.1em V\!/}c^2}\xspace}
\newcommand{\gevgevcccc}{\ensuremath{\gev^2\!/c^4}\xspace} %

\def\fb   {\ensuremath{\aunit{fb}}\xspace}
\def\invfb   {\ensuremath{\fb^{-1}}\xspace}

\newcommand{\orcidicon}[1]{\href{https://orcid.org/#1}{\hspace*{0.1em}\raisebox{-0.45ex}{\includegraphics[width=1em]{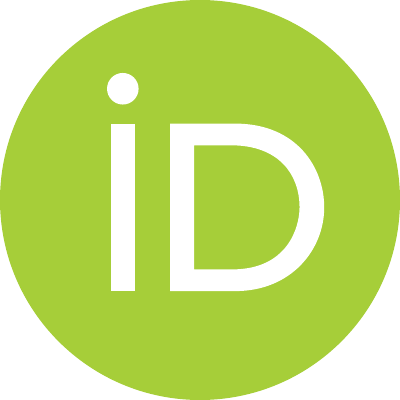}}}}

\hypersetup{
  colorlinks   = true, %
  urlcolor     = blue, %
  linkcolor    = blue, %
  citecolor    = red   %
}

\ifEnableSectionTOCLinks
    \usepackage[explicit]{titlesec} 
    
    \let\oldcontentsline\contentsline
    \renewcommand

    \titleformat{\section}{\normalfont\Large\bf}{\hyperlink{tocsection.\thesection}{{\thesection} \parbox[t]{\dimexpr\textwidth-1pc}{#1}}}{1pc}{}

    \titleformat{\subsection}{\normalfont\bf}{\hyperlink{tocsubsection.\thesubsection}{{\thesubsection} \parbox[t]{\dimexpr\textwidth-1pc}{#1}}}{1pc}{}

    \titleformat{name=\section,numberless}[display]{}{}{0pt}{\normalfont\Huge\bfseries #1}
\fi

\usepackage{cite}
\usepackage{mciteplus}
\usepackage{cleveref}
\crefname{figure}{Fig.}{Figs.}
\Crefname{figure}{Fig.}{Figs.}

\crefname{table}{Tab.}{Tabs.}
\Crefname{table}{Tab.}{Tabs.}

\crefname{equation}{Eqn.}{Eqns.}
\Crefname{equation}{Eqn.}{Eqns.}

\usepackage{placeins} \usepackage{longtable}

\def\paperauthors{
Sebastian Schmitt$^1$\orcidicon{0000-0002-6394-1081}, 
Amr Elmarassy$^1$\orcidicon{0009-0002-9592-5286}, 
Michele Atzeni$^1$\orcidicon{0000-0002-3208-3336}, 
Eluned Smith$^1$\orcidicon{0000-0002-9740-0574}\\
} 

\begin{document}

\renewcommand{\thefootnote}{\fnsymbol{footnote}}
\setcounter{footnote}{1}

\begin{titlepage}
\pagenumbering{roman}

\vspace*{-1.5cm}
\vspace*{1.cm}
\noindent
\begin{tabular*}{\linewidth}{lc@{\extracolsep{\fill}}r@{\extracolsep{0pt}}}
\\
 & & \today \\
 & & \\
\end{tabular*}

\vspace*{1.0cm}

{\normalfont\bfseries\boldmath\huge
\begin{center}
  \papertitle 
\end{center}
}

\vspace*{1.5cm}

\begin{center}
\paperauthors
\vspace{1em}
$^1$Massachusetts Institute of Technology, Cambridge, MA, United States

\end{center}

\vspace{\fill}

\begin{abstract}
  \noindent
This work investigates the prospects for performing a time-dependent angular analysis of 
$B_s^0 \rightarrow \phi \mu^+\mu^-$ decays at hadron colliders, and introduces new 
optimised angular observables associated with $B_s^0$-mixing in the decay rate.  
The time-dependent normalised decay rate and corresponding probability density function is presented both for when the flavour of the $B_s^0$ meson at production is tagged and when it is untagged. 
The normalised angular terms linked to $B_s^0$-mixing in the tagged (untagged) case are denoted $\mathcal{Z}_i$ ($\mathcal{H}_i$), and their optimised counterparts as $Q_i$ ($\mathcal{M}_i$).  

The expected sensitivities of these observables at the end of Run~3, Run~4, and Run~5 of the LHC are determined using pseudoexperiments generated with a decay-time resolution, background level, and signal yield similar to those reported by the LHCb collaboration. 
It is found that all observables can be extracted with Run~3 statistics, and that the $\mathcal{H}_i$ and $\mathcal{Z}_i$ observables have similar sensitivity, despite the former being suppressed by mixing terms. 
Moreover, the angular observables only accessible via flavour tagging, such as the equivalent of $P_5^{\prime}$ in the $B^0_s$ system, are found to exhibit sensitivities comparable to current measurements in $B^0 \rightarrow K^{\ast 0} \mu^+\mu^-$ decays once Run~5 datasets are available. 
Fits to the observables to extract the Wilson coefficients show a significant increase in precision when either the observables accessible via time-dependent analysis, or flavour tagging, are included. 
A marked increase in sensitivity to $CP$-violating short-distance effects is observed for a subset of the new optimised $M_i$ and $Q_i$ observables.

\end{abstract}

\vspace*{2.0cm}

\begin{center}
 \href{https://doi.org/10.1103/4wpp-rrxk}{Published in Physical Review D 113, 036001}
\end{center}

\vspace{\fill}

{\footnotesize 
\centerline{\copyright~\papercopyright. \href{\paperlicenceurl}{\paperlicence}.}}
\vspace*{2mm}

\end{titlepage}

\newpage
\setcounter{page}{2}
\mbox{~}
 
\renewcommand{\thefootnote}{\arabic{footnote}}
\setcounter{footnote}{0}

\cleardoublepage

\pagestyle{plain}
\setcounter{page}{1}
\pagenumbering{arabic}

\section{Introduction}
\label{sec:introduction}

In the Standard Model of Particle Physics (SM) transitions of \bquark-quarks to \squark-quarks are forbidden at tree-level and may only proceed via higher order diagrams. 
This absence of flavour changing neutral currents at tree-level makes \btosll transitions excellent probes for New Physics (NP) beyond the SM. 
In recent years, several tensions have been observed in \btosll transitions that emerge in branching fractions~\cite{LHCb-PAPER-2014-006,LHCb-PAPER-2015-009,LHCb-PAPER-2016-025,LHCb-PAPER-2021-014,LHCb-PAPER-2023-033}, angular distributions~\cite{Belle:2016fev,ATLAS:2018gqc,LHCb-PAPER-2020-002,LHCb:2020gog,LHCb-PAPER-2021-022,CMS:2024atz}, and lepton universality tests~\cite{BaBar:2012mrf,BELLE:2019xld,LHCb-PAPER-2019-040,LHCb-PAPER-2021-038,LHCb-PAPER-2022-045,LHCb-PAPER-2022-046,LHCb-PAPER-2024-032}. 
While the latter currently agree with the SM within their $5\,\%$ precision, the tensions in the angular distributions and branching fractions pose challenging questions to the SM, which are reflected by the discussion of the measurements in the theory community~\cite{Cornella:2021sby,Alguero:2021anc,Gubernari:2022hxn,Crivellin:2023zui,Capdevila:2023yhq,Athron:2023hmz,Allanach:2023uxz,Davighi:2023evx,Isidori:2023unk,Ciuchini:2022wbq,Greljo:2022jac,SinghChundawat:2022zdf,Parrott:2022zte,FernandezNavarro:2022gst,SinghChundawat:2022ldm,Allanach:2022iod,Bause:2022rrs, Ciuchini:2021smi,Alguero:2023jeh}. 
Of particular note is the angular distribution of \decay{\Bz}{\Kstarz(\rightarrow \Kp \pi^-) \mumu} and especially the observable $P_5^\prime$. 
While the \Bz decays are currently more powerful in constraining NP, due to their larger sample size, \decay{\Bs}{\Pphi(\rightarrow \Kp\Km) \mumu} decays may be more precisely calculable in the SM, as the frequently employed narrow-width approximation for the hadron-system holds to a better degree~\cite{Descotes-Genon:2019bud}. 
As a result, with a larger sample size of collected \bquark-quark decays, the prospects for the \Bs decays are increasingly more interesting. 

The decay \decay{\Bs}{\phi\mumu}\footnote{In the following the decay of the \Pphi into two charged kaons is implied. } is studied in detail by the \lhcb collaboration regarding its branching fraction~\cite{LHCb-PAPER-2021-014}, angular distribution~\cite{LHCb-PAPER-2021-022}, and lepton universality~\cite{LHCb-PAPER-2024-032}. 
Due to neutral meson mixing and the decay into a \CP-conjugate state, the decay offers the possibility to measure time-dependent \CP-violation phenomena, as for example discussed in Refs.~\cite{Bobeth:2008ij,Descotes-Genon:2015hea,Kwok:2025fza}. 
As the final state is not self-tagging, many observables, such as $P_5^\prime$, are only accessible if flavour-tagging\footnote{Flavour-tagging refers to the determination of the \Bs flavour at production. } can be performed. 
In this work, the projected sensitivity to the observables of the decay rate are explored for both the tagged and untagged time-dependent and time-independent angular analyses of \decay{\Bs}{\Pphi\mumu} decays at hadron colliders. We also present new observables which exhibit reduced dependence on the hadronic form-factors, similar to the observables $Q_8^-$ and $Q_9$ discussed in Ref.~\cite{Descotes-Genon:2015hea}. 
As a benchmark for the capabilities  of a dedicated heavy-flavour experiment, the expected performance of the \lhcb upgrade I is used frequently throughout this work. 
Our extrapolations show that a flavour-tagged and time-dependent angular analysis is feasible at hadron colliders, in contrast to previous suggestions in the literature~\cite{Kwok:2025fza}. 
To explore constraints on short-distance new physics from these new angular coefficients, we employ an effective field theory (EFT) framework, in which the effective Hamiltonian is expressed as a set of local dimension-six operators, $\mathcal{O}_i$, each multiplied by a corresponding Wilson coefficient, $\mathcal{C}_i$, representing the effective coupling strength. 
Using this framework we show that the angular observables obtained provide additional constraints of the Wilson Coefficients. 

For the first time, we do a full analysis of the time-dependent angular structure of \decay{\Bs}{\Pphi \mumu} decays, spelling out the full probability density function (PDF) which is altered by mixing effects, including its normalisation in different regions of the invariant mass squared of the di-lepton system ($q^2$). 

This paper is structured in the following way, in Section~\ref{sec:differential-decay-rate} the parameterisation of the decay rate and the definition of observables is given. 
A new set of optimised observables with reduced dependence on hadronic form-factors is introduced in Section~\ref{sec:optimised-observables}. 
The setup of the pseudoexperiments used to derive the sensitivity on the observables is described in Section~\ref{sec:pseudoexperiments}. 
The results are laid out in Section~\ref{sec:results}, depending on the expected number of collected \Bs decays and the presumed tagging efficiency. 

\section{Differential Decay Rate}
\label{sec:differential-decay-rate}

\begin{figure}[tb!]

    \centering
    \includegraphics[width=0.5\textwidth]{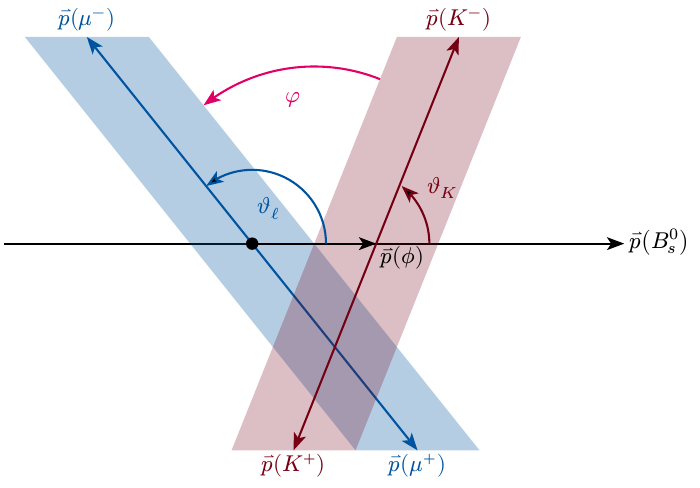}
    \caption{Visualisation of the helicity angles describing the \decay{\Bs}{\Pphi\mumu} decay. \tk marks the angle between the momentum vectors of the negatively charged kaon and the \Bs in the di-kaon rest frame. The angle between the momentum vector of the negatively charged muon and the \Bs in the di-muon rest frame is called \tl. The angle between the planes spun by the di-kaon and di-muon system is called $\varphi$. }
    \label{fig:helicity-angles}

\end{figure}

The differential decay rate for \decay{\PB}{$V$ \ell^+\ell^-} decays can be expressed in terms of four kinematic quantities: \qsq and  three helicity angles which we call \tk, \tl, and $\varphi$~\cite{Descotes-Genon:2015hea}. 
Here we disregard any dependence on the di-hadron mass which is valid considering the narrow width of the \Pphi meson.  
The helicity angles are depicted in Figure~\ref{fig:helicity-angles}. 
The angle \tk (\tl) is given by the angle between the momenta of the negatively charged kaon (muon) and the \Bs in the di-kaon (di-muon) rest frame. 
The angle between the planes formed by the di-kaon and di-muon systems in the rest frame of the \PB is called $\varphi$. 
This convention follows for example Refs.~\cite{LHCb-PAPER-2021-022,Descotes-Genon:2015hea}. 

In the special case for the \decay{\Bs}{\Pphi \mumu} decays under study, the decay-time is also needed to describe the decay-rate due to time-dependent \CP-violation effects that arise because the final state is a \CP-conjugate state. 
Following Refs.~\cite{Descotes-Genon:2015hea,Gratrex:2015hna}, the spin-summed decay-rate depending on \qsq, the helicity-angles and the decay-time, $t$, can be expressed using the angular coefficients $J_i$, $h_i$, and $s_i$ as
\begin{equation}
  \begin{aligned}
    \frac{d^5 \Gamma(\decay{\BsorBsbar}{\Pphi\mumu})}{d\qsq d\ctk d\ctl d\varphi dt} & = \frac{9e^{-\Gamma t}}{64\pi} \sum_i \Big \{ [J_i(\qsq) + \zeta_i\bar{J}_i(\qsq)] \cosh(y \Gamma t) - h_i(\qsq) \sinh(y \Gamma t) \\
    & \hspace{-3.5em} \varpm \left( [J_i(\qsq) - \zeta_i\bar{J}_i(\qsq) ]  \cos(x \Gamma t) - s_i(\qsq) \sin(x \Gamma t) \right) \Big \} f_i(\ctk, \ctl, \varphi),
  \end{aligned}
\end{equation}
where the sum runs  over $i \in [1s, 1c, 2s, 2c, 3, 4, 5, 6s, 6c, 7, 8, 9]$ and the mixing parameters $x$ and $y$ are defined as  $x = \Delta m / \Gamma$, $y = \Delta\Gamma/\Gamma$, where $\Delta m$ and $\Delta\Gamma$ refer to the difference in mass and width respectively between the heavy and light mass eigenstates. 
The \CP-conjugated coefficients $\bar{J}_i$ are obtained by complex conjugation of all weak-phases. 
Due to the same definition of the helicity angles\footnote{The negatively charged lepton and kaon are used to define the helicity angles for both \Bs and \Bsb. } for the \Bs and \Bsb, the factor $\zeta$ is introduced in the decay-rate, which is given by $\zeta_i = -1$ for $i = 5, 6s, 6c, 8, 9$ and $\zeta_i = 1$ otherwise. 
The angular functions $f_i$ are given by
\begin{alignat*}{2}
    f_{1s} &= \sin^2\tk, &\qquad f_{1c} &= \cos^2\tk, \\
    f_{2s} &= \sin^2\tk \cos2\tl, &\qquad f_{2c} &= \cos^2\tk \cos2\tl, \\
    f_3 &= \sin^2\tk\sin^2\tl\cos2\varphi, &\qquad f_4 &= \sin2\tk \sin2\tl \cos\varphi, \\
    f_5 &= \sin2\tk \sin\tl \cos\varphi, &\qquad &\\
    f_{6s} &= \sin^2\tk \cos\tl, &\qquad f_{6c} &= \cos^2\tk \cos\tl, \\
    f_7 &= \sin2\tk \sin\tl \sin\varphi, &\qquad f_8 &= \sin2\tk\sin2\tl\sin\varphi, \\
    f_9 &= \sin^2\tk \sin^2\tl \sin2\varphi. &\qquad &
\end{alignat*}

A parameterisation of the $J_i$, $h_i$ and $s_i$ coefficients using the transversity amplitude formalism is given in Appendix~\ref{app:angular-amplitudes}, where we follow Ref.~\cite{Descotes-Genon:2015hea}.
We define the normalised angular coefficients\textbf{}
\begin{align}\label{eq:obses-tdep}
    &\begin{aligned}
    S_i^t & = \frac{J_i + \bar{J}_i}{\mathcal{I}}, \\
    \mathcal{H}_i & = \frac{h_i}{\mathcal{I}}\text{, and} \\
    \end{aligned}
    \hspace{2cm}
    \begin{aligned}
    A_i^t & = \frac{J_i - \bar{J}_i}{\mathcal{I}}, \\
    \mathcal{Z}_i & = \frac{s_i}{\mathcal{I}},
    \end{aligned}
\end{align}
where $\mathcal{I}$ is a normalisation constant that is related to the total decay-rate of both \Bs and \Bsb in a given \qsq-region
\begin{equation}
    \begin{aligned}
    \frac{d\Gamma + \overline{\Gamma}}{d\qsq} & = \iiiint \frac{d^5 \Gamma(\decay{\Bs}{\Pphi\mumu}) + d^5 \Gamma(\decay{\Bsb}{\Pphi \mumu})}{d\qsq d\ctk d\ctl d\varphi dt} d\ctk d\ctl d\varphi dt \\
    & = \frac{1}{4\Gamma(1-y^2)}\left( 6 (J_{1s} + \bar{J}_{1s}) + 3 (J_{1c} + \bar{J}_{1c}) - 2 (J_{2s} + \bar{J}_{2s}) - (J_{2c} + \bar{J}_{2c}) \right. \\
    & \left. \hspace{2.65cm} - y( 6 h_{1s} + 3 h_{1c} - 2 h_{2s} - h_{2c} ) \right) \\
    & \equiv \frac{\mathcal{I}}{\Gamma(1 -y^2)}. 
    \end{aligned}
\end{equation}
We therefore write the time-dependent normalised differential decay rates for \BsorBsbar in the presence of mixing as 
\begin{equation}\label{eq:full-tdep-flavourdep-decay-rate}
  \begin{aligned}
    & \frac{1}{d(\Gamma + \overline{\Gamma})/d\qsq} \frac{d^5 \Gamma(\decay{\BsorBsbar}{\Pphi\mumu})}{d\qsq d\ctk d\ctl d\varphi dt} = \frac{(1-y^2)9\Gamma e^{-\Gamma t}}{64\pi} \Bigg\{ \\
    & \sum_{i\in[\zeta_i = +1]} \Big [ S_i^t \cosh(y \Gamma t) - \mathcal{H}_i \sinh(y \Gamma t) \varpm ( A_i^t \cos(x \Gamma t) - \mathcal{Z}_i \sin(x \Gamma t)) \Big ] \\ 
    & \hspace{1.5cm} \cdot f_i(\ctk, \ctl, \varphi) \\
    & + \sum_{i\in[\zeta_i = -1]}  \Big [ A_i^t \cosh(y \Gamma t) - \mathcal{H}_i \sinh(y \Gamma t) \varpm ( S_i^t \cos(x \Gamma t) - \mathcal{Z}_i \sin(x \Gamma t)) \Big ] \\ 
    & \hspace{1.9cm} \cdot f_i(\ctk, \ctl, \varphi) \Bigg\}.
  \end{aligned}
\end{equation}

Integrating over the decay-time reduces the number of coefficients resulting in 
\begin{equation}
    \frac{d^4 \Gamma(\decay{\BsorBsbar}{\Pphi\mumu})}{d\qsq d\ctk d\ctl d\varphi} = \frac{9}{32\pi}\sum_{i} \opttilde{I}_i(\qsq) f_i(\ctk, \ctl, \varphi),
\end{equation}
with $I_i$ and $\tilde{I}_i$ denoting time-integrated angular coefficients, where once again $\tilde{I}_i$ is  defined in terms of the \CP-conjugate terms, $\bar{I}$, as  $\tilde{I}_i = \zeta_i \bar{I}_i$. 
The LHCb collaboration has measured the untagged sum of the time-integrated decay-rates~\cite{LHCb-PAPER-2021-022}, which yields a mixture of \CP-symmetric and asymmetric terms, denoted $S_i \propto I_i + \bar{I}_i$ and $A_i \propto I_i - \bar{I}_i$, respectively. 
Following Ref.~\cite{Descotes-Genon:2015hea} the $S_i$ and $A_i$ observables can be expressed using the $J_i$, $h_i$, and $s_i$ coefficients as
\begin{align}
    \langle S_i \rangle = \frac{\langle I_i + \bar{I}_i \rangle}{\langle d(\Gamma + \overline{\Gamma})/d\qsq \rangle} & = \frac{1}{\langle\mathcal{I}\rangle} \begin{cases} 
    \langle J_i + \bar{J}_i - yh_i\rangle, \text{ for } i \in [1s,1c,2s,2c,3,4,7] \\
    \frac{1 - y^2}{1 + x^2}\langle J_i + \bar{J}_i - xs_i \rangle, \text{ for } i \in [5,6s,6c,8,9]
    \end{cases}\text{, and} \\
    \langle A_i \rangle = \frac{\langle I_i - \bar{I}_i \rangle}{\langle d(\Gamma + \overline{\Gamma})/d\qsq \rangle} & = \frac{1}{\langle\mathcal{I}\rangle} \begin{cases}
    \frac{1 - y^2}{1 + x^2}\langle J_i - \bar{J}_i - xs_i \rangle, \text{ for } i \in [1s,1c,2s,2c,3,4,7] \\
    \langle J_i - \bar{J}_i - yh_i \rangle, \text{ for } i \in [5,6s,6c,8,9]
    \end{cases}.
\end{align}

Whereas the $S_i^t$ coefficients, introduced in Eq.~\ref{eq:obses-tdep}, describe the sum of the time-independent $J_i$ and $\bar{J}_i$ coefficients, similar to the observables in the decay \decay{\Bz}{\Kstarz \mumu}, the $S_i$ observables are related to the integral over the time-dependent terms, including the $J_i$, $\bar{J}_i$, $h_i$, and $s_i$ coefficients. 

Another common simplification of the decay-rate  description can be  obtained by neglecting the masses of the leptons, i.e. assuming $m_\ell^2 \ll \qsq$, and by assuming there are no scalar or tensor amplitudes. 
This yields the following relations between the angular coefficients
\begin{align}\label{eq:massless-approx}
    &\begin{aligned}
    J_{1s} & = 3 J_{2s}, \\
    h_{1s} & = 3 h_{2s}, \\
    s_{1s} & = 3 s_{2s}\text{, and }      
    \end{aligned} 
    \hspace{1cm}
    \begin{aligned}
    J_{1c} & = - J_{2c}, \\
    h_{1c} & = - h_{2c}, \\
    s_{1c} & = - s_{2c},
    \end{aligned}
\end{align}
and has the additional corollary that all $6c$ coefficients are zero. The same relations hold for the $S_i^{(t)}$, $W_i^{(t)}$, $\mathcal{H}_i$, and $\mathcal{Z}_i$ observables. 
Using these approximations simplifies the parameter space, and reduces the number of highly correlated parameters present in the analysis. 
Requiring that the total decay-rate is normalised imposes an additional constraint on the angular coefficients 
\begin{equation}
    \begin{aligned}
        1 & = \frac{1}{d(\Gamma + \overline{\Gamma})/d\qsq}\iiiint \frac{d^5 \Gamma(\decay{\Bs}{\Pphi \mumu}) + d^5\Gamma(\decay{\Bsb}{\Pphi \mumu})}{d\qsq d\ctk d\ctl d\varphi dt} d\ctk d\ctl d\varphi dt \\
        & = \frac{1}{4} \left[ 6 S^t_{1s} + 3 S^t_{1c} - 2 S^t_{2s} - S^t_{2c} -y ( 6 \mathcal{H}_{1s} + 3 \mathcal{H}_{1c} - 2 \mathcal{H}_{2s} - \mathcal{H}_{2c} ) \right],
    \end{aligned}
\end{equation}

which simplifies in the massless lepton case to
\begin{equation}\label{eq:normalisation-condition}
    1 = \left[ \frac{4}{3} S^t_{1s} + S^t_{1c} - y\left( \frac{4}{3} \mathcal{H}_{1s} + \mathcal{H}_{1c} \right) \right]. 
\end{equation}
Using this normalisation condition, we can express one of the parameters in terms of the others. 
We choose to parameterise $S^t_{1s}$ as a function of $S^t_{1c}$, $\mathcal{H}_{1s}$, and $\mathcal{H}_{1c}$. 
This choice preserves the parameter $S^t_{1c}$, which, under the assumption of massless leptons, is directly related to the longitudinal polarisation fraction $F_L$. 

The \CP-asymmetry $A_\CP$ is defined as the difference between the differential decay rates of the \Bs and \Bsb mesons normalised to the \CP-averaged rate. 
Because the \Bs meson undergoes quantum mechanical oscillations and the studied decay mode has a \CP-conjugate final state, this asymmetry can be determined solely from the angular distributions. 
This arises because both the $A_X$ and $\tilde{A}_X$ decay amplitudes enter explicitly in the decay rates of \Bs and \Bsb mesons, as can be seen from Eq.~\ref{eq:full-tdep-flavourdep-decay-rate}.
The corresponding expression for the integrated differential decay-rate of the \Bs and \Bsb are given by
\begin{equation}
    \begin{aligned}
    \frac{1}{d(\Gamma + \overline{\Gamma})/d\qsq}\iiiint &\frac{d^5\Gamma(\decay{\BsorBsbar}{\Pphi \mumu})}{d\qsq d\ctk d\ctl d\varphi dt}d\ctk d\ctl d\varphi dt = 
    \\ & = \frac{1}{2} \left\{ \left[ \frac{4}{3} S^t_{1s} + S^t_{1c} - y\left( \frac{4}{3} \mathcal{H}_{1s} + \mathcal{H}_{1c} \right) \right] \right. \\ 
    & \hspace{1.725em} \left. \varpm \frac{1-y^2}{1 + x^2} \left[ \frac{4}{3} A^t_{1s} + A^t_{1c} - y \left( \frac{4}{3}\mathcal{Z}_{1s} + \mathcal{Z}_{1c} \right) \right] \right\}
    \end{aligned}
\end{equation}
in the massless limit. 
From the difference of the integrals of the \Bs and \Bsb decay rates we identify
\begin{equation}
    A_\CP = \frac{1-y^2}{1 + x^2} \left[ \frac{4}{3} A^t_{1s} + A^t_{1c} - y \left( \frac{4}{3}\mathcal{Z}_{1s} + \mathcal{Z}_{1c} \right) \right].
\end{equation}
The same is not true for a decay into a \CP specific final state such as \decay{\Bz}{\Kstarz(\rightarrow K^+\pi^-)\mumu}, for which $A_\CP$ needs to be determined in an extended maximum-likelihood fit including the signal yields. 
Note that we assume that \Bs and \Bsb mesons are produced in equal quantities, which is the case at the LHC~\cite{LHCB-PAPER-2016-062}. 
For small production asymmetries, experimental measures can be implemented to mitigate the effect on the observables defined above. 

\section{Optimised Observables}\label{sec:optimised-observables}

The angular coefficients depend on transversity amplitudes, which in turn depend on hadronic form-factors describing the  $\Bs \rightarrow \Pphi$ transition. 
As the calculation of these form-factors is associated with sizeable uncertainties, it can potentially be beneficial to define so-called ``optimised'' observables, which have a reduced dependence on the form-factors. 
This is done for the \decay{\Bz}{\Kstarz\mumu} decays where the $J_i$ are redefined in such a way that the form-factors cancel at leading order, as discussed in Refs.~\cite{Descotes-Genon:2012isb,Matias:2012xw,Egede:2008uy,Boer:2014kda,Descotes-Genon:2015hea}. 
For the time-dependent angular coefficients $h_i$ and $s_i$ a discussion of optimised observables is given in Ref.~\cite{Descotes-Genon:2015hea}, where two such observables are identified as particularly promising
\begin{align}
    Q_8^- & = \frac{s_8}{\sqrt{-2\langle J_{2c} + \bar{J}_{2c} \rangle [ 2 \langle J_{2s} + \bar{J}_{2s} \rangle - \langle J_3 + \bar{J}_3 \rangle ]}}, \\
    Q_9 & = \frac{s_9}{2\langle J_{2s} + \bar{J}_{2s} \rangle}. 
\end{align}

The methodology of constructing the optimised observables relies on finding a normalisation with a similar dependence on the form-factors, cancelling their contributions at leading order. 
In other words, the dependence of the coefficients on the different transversity amplitudes dictates how the optimised observables will be constructed. 
A more complete discussion of the transversity amplitudes and the angular coefficients can be found in Appendix~\ref{app:angular-amplitudes}, but the general structure of the dependence is given by
\begin{align}
    h_{1s} & \propto \mathcal{R}\{ \tilde{A}_\perp A_\perp^* + \tilde{A}_{\parallel} A_\parallel^* \} + \mathcal{O}(m_\ell^2/q^2), \hspace{6em} h_{2s} \propto \mathcal{R}\{ \tilde{A}_\perp A_\perp^* + \tilde{A}_\parallel A_\parallel^* \} ,\\
    h_{1c} & \propto \mathcal{R}\{ \tilde{A}_0 A_0^* \} + \mathcal{R}\{ \tilde{A}_S A_S^*\} + \mathcal{O}(m_\ell^2/q^2), \hspace{4.35em} h_{2c} \propto \mathcal{R}\{ \tilde{A}_0 A_0^* \} ,\\
    h_3 & \propto \mathcal{R}\{ \tilde{A}_\perp A_\perp^* - \tilde{A}_\parallel A_\parallel^* \}, \hspace{11.85em} h_4 \propto \mathcal{R}\{ \tilde{A}_0 A_\parallel^* \} ,\\
    h_5 & \propto \mathcal{R}\{ \tilde{A}_0 A_\perp^* \} + \mathcal{O}(m_\ell/\sqrt{q^2}), \hspace{8.8em} h_{6s} \propto \mathcal{R}\{ \tilde{A}_\parallel A_\perp^* \} ,\\
    h_7 & \propto \mathcal{I}\{ \tilde{A}_0 A_\parallel \} + \mathcal{O}(m_\ell/\sqrt{q^2}), \hspace{9.5em} h_8 \propto \mathcal{I}\{ \tilde{A}_0 A_\perp^* \} ,\\
    h_9 & \propto \mathcal{I}\{ \tilde{A}_\parallel A_\perp^* \},
\end{align}
with similar dependence for the $s_i$ where $\mathcal{R}$ is exchanged with $\mathcal{I}$ and vice-versa\footnote{This is a simplified version of the transformation between the $h_i$ and $s_i$ with the more accurate version also discussed in Appendix~\ref{app:angular-amplitudes}. }. 
The $J_i$ share the same general dependence and would therefore qualify to cancel the form-factor dependence well. 
Apart from the coefficients $1$ and $2$, the $J_i$ exhibit zero points along \qsq, leading to a divergence of an observable normalised using them. 
Therefore, we construct the following observables where the form-factor dependence cancels at leading order and the normalisation is strictly positive across \qsq
\begin{align}\label{eq:opt-obses}
    \mathcal{M}_{1s} & = \frac{h_{1s}}{\langle J_{2s} + \bar{J}_{2s}\rangle }, \qquad \mathcal{M}_{2s} = \frac{h_{2s}}{\langle J_{2s} + \bar{J}_{2s}\rangle}, \\
    \mathcal{M}_{1c} & = \frac{-h_{1c}}{\langle J_{2c} + \bar{J}_{2c}\rangle}, \qquad \mathcal{M}_{2c} = \frac{-h_{2c}}{\langle J_{2c} + \bar{J}_{2c}\rangle}, \\
    \mathcal{M}_{3} & = \frac{h_3}{2\langle J_{2s} + \bar{J}_{2s}\rangle}, \qquad \mathcal{M}_{4}^- = \frac{-h_4}{\sqrt{-\langle J_{2c} + \bar{J}_{2c}\rangle [2 \langle J_{2s} + \bar{J}_{2s}\rangle - \langle J_3 + \bar{J}_3\rangle ] }}, \\
    \mathcal{M}_5^- & = \frac{h_5}{\sqrt{\langle J_{2c} + \bar{J}_{2c}\rangle [\langle J_3 + \bar{J}_3\rangle - 2 \langle J_{2s} + \bar{J}_{2s}\rangle] } }, 
    \\
    \mathcal{M}_{6s} & = \frac{h_{6s}}{2 \langle J_{2s} + \bar{J}_{2s}\rangle}, \text{ and} \quad Q_7^- = \frac{-s_7}{\sqrt{ -\langle J_{2c} + \bar{J}_{2c}\rangle (2 \langle J_{2s} + \bar{J}_{2s}\rangle - \langle J_3 + \bar{J}_3\rangle] }}.
\end{align}
Here the brackets indicate how the denominator is formed for a measurement which integrates over a given \qsq-region. 

The set of observables is completed by the coefficients proportional to the imaginary part of the amplitudes, which are defined equivalently, yielding a $\mathcal{M}_i$ for each $h_i$ and a $Q_i$ for every $s_i$. 
In addition to the observables defined in Eq.~\ref{eq:opt-obses}, also the adapted observables $\mathcal{M}_{4,5,7,8}^{\prime}$ and $Q_{4,5,7,8}^\prime$ are studied which are defined as 
\begin{align}\label{eq:opt-obses-prime}
    M_{4,8}^\prime & = \frac{h_{4,8}}{\sqrt{F_L F_T}} = \frac{h_{4,8}}{\sqrt{-\langle J_{2c} + \bar{J}_{2c}\rangle\langle J_{2s} + \bar{J}_{2s}\rangle}}, \quad \quad Q_{4,8}^\prime = \frac{s_{4,8}}{\sqrt{F_L F_T}}, \\
    M_{5}^\prime & = \frac{h_5}{2\sqrt{F_L F_T}} = \frac{h_5}{2\sqrt{-\langle J_{2c} + \bar{J}_{2c}\rangle\langle J_{2s} + \bar{J}_{2s}\rangle}}, \quad \quad Q_5^\prime = \frac{s_5}{2\sqrt{F_L F_T}}, \\
    M_{7}^\prime & = \frac{-h_7}{2\sqrt{F_L F_T}} = \frac{-h_7}{2\sqrt{-\langle J_{2c} + \bar{J}_{2c}\rangle\langle J_{2s} + \bar{J}_{2s}\rangle}}, \text{ and } Q_7^\prime = \frac{-s_7}{2\sqrt{F_L F_T}}.
\end{align}
This primed observable basis takes inspiration from Ref.~\cite{Descotes-Genon:2012isb}. 
In addition to the optimised observables for the time-dependent observables $\mathcal{H}_i$ and $\mathcal{Z}_i$, also optimised observables for the $S^t_i$ and $A^t_i$ can be constructed. 
These also follow the work in Ref.~\cite{Descotes-Genon:2012isb}, where \decay{\Bz}{\Kstarz\mumu} decays are discussed, up to the difference that we define \CP-symmetric and \CP-asymmetric versions, $P^t_i$ and $_AP^t_i$, respectively, and the fact these observables are \emph{not} time-integrated for the case of \decay{\Bs}{\phi}{\mumu}, which we emphasis with the $t$ superscript.  
We define 
\begin{alignat}{2}\label{eq:opt-obses-pi}
    P_1^t &= \frac{J_3 + \bar{J}_3}{2 \langle J_{2s} + \bar{J}_{2s}\rangle}, &\qquad P_4^{t\prime} &= \frac{ J_4 + \bar{J}_4}{\sqrt{-\langle J_{2c} + \bar{J}_{2c}\rangle \langle J_{2s} + \bar{J}_{2s}\rangle}}, \\
    P_2^t &= \frac{J_{6s} + \bar{J}_{6s}}{8 \langle J_{2s} + \bar{J}_{2s}\rangle}, &\qquad P_5^{t\prime} &= \frac{J_{5} + \bar{J}_{5}}{2\sqrt{-\langle J_{2c} + \bar{J}_{2c}\rangle\langle J_{2s} + \bar{J}_{2s}\rangle}}, \\
    P_3^t &= \frac{J_9 + \bar{J}_9}{4 \langle J_{2s} + \bar{J}_{2s} \rangle},  &\qquad P_6^{t\prime} &= \frac{-(J_7 + \bar{J}_7)}{2\sqrt{-\langle J_{2c} + \bar{J}_{2c}\rangle \langle J_{2s} + \bar{J}_{2s}\rangle}}, \\
    P_S^t &= \frac{J_{6c} + \bar{J}_{6c}}{\langle J_{2s} + \bar{J}_{2s}\rangle}\text{, and}  &\qquad 
    P_8^{t\prime} &= \frac{J_8 + \bar{J}_8}{\sqrt{-\langle J_{2c} + \bar{J}_{2c}\rangle \langle J_{2s} + \bar{J}_{2s}\rangle}},
\end{alignat}
and the corresponding $_AP_i^t$ which are obtained by exchanging the plus in the numerator with a minus. 
While the asymmetries are close to zero in the SM, their value can change if for example the imaginary parts of the Wilson coefficients are altered. 
The symmetric $P^{t}_i$ observables in \decay{\Bs}{\Pphi \mumu} behave similar to the $P_i$ for the \decay{\Bz}{\Kstarz \mumu} decays, introduced in Ref.~\cite{Descotes-Genon:2012isb}, and are similarly sensitive to variations in the vector coupling. 
This similarity, however, only holds when a time-dependent analysis is performed, as the final-state of the \Bs decay is a \CP-conjugate state. 
A tagged, time-integrated analysis of \decay{\Bs}{\Pphi \mumu} decays would  yield, for example
\begin{align}
  P^{\prime}_5 \propto S_5 \propto   \int dt [ (J_5 + \bar{J}_5) e^{-\Gamma t} \cos(x \Gamma t) + s_5 e^{-\Gamma t} \sin(x \Gamma t) ],
\end{align} from which is can be seen that, due the large values of $x$ in the \Bs system, the oscillatory parts integrate to a small value. 
This results in a strong suppression of the sensitivity to underlying short-distance effects. 
In contrast, in a time-dependent analysis, yielding either the $S^t_5$  (or $P^{t\prime}_5$) observable, results in a sensitivity to short-distance effects similar to that of $S_5$ (or $P^{\prime}_5$) for \decay{\Bz}{\Kstarz\mumu}. 
Note in a time-independent analysis, the $P_i$ and $_AP_i$ are obtained similarly as in Eq.~\ref{eq:opt-obses-pi}, but exchanging the $J_i$ with the $I_i$. 
The sensitivity to these and the other angular coefficients are evaluated using realistic pseudoexperiments, as described in the next section. 

\section{Realistic Projections using Pseudoexperiments} 
\label{sec:pseudoexperiments}

Our extrapolations assume a dedicated heavy flavour experiment at a proton-proton collider like the LHC, and we therefore use the performance of the \lhcb and \lhcb upgrade I detectors as a concrete performance benchmark. 
All the performance numbers used for this work are extracted from publicly-available sources. 

The sensitivity to the different angular observables are determined using a large ensemble of pseudoexperiments where pseudodata are generated according to the normalised decay-rate defined in Eq.~\ref{eq:full-tdep-flavourdep-decay-rate}. 
In addition to the helicity angles, \qsq, and the decay-time, the invariant four-body mass is generated following a Gaussian with mean of $5366\mevcc$ and width of $15\mevcc$ which closely resembles the distributions observed in Refs.~\cite{LHCb-PAPER-2021-014,LHCb-PAPER-2021-022,LHCb-PAPER-2024-032}. 
The pseudodata are generated using jax~\cite{jax2018github} with an accept-reject method. 
The expectation of the $S^t_i$, $A^t_i$, $\mathcal{H}_i$, and $\mathcal{Z}_i$ are computed using flavio~\cite{straub:2018flavio}, where we added the $S^t_i$, $A^t_i$, $\mathcal{H}_i$, and $\mathcal{Z}_i$ observables for this work. 
We assume that the dominating background source to the analysis in the signal region $5340 < m(K^+K^-\mumu) < 5400\mevcc$ is at the $10\,\%$ level and formed by random combinations of final state particles in the detector, so-called combinatorial background. 
This is consistent with what is observed in Refs.~\cite{LHCb-PAPER-2021-014,LHCb-PAPER-2021-022,LHCb-PAPER-2024-032}. 
Therefore, the background distributions are generated according to flat distributions in the helicity angles assuming factorisation between the angles and an exponential function in the invariant \Kp\Km\mumu mass. 
Again, the assumption that the angles factorise is consistent with previous analyses of this mode by the \lhcb collaboration~\cite{LHCb-PAPER-2021-014,LHCb-PAPER-2021-022,LHCb-PAPER-2024-032}. 
The background in the decay-time is assumed to factorise from the invariant four-body mass and the helicity angles and is modelled using the sum of two exponential functions. 
The sum of two exponential functions accounts for both charm and beauty components of the combinatorial background, where we assume that the ratio of charm to beauty background is three-to-one. 

Flavour-tagging at hadronic heavy-flavour experiments is usually implemented in two separate algorithms, a same-side (SS) and opposite-side (OS) tagger, see for example Refs.~\cite{LHCb-PAPER-2011-027,LHCb-PAPER-2015-056}. 
The same-side algorithm studies kaons produced in the hadronisation process, indicating the flavour of the \Bs meson at production. 
On the other hand, the opposite-side tagger relies on reconstructing the other \PB decay, inferring its flavour, and therefore the flavour of the signal \Bs at production. 
The flavour-tag decision is randomly sampled following a tagging efficiency, $\epsilon_{\rm tag}$ and mistag-probability, $\omega$. 
The tagging efficiency refers to the fraction of tagged candidates and the mistag-probability gives the number of wrongly tagged candidates relative to all tagged candidates. 
The values used are summarised in Table~\ref{tab:tagging-performance} and follow the general \lhcb performance observed in Run~1 and 2~\cite{LHCb:2025okz,Prouve:2024xgq,LHCb-PAPER-2011-027,LHCb-PAPER-2015-056}. 
With these numbers, the OS and SS taggers are each given an effective tagging power of approximately $\epsilon_{\rm eff} = \epsilon_{\rm tag} (1 - 2 \omega)^2 \approx 3\,\%$, leading to a combined tagging power at the $6\,\%$ level. 
As the mistag-probability usually declines with larger particle density, for our extrapolations we also explore larger mistag-probabilities, which lead to an effective tagging power at the $4\,\%$ level. 

\begin{table}[tb!]
    \caption{Values used for the tagging efficiency and mistag-probability for the data generation, extracted from~\cite{LHCb:2025okz,Prouve:2024xgq,LHCb-PAPER-2011-027,LHCb-PAPER-2015-056}. The tagging efficiency is kept constant and two scenarios for the mistag-probabilities are explored leading to two different effective tagging powers. The mistag-probability for the OS and SS tag are assumed to be uncorrelated. }
    \label{tab:tagging-performance}
    \centering
    \begin{tabular}{lccc}
    \toprule
        Algorithm & $\epsilon_{\rm tag}$ & $\omega$ & $\epsilon_{\rm eff}$ \\ \midrule
        OS-tag    & $40\,\%$             & $36\,\%$ ($39\,\%$) & $3\,\%$ ($2\,\%$) \\
        SS-tag    & $80\,\%$             & $40\,\%$ ($42\,\%$) & $3\,\%$ ($2\,\%$) \\ \bottomrule
    \end{tabular}
\end{table}
As decay-time measurements rely on the determination of the \Bs decay-vertex, which is also typically used for candidate selection, the measured \Bs decay-time distribution is expected to be morphed and no longer exponential. 
This phenomenon is described in this work using an acceptance function which is multiplied by the normalised decay-rate. 
We derive a decay-time acceptance which closely resembles the acceptance in Ref.~\cite{LHCb-PAPER-2020-042} using an analytical function of the form
\begin{equation}\label{eq:time-acceptance}
    \epsilon(t) = \frac{A}{1 + B e^{-\delta_t t}},
\end{equation}
where $A = 1.05$, $B = 13.3$, and $\delta_t=8.1\,\mathrm{ps}^{-1}$. 
This process is further explained in Appendix~\ref{app:time-acceptance}. 
Similar to the morphing of the decay-time acceptance, there can be detector effects or selection requirements which impact the acceptance of the helicity angles. 
Given that modelling these effects is non-trivial and would require very detector-specific assumptions, for this work, we assume a flat acceptance in the helicity angles. 

In order to account for the finite resolution of the decay-time measurement, an additional dilution factor is introduced. 
The dilution is written as
\begin{equation}\label{eq:time-dilution}
    \mathcal{D}(t, \sigma_t) = \frac{1}{\sqrt{2\pi\sigma_t^2}}e^{-\frac{t^2}{2\sigma_t^2}},
\end{equation}
where the decay-time, $t$, and the estimated decay-time uncertainty, $\sigma_t$, are used. 
While the decay-time resolution varies across different experiments, we will again use the typical \lhcb performance as a benchmark. 
In typical LHCb analyses, decay-time resolutions of $40\,\mathrm{fs}$ are achieved~\cite{LHCb-PAPER-2013-006}. 
Usually, smaller decay-times have better decay-time resolution, as for large decay-times the momentum resolution contributes more to the uncertainty budget. 
This is taken into account by making the width of the Gaussian in Eq.~\ref{eq:time-dilution} linearly dependent on the decay-time
\begin{equation}\label{eq:time-res-dependence}
    \sigma_t = 37.5\,\mathrm{fs} + \frac{t\,[\mathrm{ps}]}{2\times 10^{3}}\,\mathrm{fs}.
\end{equation} 
Note, this corresponds to a reparameterisation of $\sigma_t$ in Eq.~\ref{eq:time-dilution}, not a per-candidate resolution, therefore it cannot induce a Punzi-bias~\cite{Punzi:2003wze}.
We stress that Eq.~\ref{eq:time-res-dependence} is based on publicly available data from the \lhcb collaboration and only acts as an approximation of the typical time-resolution expected with the \lhcb experiment. 

\begin{figure}[tb!]
    \centering
    \includegraphics[width=\textwidth]{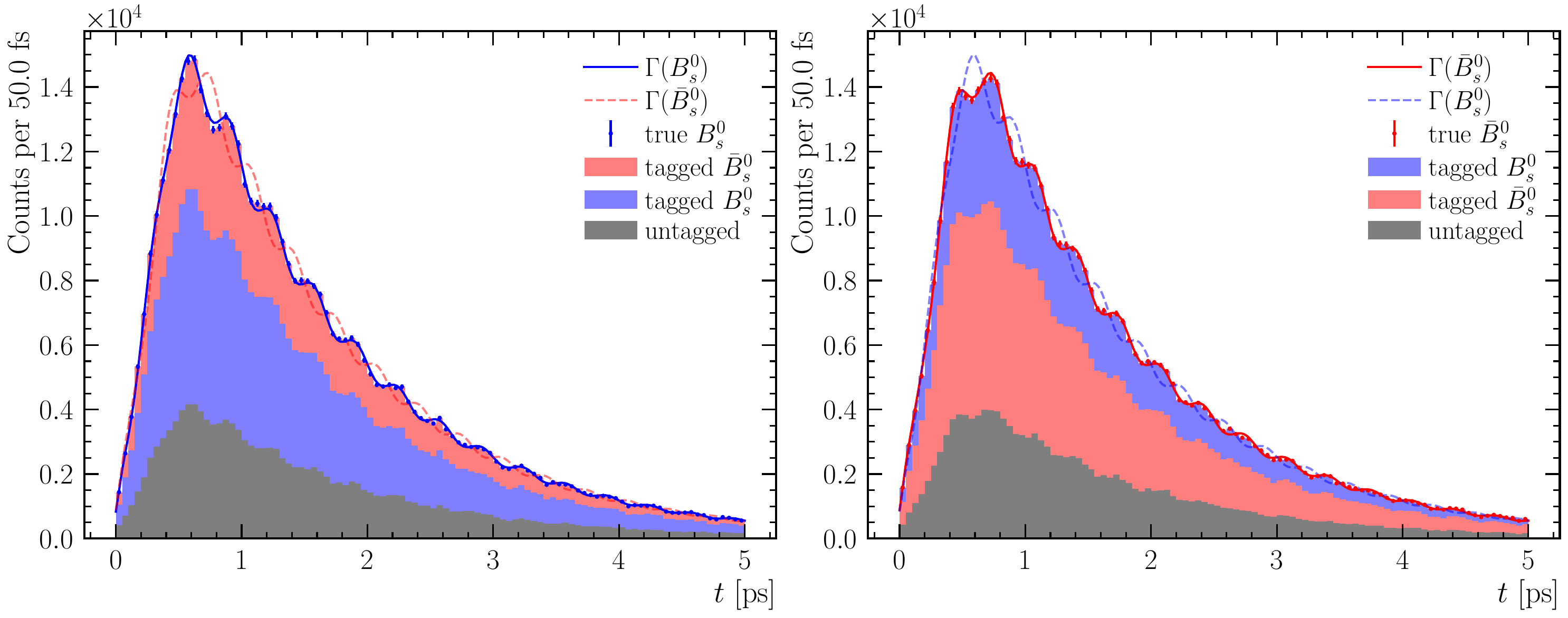}

    \caption{Example of the generated decay-time distribution for the \Bs (left) and \Bsb (right) candidates with an overlay of the corresponding decay-rate. The simulated tagging decision is indicated in the histograms, showing a tag for the \Bs (blue) or the \Bsb (red) or no conclusive decision (black). The category tagged \Bs (\Bsb) is made up from events, where both taggers agree on the classification of the event. Untagged events and events where the tagging decisions disagree make up the "untagged" category. }
    \label{fig:pseudo-data}

\end{figure}

An example of the generated decay-time distribution is shown in Fig.~\ref{fig:pseudo-data} for the \Bs and the \Bsb candidates. 
In addition, our simulated tagging decisions are illustrated, where tagged \Bs refers to a \Bs tag from either the SS or OS tagging algorithms, where the other tagger either agrees or yields no decision. 
Candidates without a tagging decision, or with differing  decisions between the two taggers, are labelled as ``untagged''. 
The characteristic oscillations of the \Bs and the \Bsb decay-rates can be seen in the figure, where $10^{5}$ signal candidates are displayed without any background pollution. 

For the pseudoexperiments used to extract the observable sensitivity, we estimate the amount of expected signal candidates by scaling the yields reported in Ref.~\cite{LHCb-PAPER-2021-014} with the expected increase in luminosity from Run~3 and Run~4, quoted in Ref.~\cite{LHCb-DP-2022-002}. 
As the \lhcb experiment was upgraded for Run~3 and beyond, removing the hardware-trigger, we assume an increase in the efficiency of $20\,\%$ with respect to the Run~2 analysis. 
This assumption is motivated by a preliminary study comparing the observed yields per \invfb for the decays \decay{\Bp}{\Kp\mumu} and \decay{\Bz}{\Kstarz\mumu} in early 2024 data~\cite{LHCB-FIGURE-2024-022} with those reported at the end of Run~2 after the full selection~\cite{LHCb-PAPER-2022-045}, indicating improvements in the range of $5\text{--}25\,\%$.

We also study the prospects after the conclusion of the operation of the LHC, where the LHCb upgrade II aims to have collected a dataset corresponding to an integrated luminosity of $300\invfb$~\cite{LHCb-TDR-023}, leading to an unprecedented sample of recorded \PB decays. 
The generated signal yields are listed in Table~\ref{tab:generated-yields} for the different scenarios and LHC Runs. 

\begin{table}[tb!]

    \caption{Generated signal yields for the two different luminosity scenarios at the end of Run~3, Run~4, and Run~5 used to estimate the sensitivity on the angular observables. 
    Note that all \qsq-regions also includes the region from $[6.0, 8.0]$ and $[11.0, 12.5]\,\mathrm{GeV}^2/c^4$. As explained in the text, the numbers are extrapolated following the yields quoted in Ref.~\cite{LHCb-PAPER-2021-014}. }
    \label{tab:generated-yields}
    \centering
\begin{tabular}{cccc}
\toprule
  $q^2$-Region [$\gevgevcccc$]  &  $\mathcal{L}_{\mathrm{int.}} = 23\,\mathrm{fb}^{-1}$  &  $\mathcal{L}_{\mathrm{int.}} = 50\,\mathrm{fb}^{-1}$  &  $\mathcal{L}_{\mathrm{int.}} = 300\,\mathrm{fb}^{-1}$  \\
\midrule
      $\phantom{1}0.1 - 0.98$       &                     $760$                     &                    $1650$                    &                    $9900$                     \\
    $\phantom{1}1.1 -6.0\phantom{0}$     &                    $1550$                    &                    $3400$                    &                   $20000$                    \\
      $15.0 -18.9$      &                    $2030$                    &                    $4400$                    &                   $26700$                    \\
      all       &                    $6140$                    &                   $13350$                   &                   $80000$                    \\
\bottomrule
\end{tabular}       

\end{table}

The sensitivity to an observable is studied by performing an unbinned maximum-likelihood fit to the pseudodata implemented with jax~\cite{jax2018github}. 
The signal PDF, $\mathcal{P}$, takes the tagging decision into account and follows
\begin{equation}\label{eq:pdftag}
\begin{aligned}
    \mathcal{P}(\ctk, & \ctl, \varphi,  t, q_{\rm SS}, q_{\rm OS}) = \\ & = \prod_{j \in [\rm SS, OS]}\Big(1 + q_{j}(1 - 2 \omega_j)\Big) f_{\Bs}(\ctk, \ctl, \varphi, t) \otimes \mathcal{D}(t,\sigma_t) \epsilon(t) \\ 
    & + \prod_{j \in [\rm SS, OS]} \Big(1 - q_j(1 - 2 \omega_j)\Big) f_{\Bsb}(\ctk, \ctl, \varphi, t) \otimes \mathcal{D}(t,\sigma_t) \epsilon(t),
\end{aligned}
\end{equation}
with $q_j$ being the tag decision by the tagger $j$ with the corresponding mistag-probability $\omega_j$, $f_{\Bs}$ and $f_{\Bsb}$ corresponding to Eq.~\ref{eq:full-tdep-flavourdep-decay-rate}, $\epsilon(t)$ denoting the acceptance introduced in Eq.~\ref{eq:time-acceptance}, and $\mathcal{D}(t,\sigma_t)$ the dilution given in Eq.~\ref{eq:time-dilution}. 
A tag decision for a \Bs(\Bsb) corresponds to $q = +1(-1)$. 
Every event contributes to the likelihood according to a weight that encompasses the mistag-probability and the corresponding tagging decision of the tagger in question. 
For untagged events, the form of the time-dependent flavour averaged measurement is recovered. 
The invariant four-body mass dimension factorises with the rest of the PDF, and for  brevity is not written out in Eq.~\ref{eq:pdftag}. The mass distribution is flavour-symmetric and does not depend on the tag decision. 

\begin{figure}[!tb]
    \centering

    \includegraphics[width=\textwidth]{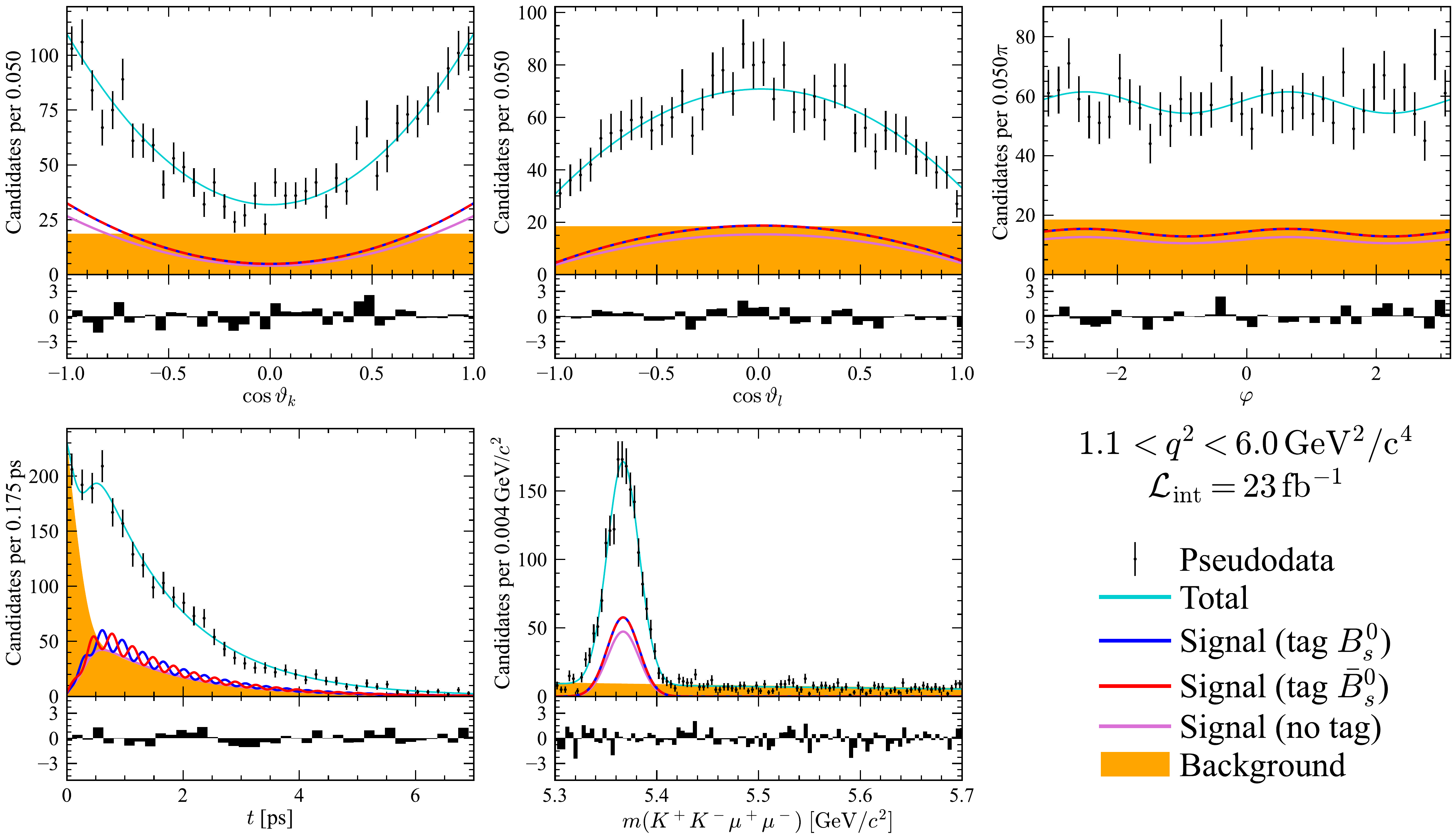}

    \caption{Example of a fit to the generated pseudodata in the region where $1.1 < \qsq < 6.0\gevgevcccc$ using the model described in the text. In addition to the data, the fit-model and its components are displayed, with the \Bs (\Bsb) shown in blue (red), the untagged component in purple, and the combinatorial background in orange. The characteristic oscillations of the \Bs-system are well visible in the individual components of the decay-time PDF in the lower left. }
    \label{fig:fit-example-central}
    
\end{figure}

\begin{figure}[!tb]
    \centering

    \includegraphics[width=\textwidth]{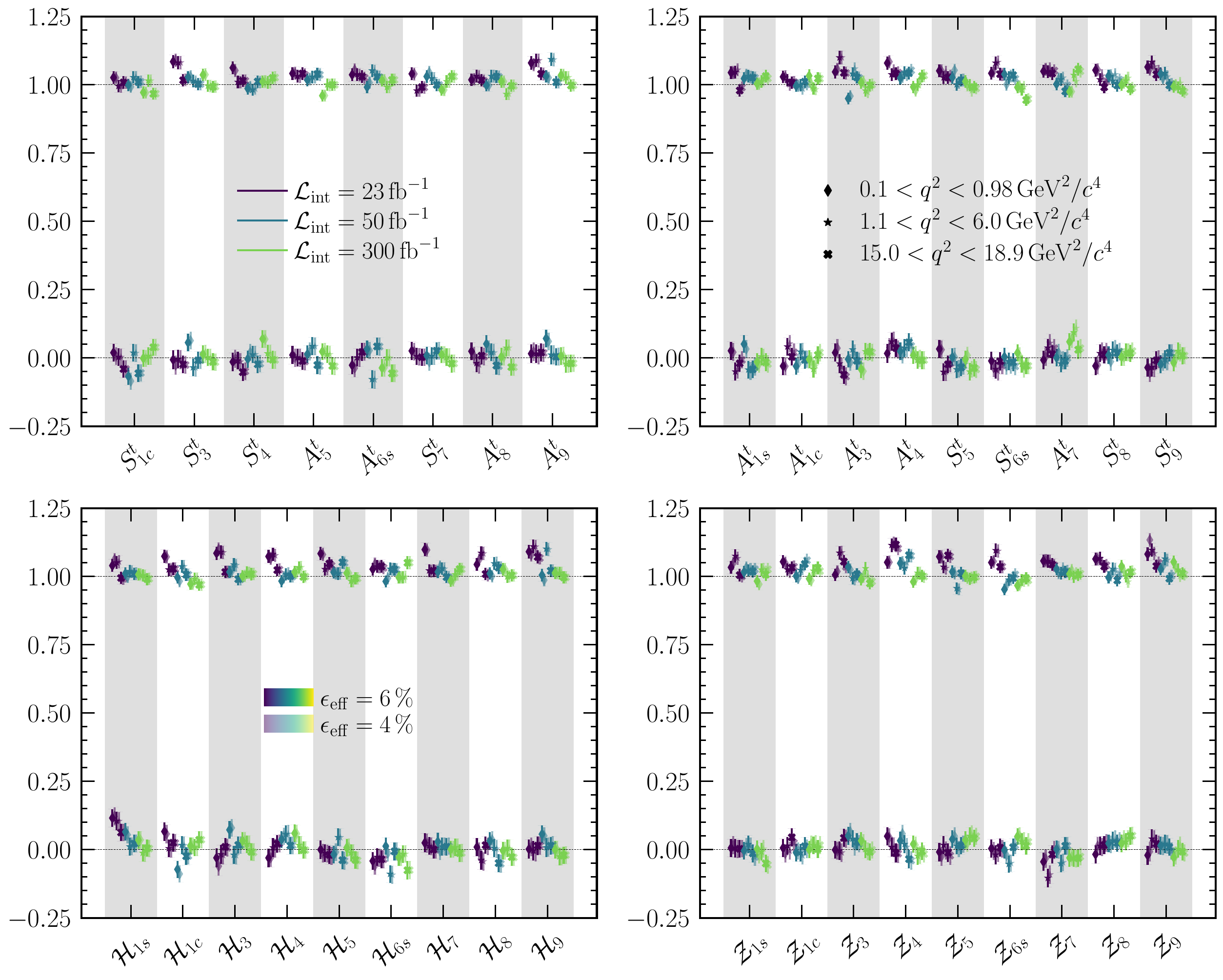}

    \caption{Distribution of the bias and coverage for the ensemble of pseudoexperiments performed for the different number of generated signal decays, \qsq-regions, and effective tagging-power. Small biases and undercoverage is observed for the smallest sample sizes which vanish in the large sample limit. }
    \label{fig:toy-studies}
    
\end{figure}

In the fit, the well-measured parameters $\Gamma$, $x$, and $y$ are fixed to the central values in Ref.~\cite{PDG2024}. 
An example for a fit to the pseudodata generated with $1.5\times10^{3}$ signal candidates is displayed in \cref{fig:fit-example-central}, where the three signal categories and the combinatorial background are shown in blue, red, purple, and orange, respectively. 
Examples of the fit projections for all \qsq-regions and fit-setups can be found in App.~\ref{app:fit-projections}. 
The fit is validated using large ensembles of pseudoexperiments with large numbers of \Bs decay candidates. 
For the $\mathcal{L}_{\rm int}=23\invfb$ scenario, we observe a bias smaller than $20\,\%$ of the statistical uncertainty in the $\mathcal{H}_{1s}$ and $\mathcal{H}_{1c}$ parameters, which disappears for larger sample sizes. 
Similarly, in the $\mathcal{L}_{\rm int} = 23\invfb$ scenario, some observables exhibit undercoverage at the $10\,\%$ level, an effect not present for larger sample sizes. 
This undercoverage does not affect the quoted sensitivities, since these are obtained directly from the distribution of fitted values across the pseudoexperiments, rather than the individual fit uncertainties. 
The bias and undercoverage are smaller in the untagged fit-setup, and also are not present for the large sample-sizes. 
Figure~\ref{fig:toy-studies} summarises the observed bias and coverage for ensembles of 1000 pseudoexperiments per configuration, with marker colour indicating the number of generated signal candidates, marker type the \qsq-region, and marker opacity the tagging power. 
An unbiased measurement is indicated by a marker compatible with zero and an accurate coverage is represented by a marker compatible with unity. 
A similar summary for the untagged time-dependent and untagged time-integrated measurements are given in App.~\ref{app:pulls-untagged}. 
The data analysed in this work are generated using the publicly available code in Ref.~\cite{code-repository-phimumu}. 

\section{Results}
\label{sec:results}

\begin{figure}[tb!]

    \centering
    \includegraphics[width=\textwidth]{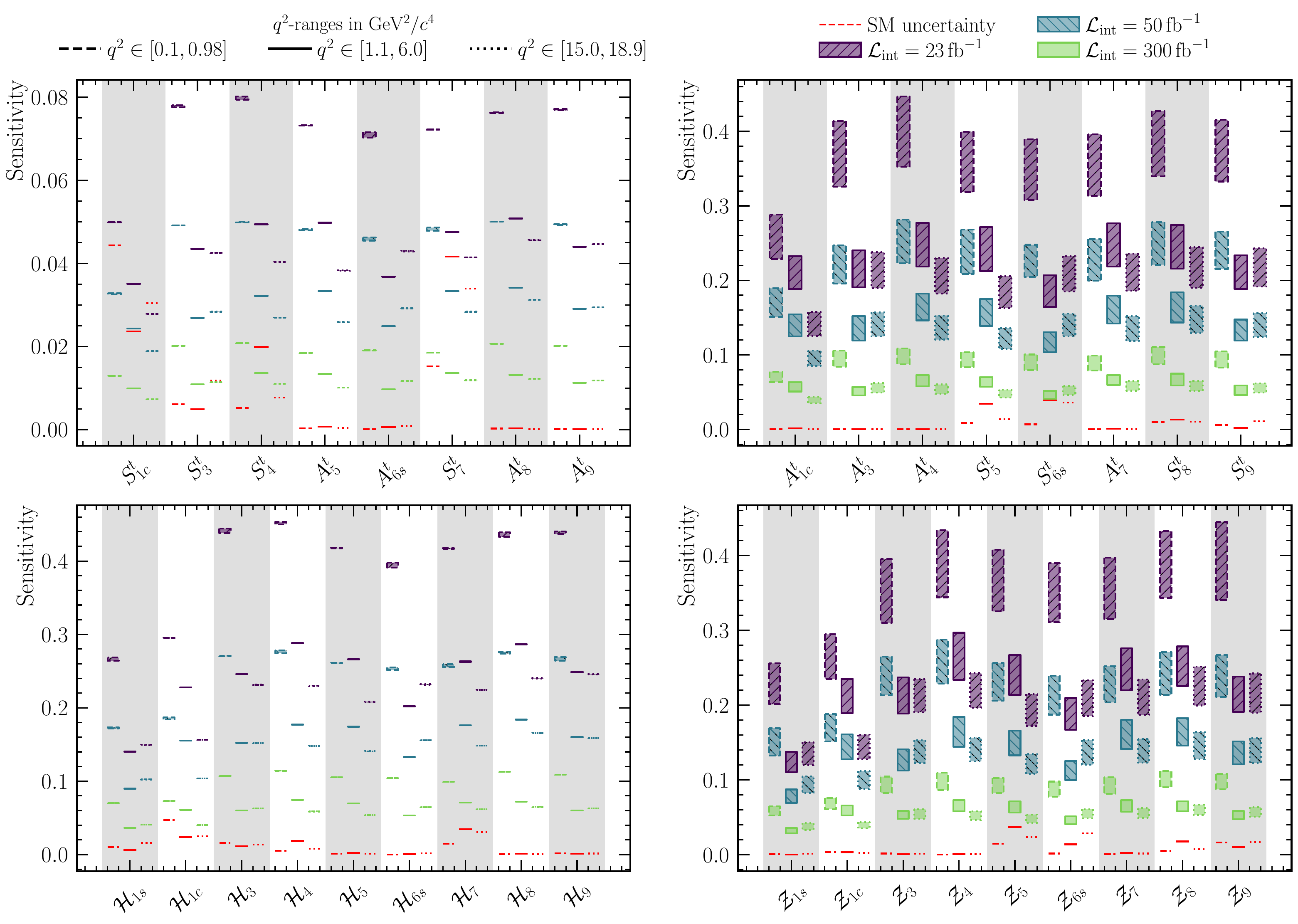}

    \caption{Extrapolated sensitivity to the \CP-averages $S^t_i$, \CP-asymmetries $A^t_i$, $\mathcal{H}_i$, and $\mathcal{Z}_i$ in comparison to the uncertainty of SM predictions evaluated using flavio~\cite{straub:2018flavio} where the local form-factors are taken from Ref.~\cite{Bharucha:2015bzk}, and the nonlocal form-factors follow the parameterisation of Ref.~\cite{Altmannshofer:2014rta}. The expected range for the sensitivity after the respective luminosity milestones is shown in the shaded area. The observables accessible in the untagged measurement are displayed on the left and the observables which require flavour-tagging are shown on the right. Depending on the assumed tagging-power, the sensitivity to the observables which require flavour-tagging varies. }
    \label{fig:sens-all-obses}

\end{figure}

The resulting sensitivity to the \CP-averaged and \CP-asymmetric observables $S^t_i$ and $A^t_i$ are shown in Fig.~\ref{fig:sens-all-obses} for the different luminosity scenarios. 
The extrapolations are compared to the precision of the SM predictions, evaluated using flavio~\cite{straub:2018flavio} with the local form-factors following Ref.~\cite{Bharucha:2015bzk} and the nonlocal form-factors are parameterised according to Ref.~\cite{Altmannshofer:2014rta}. 
The best precision is obtained for the coefficients which are accessible through the untagged sum of the \Bs and \Bsb decay-rate, shown in the left panels of Fig.~\ref{fig:sens-all-obses}. 
Observables which are accessible only using flavour-tagging are shown in the right panels. 
If a dedicated heavy-flavour experiment with a similar performance to the \lhcb upgrade I collects approximately $80 \times 10^3$ \decay{\Bs}{\Pphi\mumu} decays, as could be expected by the conclusion of Run~5, the measurement for $S^t_{1c}$ will likely be at the $10^{-2}$ level and therefore more precise than the current SM predictions. 
The other observables accessible in the untagged decay-rate which are related to $J_i + \tilde{J_i}$ will also be measured with an uncertainty at the $10^{-2}$ level. 
The observables which are only accessible using flavour-tagging could be measured up to a sensitivity of $5\times10^{-2}$, depending on the flavour-tagging performance. 
An initial measurement of these observables, with a data set corresponding to an integrated luminosity of $23\invfb$, could yield an uncertainty at the $1\text{--}4\times 10^{-1}$ level, with the exact value depending on the flavour-tagging performance of the machine. 
In this context also $P_5^{t\prime}$ could be measured for the first time in rare \Bs-decays. 
After the conclusion of Run~5, the precision on  $P_5^{t\prime}$ should be similar to that obtained by the current measurements in \Bz decays~\cite{LHCb-PAPER-2020-002}. 

The expected sensitivities to the $\mathcal{H}_i$ and $\mathcal{Z}_i$ observables, which have not yet been measured, are illustrated in Fig.~\ref{fig:sens-all-obses} in the bottom row and compared to the uncertainties computed within the SM. 
After the conclusion of Run~5, a dedicated heavy-flavour experiment with a similar performance as \lhcb could measure the new observables with a precision reaching $5 \times 10^{-2}$. 
Here, similar precision for the $\mathcal{H}_i$ and $\mathcal{Z}_i$ are obtained: the $\mathcal{H}_i$ are challenging to measure due to the proportionality to $\sinh(y\Gamma t)$ which is small in the SM, whereas the precision of the $\mathcal{Z}_i$ observables is limited by the flavour-tagging performance at hadronic machines. 
Extrapolations for the time-independent setup and a comparison to the existing measurements are discussed in App.~\ref{app:time-idenpendent-results}.

\begin{figure}[tb!]
    \centering
    \includegraphics[width=\textwidth]{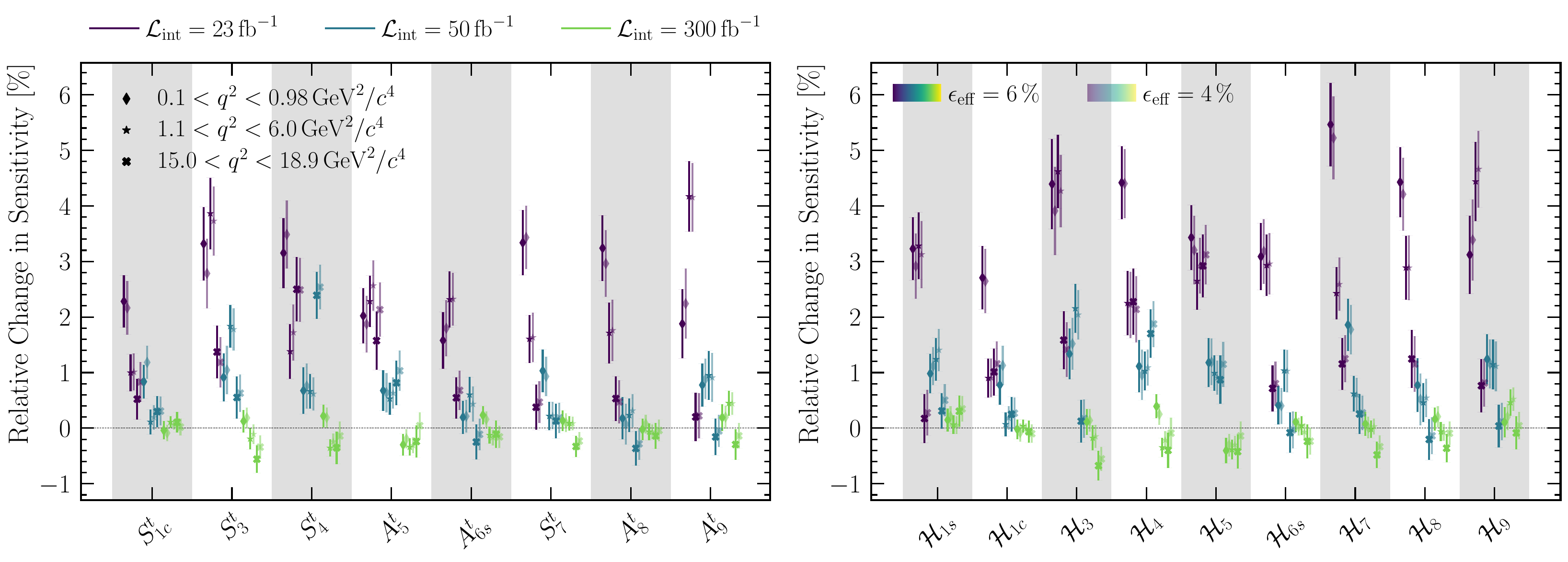}
    \caption{Gain in sensitivity obtained for the observables when switching from a tagged to an untagged time-dependent fit setup relative to the untagged setup. The luminosity scenarios are shown using different colours, the \qsq-regions are differentiated using markers, and the flavour-tagging performance is encoded by the opacity of the marker. }
    \label{fig:change-in-sens-tagged-untagged}
\end{figure}

In order to assess the relative advantages of performing  time-dependent tagged vs time-dependent untagged fits, the sensitivities for both are evaluated on the same generated data. 
In the case of an untagged fit, only the $\mathcal{H}_i$ observables along with the $S^t_i$ ($A^t_i$) observables with $\zeta_i$ =1 ($\zeta_i$ =-1) can be extracted. 
In Fig.~\ref{fig:change-in-sens-tagged-untagged} the change in sensitivity for the tagged and untagged time-dependent fits relative to the untagged sensitivity are shown for these observables. 
The gain in sensitivity when using an untagged setup ranges from zero up to six percent of the statistical uncertainty. 
The gain observed for the $\mathcal{H}_i$ observables when using an untagged approach is generally larger than for the $S^t_i$ and $A^t_i$ observables. 
For the $\mathcal{L}_{\rm int} = 50\invfb$ scenario improvements on the sensitivity reduce to the $2\,\%$ level and for the $\mathcal{L}_{\rm int} = 300\invfb$ scenario, no improvement in the sensitivity is observed any more. 
This modest gain in sensitivity is not overly surprising: the tagged setup defaults to the untagged form in the case where neither tagger produces a decision, and, in the case where the taggers disagree, the resulting weights for the \Bs and \Bsb decay-rates are similar due to the comparable mistag probabilities of the two taggers. 
In the time-independent fit setup, the comparison of sensitivity is not straight forward, as the fit-parameter now corresponds to a combination of the $S^t_i$ or $A^t_i$ and $\mathcal{H}_i$, however it is checked that there is no marked gain in precision on the other variables when fixing the $\mathcal{H}_i$ observables in the fit. 

\begin{figure}[tb!]

    \centering
    \begin{minipage}{.33\textwidth}
        \includegraphics[width=\textwidth]{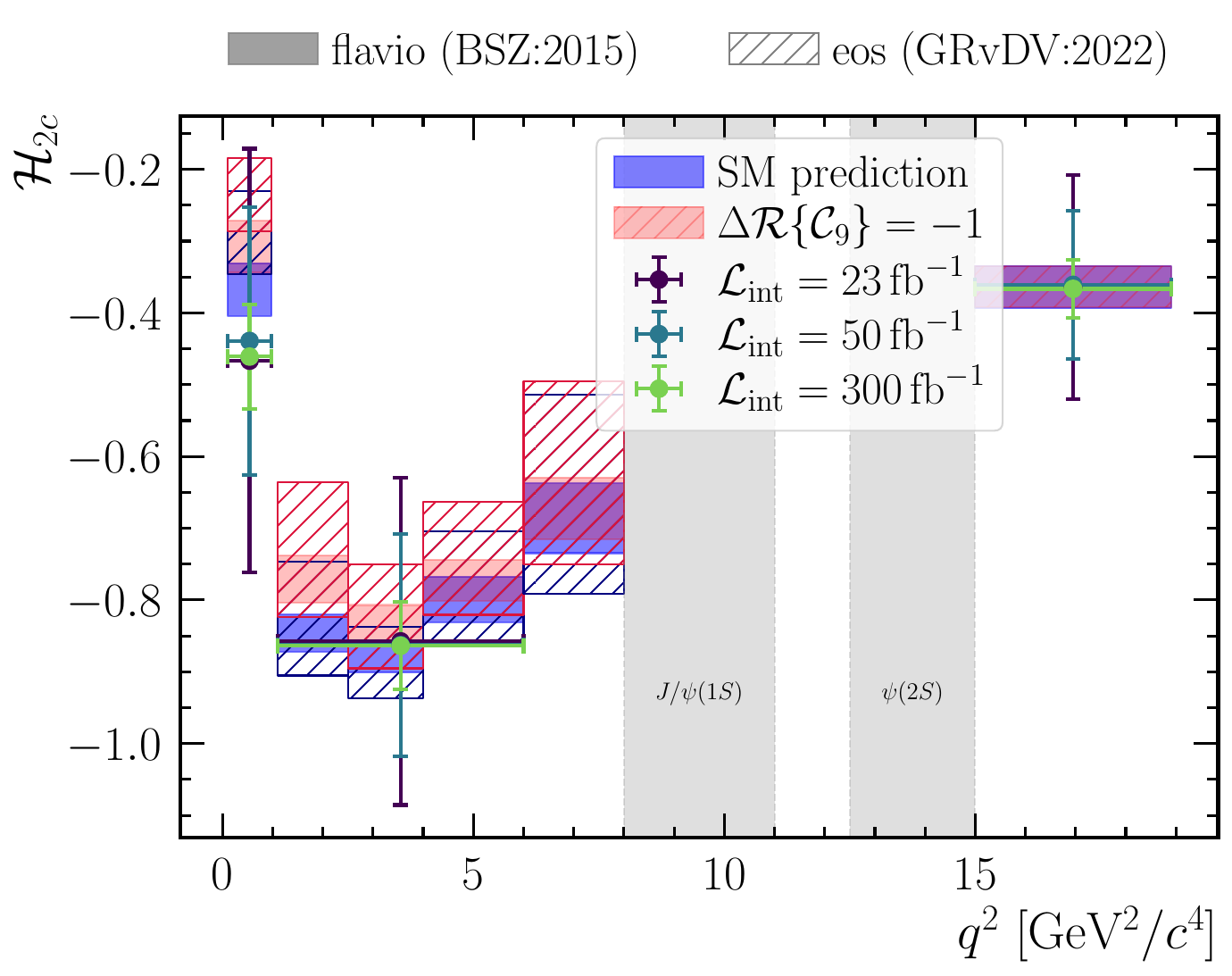}
    \end{minipage}%
    \begin{minipage}{.33\textwidth}
        \includegraphics[width=\textwidth]{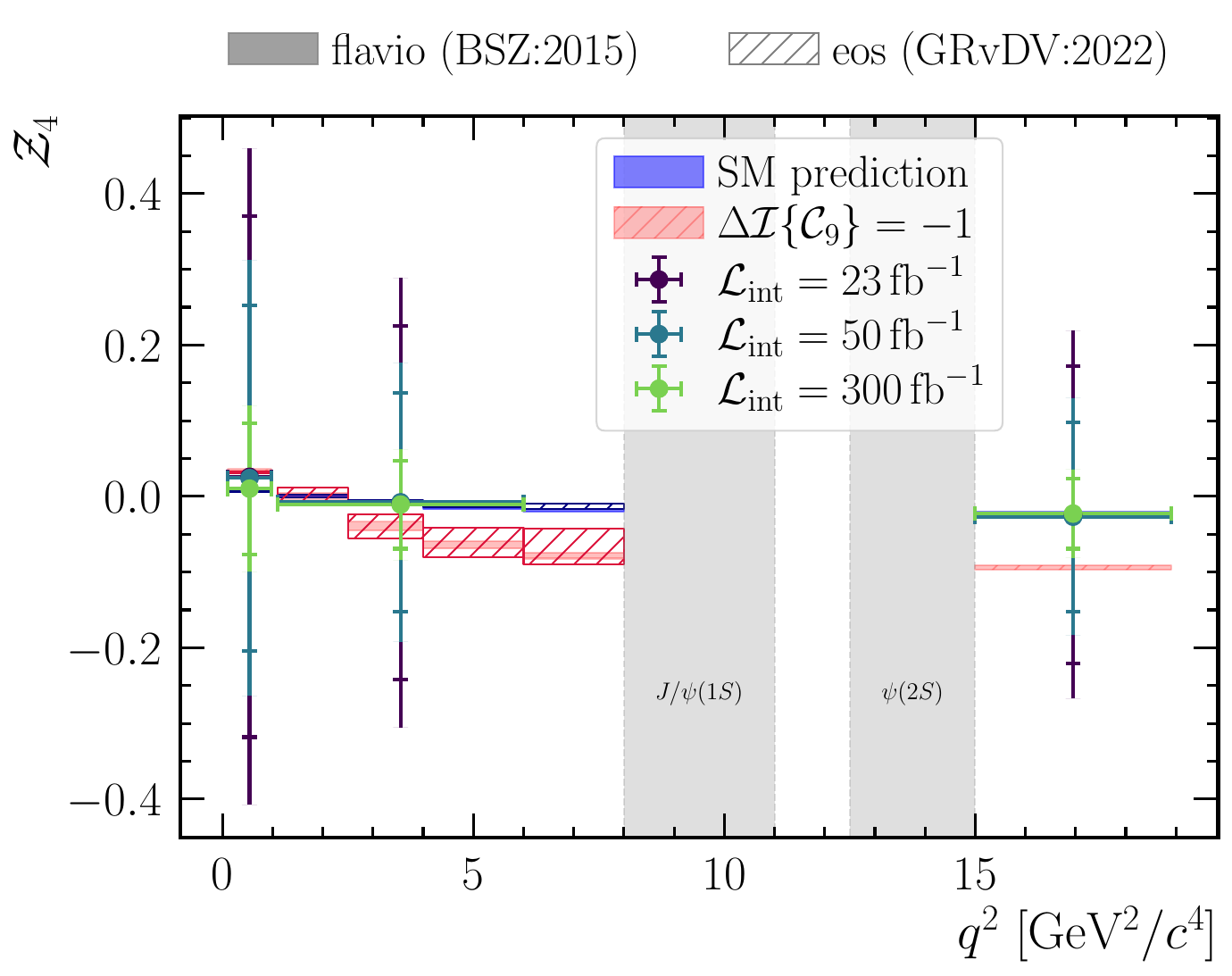}
    \end{minipage}%
    \begin{minipage}{.33\textwidth}
        \includegraphics[width=\textwidth]{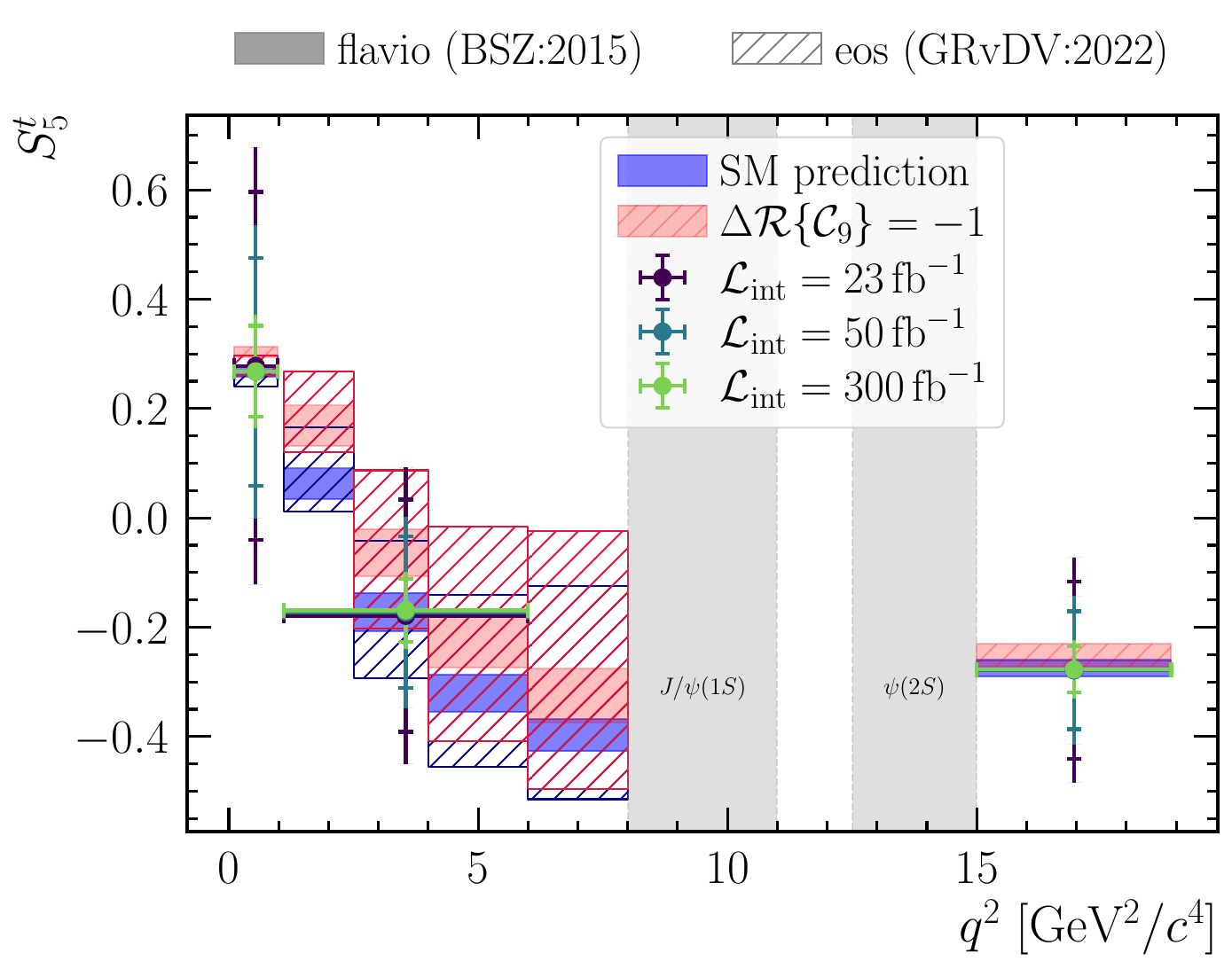}
    \end{minipage}%

    \begin{minipage}{.33\textwidth}
        \includegraphics[width=\textwidth]{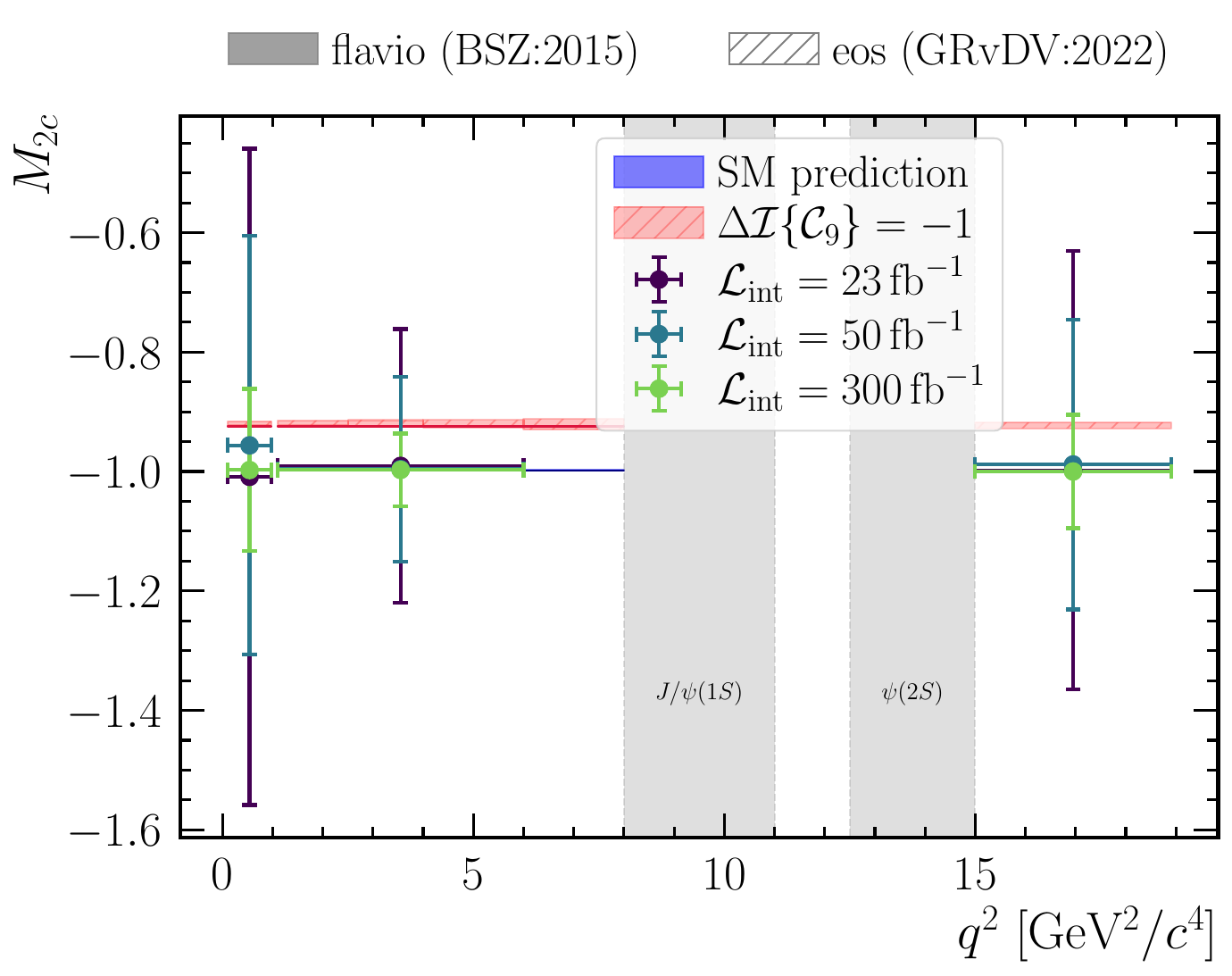}
    \end{minipage}%
    \begin{minipage}{.33\textwidth}
        \includegraphics[width=\textwidth]{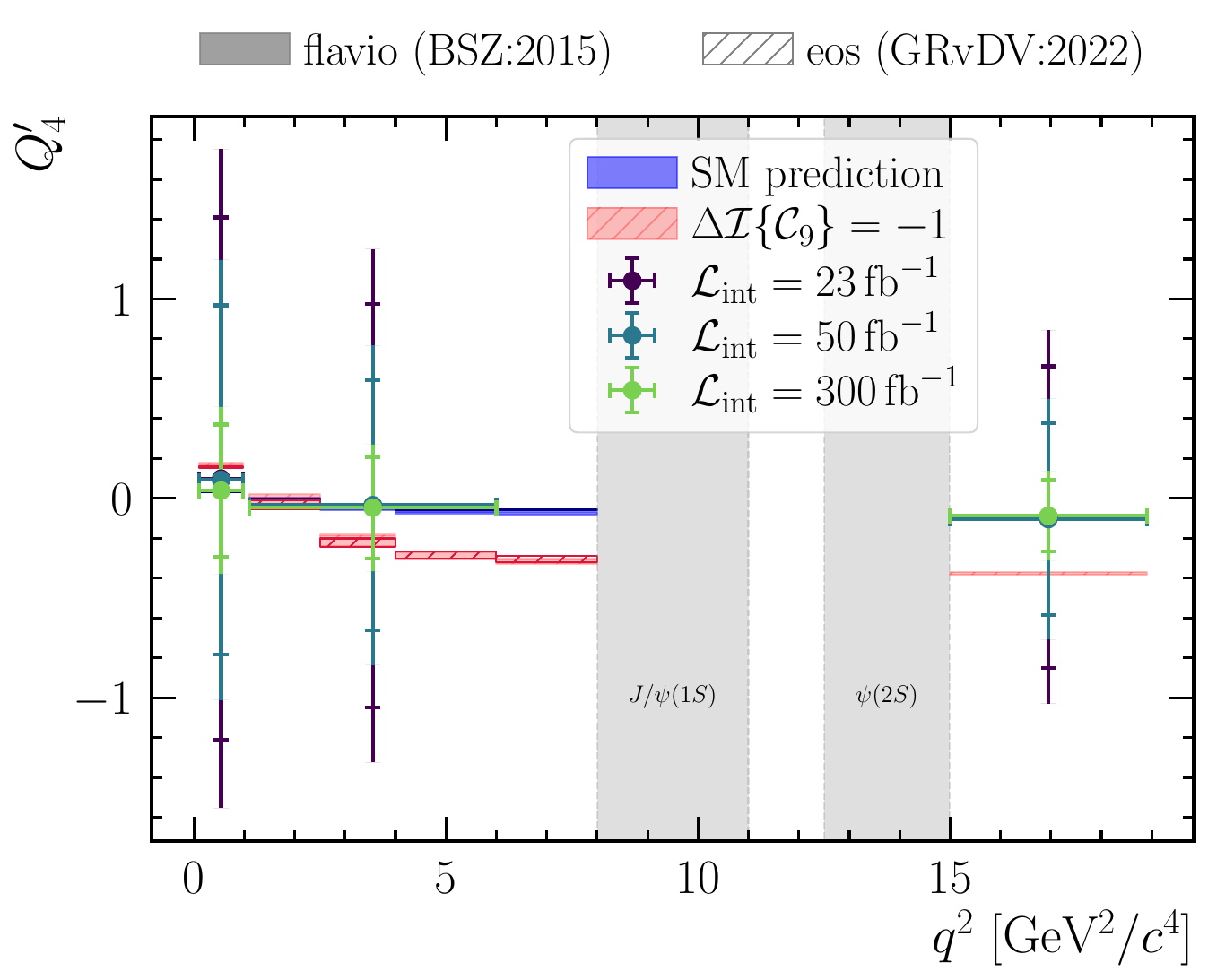}
    \end{minipage}%
    \begin{minipage}{.33\textwidth}
        \includegraphics[width=\textwidth]{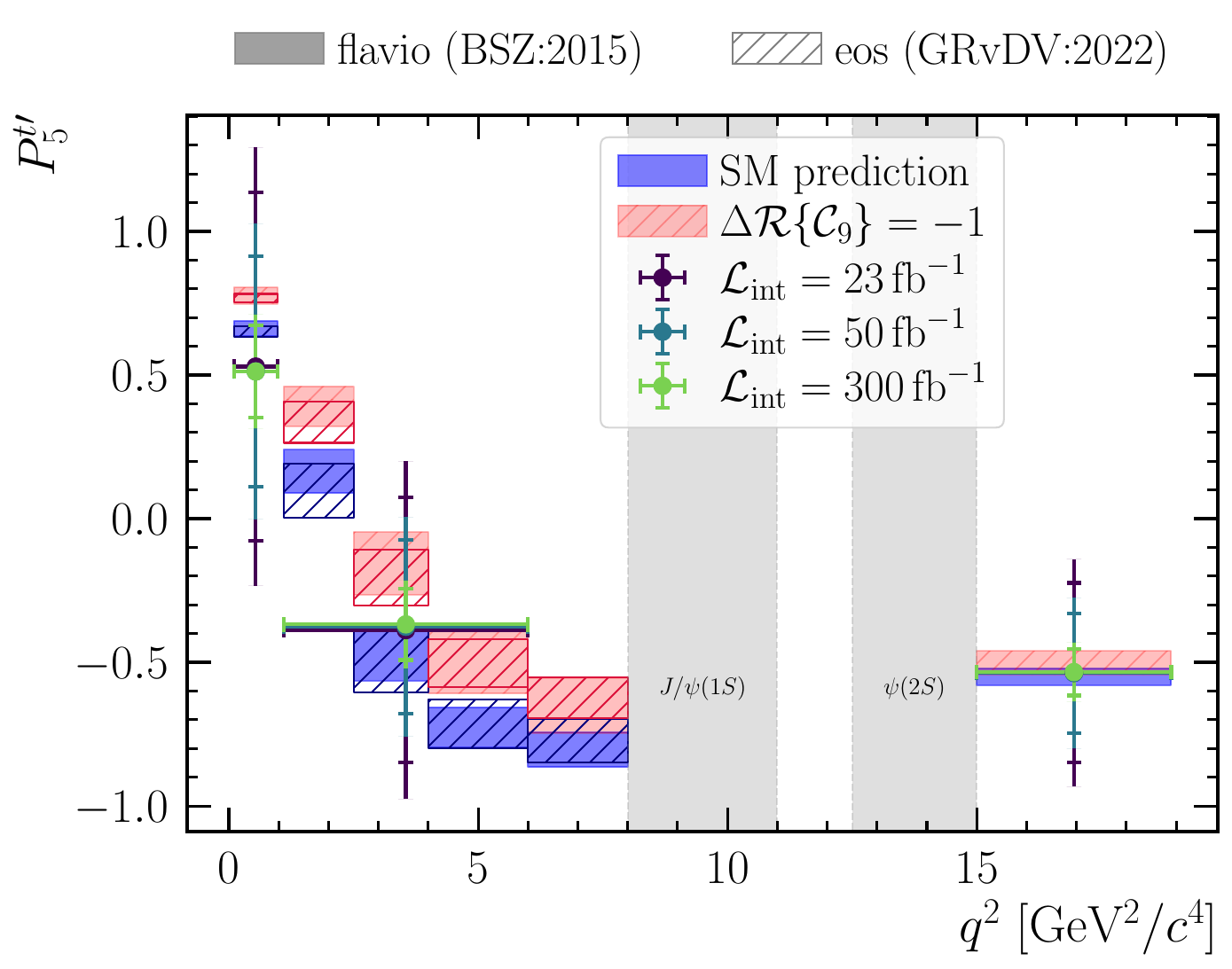}
    \end{minipage}%

    \caption{Extrapolated sensitivity to the optimised observables (from top left to bottom right) $\mathcal{H}_{2c}$, $\mathcal{Z}_4$, $S^t_5$, $Q_4^\prime$, $M_{2c}$, and $P_5^{t\prime}$ in comparison to the predictions of SM. The SM predictions are evaluated using flavio~\cite{straub:2018flavio} with form-factors from Refs.~\cite{Bharucha:2015bzk,Altmannshofer:2014rta} and eos~\cite{EOSAuthors:2021xpv} where the form factors follow Refs.~\cite{Gubernari:2018wyi,Gubernari:2022hxn,Gubernari:2023puw}. The variation of the accuracy for the different tagging-power setups within a given luminosity scenario is shown by the length between the end cap and the tip of the error bar of the same colour. This affects $\mathcal{Z}_4$, $Q_4^\prime$, $S^t_5$, and $P_5^{t\prime}$ which are only measurable in the presence of flavour-tagging. In addition to the prediction in the SM, an alternative scenario is illustrated where the imaginary or real part of $\mathcal{C}_9$ is shifted by minus one. Small discrepancies at low-\qsq are observed and attributed to the massless lepton approximation used.}
    \label{fig:sens-qobses}

\end{figure}

As outlined in Section~\ref{sec:optimised-observables}, we have proposed a set of observables which exhibit a reduced dependence on the hadronic form-factors, similar to the ones discussed in Refs.~\cite{Descotes-Genon:2012isb,Descotes-Genon:2015hea}. 
The sensitivity to a set of optimised observables are shown in Fig.~\ref{fig:sens-qobses} showing a comparison to the default observable. 
Two observables which are particularly sensitive to changes in the imaginary part of the vector coupling $\mathcal{C}_9$ are shown, $M_{2c}$ and $Q_4^\prime$. 
In addition $P_5^{t\prime}$, which is particularly sensitive to changes in the real part of the vector coupling, is displayed. 
The SM predictions are computed using both flavio~\cite{straub:2018flavio} where the form-factors from Refs.~\cite{Bharucha:2015bzk,Altmannshofer:2014rta} are used and eos~\cite{EOSAuthors:2021xpv} using the parameterisation from Refs.~\cite{Gubernari:2018wyi,Gubernari:2022hxn,Gubernari:2023puw}. 
The reduced uncertainties of the optimised observables within the SM are clearly visible when comparing the left column with the right. 
In order to visualise the sensitivity to NP altering the short-distance effects entering the Wilson coefficients, also predictions with altered values for the Wilson coefficients are displayed. 
As a benchmark for this comparison we change the imaginary part of $\mathcal{C}_9$ or the real part of $\mathcal{C}_9$ by minus one. 
For the imaginary part, this is approximately the lower edge of the $68\,\%$ confidence interval, while for the real part it coincides with the best-fit value, both as reported in Ref.~\cite{Altmannshofer:2021qrr}.
Small discrepancies between the SM predictions and the measurement positions are visible, especially for small values of the invariant di-muon mass. 
This is attributed to the massless-lepton approximation being used in our fits. 
Additional comparisons for more observables can be found in Appendix~\ref{app:optimised-observable-plots}. 

In order to assess how the additional mixing-observables improve the knowledge on the underlying short-distances physics mediating \decay{\Bs}{\phi}{\mumu}, we perform a global analysis of the measurements using the EFT framework. 
The analysis is similar to the global fits presented in Ref.~\cite{Altmannshofer:2021qrr}, which employs a weak effective Hamiltonian comprising a set of local dimension-six operators and effective coupling constants, the Wilson coefficients. 
In this work we focus on the four fermion $bs \mu\mu$ vector and axial-vector operators, as contributions to the other operators are better constrained by measurements using different decay modes. 
Additionally, whereas we focus on the left-chiral structure in this work, we expect similar constraining power for the right-chiral operators. 

In order to evaluate the constraints on the parameter space, the observables are parameterised depending on the Wilson coefficients and we profile the metric
\begin{align}
    \chi^2 = (\vec{O}_{\rm exp.} - \vec{O}_{\rm th.}(\mathcal{C}_i)) (\underline{C}^{\rm exp.} + \underline{C}^{\rm th.}(\mathcal{C}_i))^{-1} (\vec{O}_{\rm exp.} - \vec{O}_{\rm th.}(\mathcal{C}_i)),
\end{align}
where $\vec{O}$ denotes the observable vector for the experiment and theory and $\underline{C}$ is the respective covariance matrix. 
We assume that there are no correlations between the experimental observables in the different \qsq-regions and determine $\vec{O}_{\rm th.}(\mathcal{C}_i)$ and $\underline{C}^{\rm th.}(\mathcal{C}_i)$ using flavio, evaluating the Wilson coefficients at the renormalisation scale $\mu = 4.8\gev$. 

\begin{figure}[tb!]
    \centering
    \begin{minipage}{.49\textwidth}
        \includegraphics[width=\textwidth]{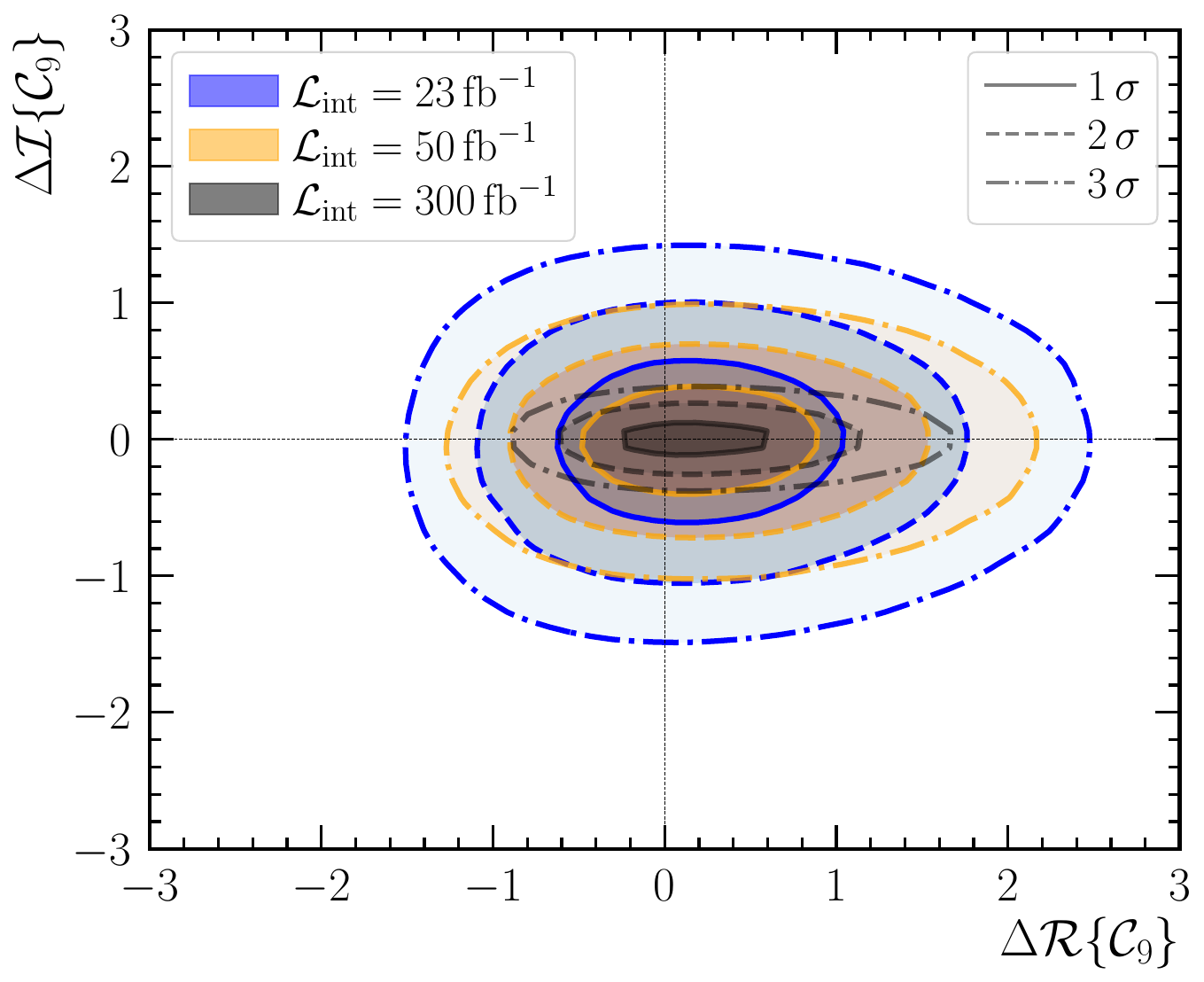}
    \end{minipage}\hspace{.019\textwidth}%
    \begin{minipage}{.49\textwidth}
        \includegraphics[width=\textwidth]{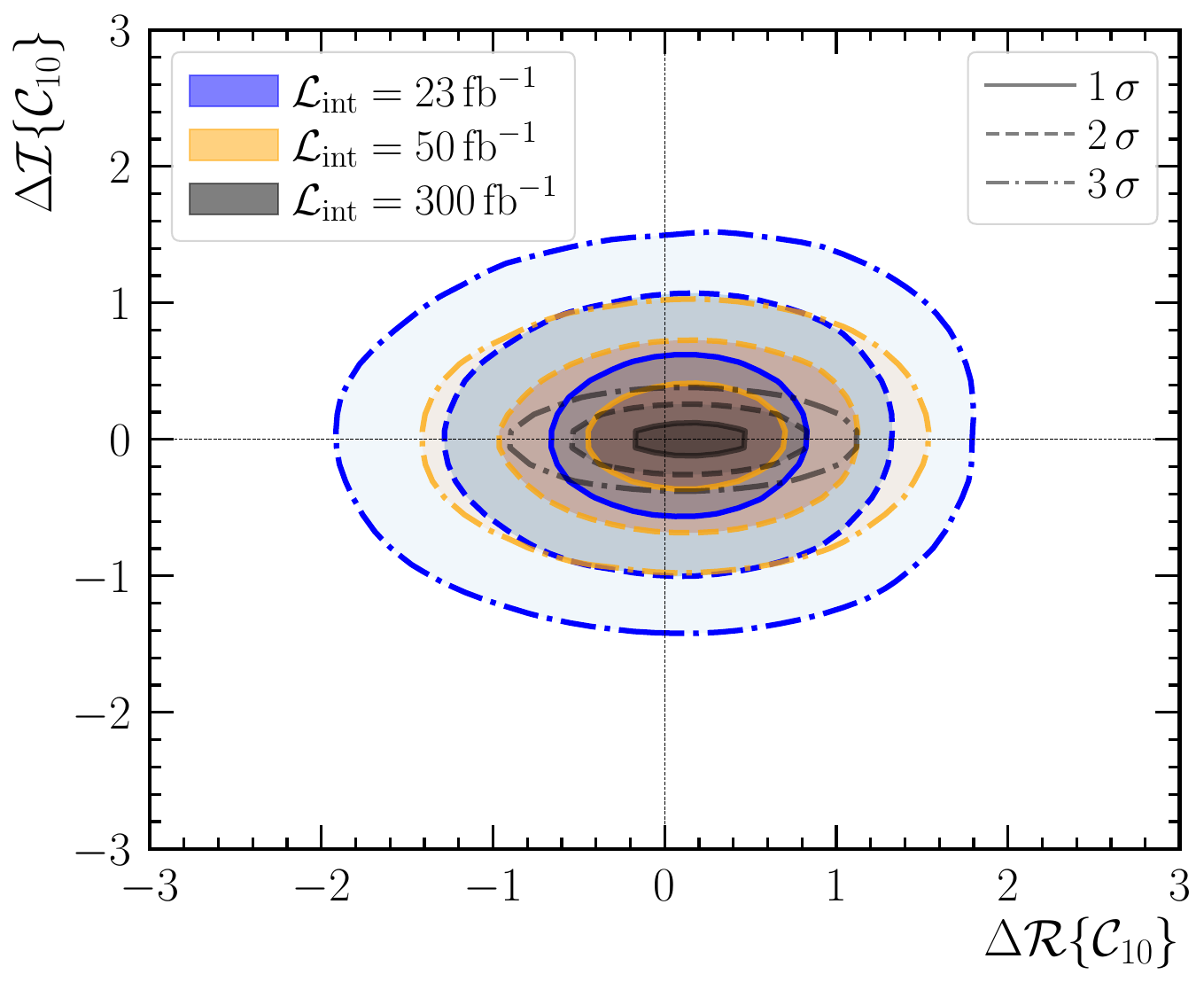}
    \end{minipage}%

    \caption{Coverage regions for fits for the Wilson coefficient $\mathcal{C}_9$ (left) and $\mathcal{C}_{10}$ (right). The change in the real (imaginary) part of the Wilson coefficient is displayed on the $x$-axis ($y$-axis). All observables are included in the fit and a comparison of the different luminosity scenarios is performed. }
    \label{fig:np-fits-2d-lumi-comp}
\end{figure}

Using this setup, we can compare the constraints of the Wilson-coefficients parameter space for the different luminosity milestones and for the different sets of observables present when including  tagging or time-dependence in the PDF. 
The fits are performed using our fit results obtained in three \qsq-regions from $0.1 - 0.98$, $1.1 - 6.0$, and $15.0 - 18.9\gevgevcccc$. 
Adding more \qsq-regions will improve the constraints further. 
The constraints obtained when measuring the full time-dependent tagged decay rate for the different luminosity milestones are compared in Fig.~\ref{fig:np-fits-2d-lumi-comp} assuming a $6\,\%$ flavour-tagging power. 
Notably the imaginary part of the Wilson coefficients can be better constrained than the real part when including the full-set of \decay{\Bs}{\Pphi \mumu} observables. 

\begin{figure}[tb!]
    \centering
    \begin{minipage}{.49\textwidth}
        \includegraphics[width=\textwidth]{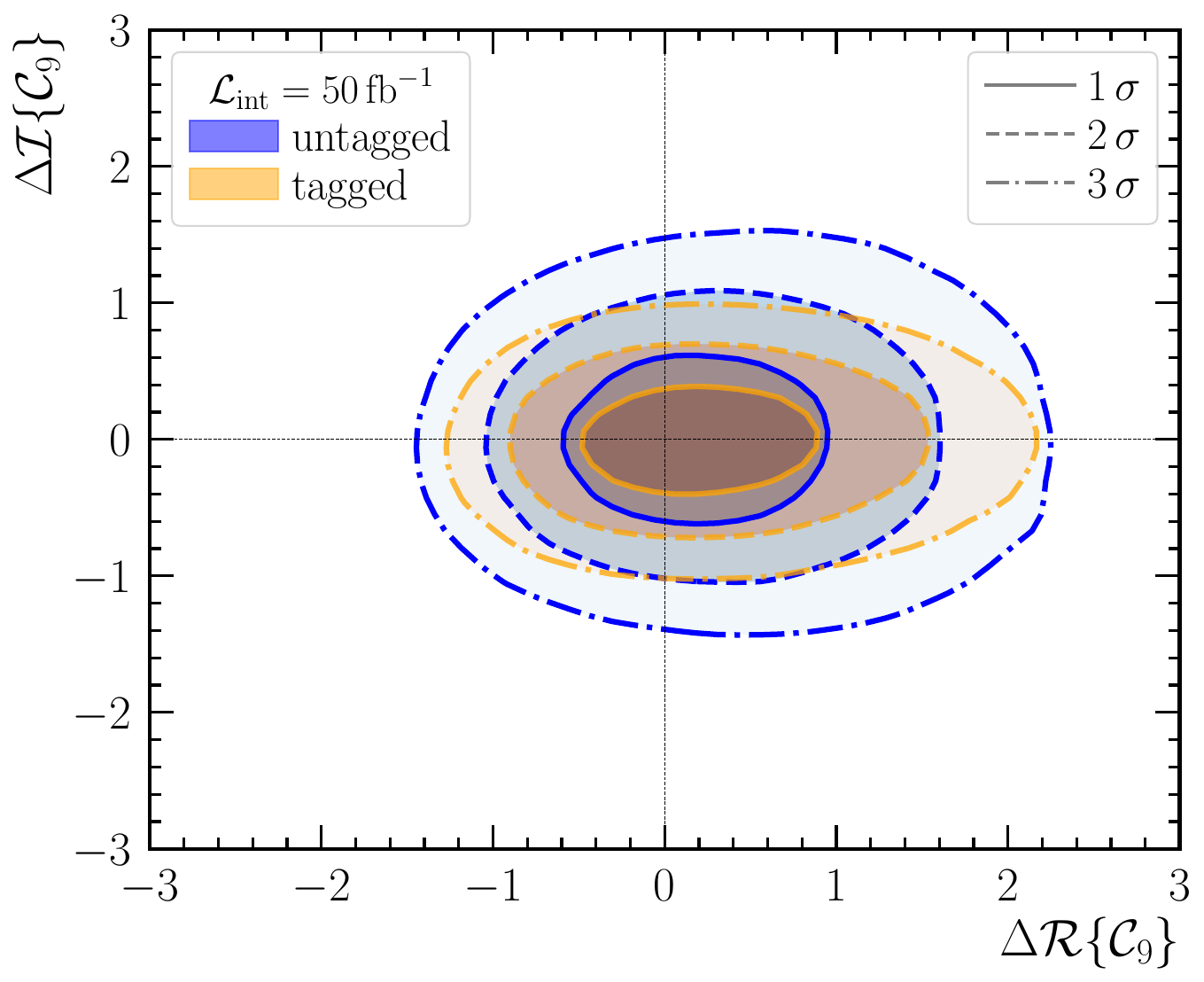}
    \end{minipage}\hspace{.019\textwidth}%
    \begin{minipage}{.49\textwidth}
        \includegraphics[width=\textwidth]{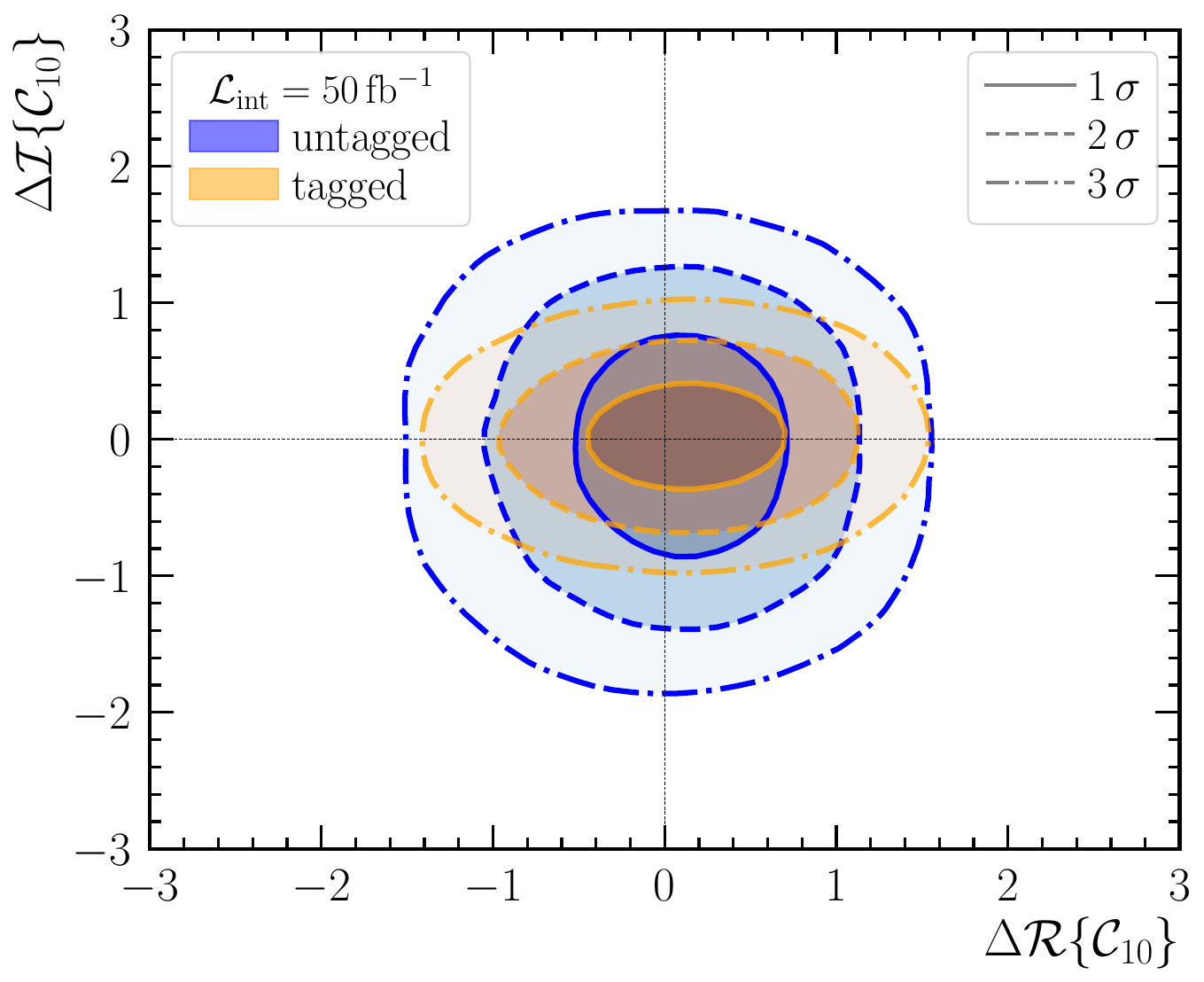}
    \end{minipage}%

    \begin{minipage}{.49\textwidth}
        \includegraphics[width=\textwidth]{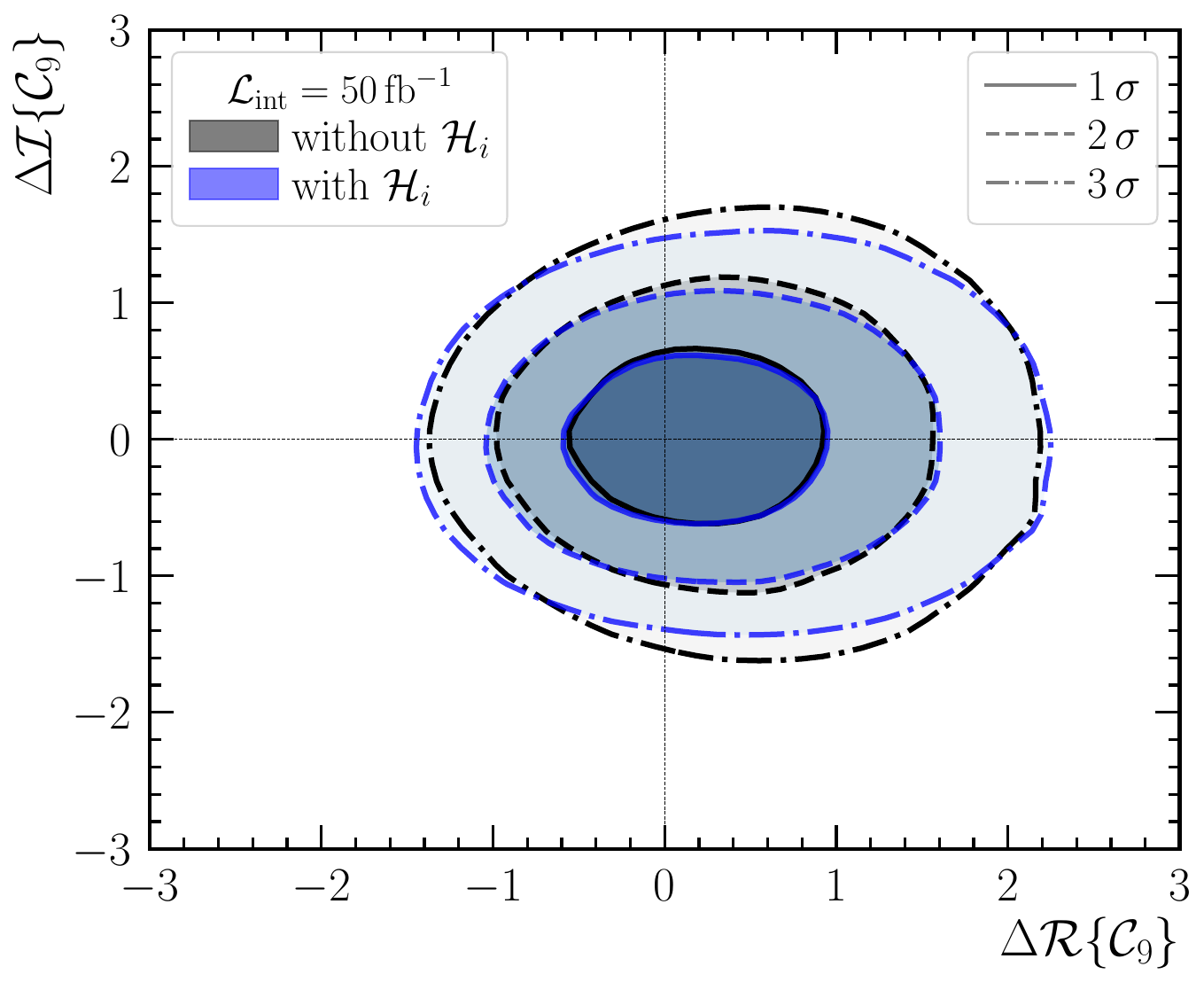}
    \end{minipage}\hspace{.019\textwidth}%
    \begin{minipage}{.49\textwidth}
        \includegraphics[width=\textwidth]{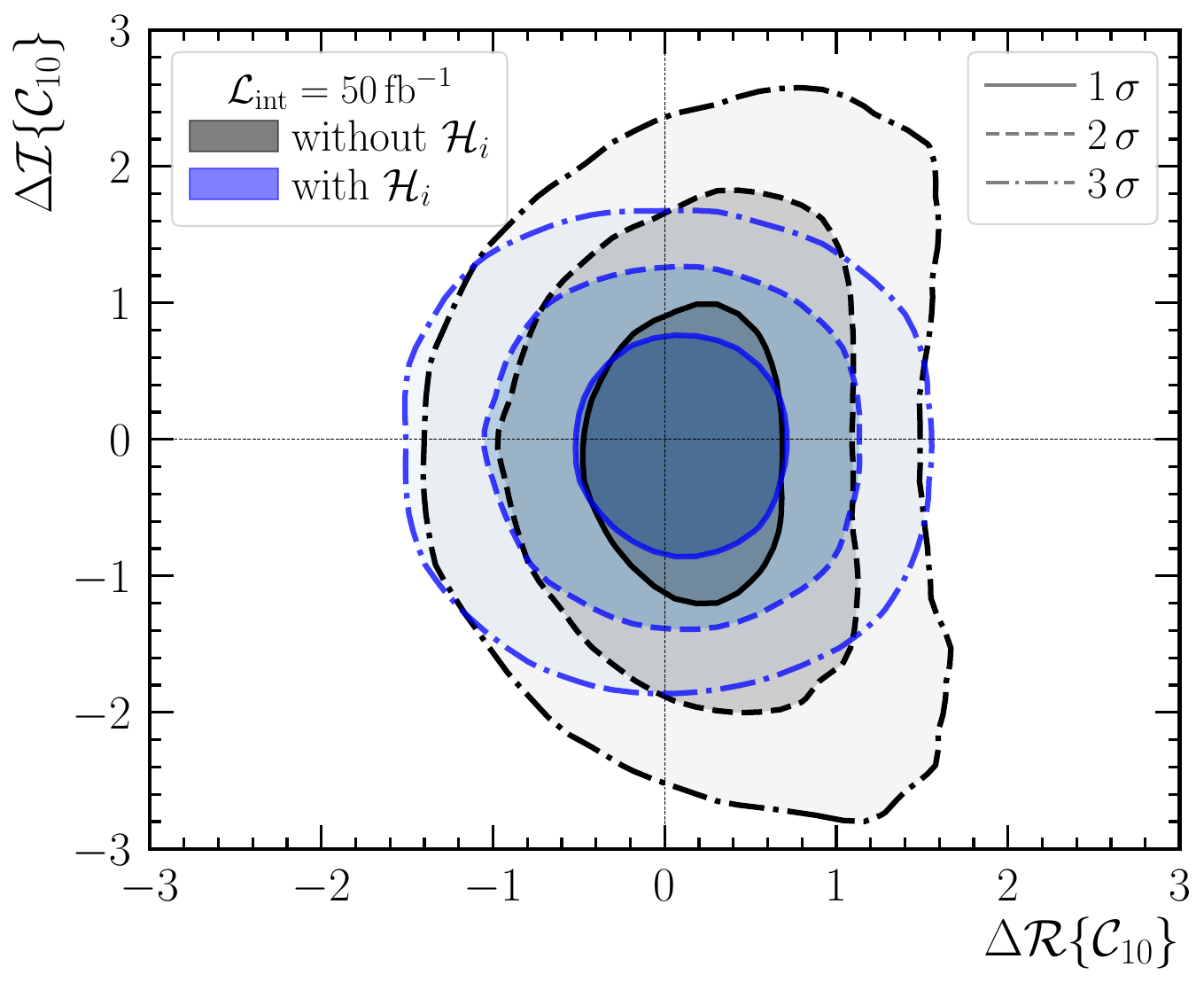}
    \end{minipage}%

    \caption{Coverage regions for fits for the Wilson coefficient $\mathcal{C}_9$ (left) and $\mathcal{C}_{10}$ (right). The change in the real (imaginary) part of the Wilson coefficient is displayed on the $x$-axis ($y$-axis). A comparison of the constraints of the Wilson coefficients given the observables measurable in the different fit setups is performed. At the top, a comparison of the flavour-tagged and untagged setups is performed, in the bottom row a comparison of the time-integrated to the time-dependent untagged setup is shown. }
    \label{fig:np-fits-2d-obs-comp}
\end{figure}

\begin{figure}[tb!]
    \centering
    \begin{minipage}{0.49\textwidth}
    \includegraphics[width=\textwidth]{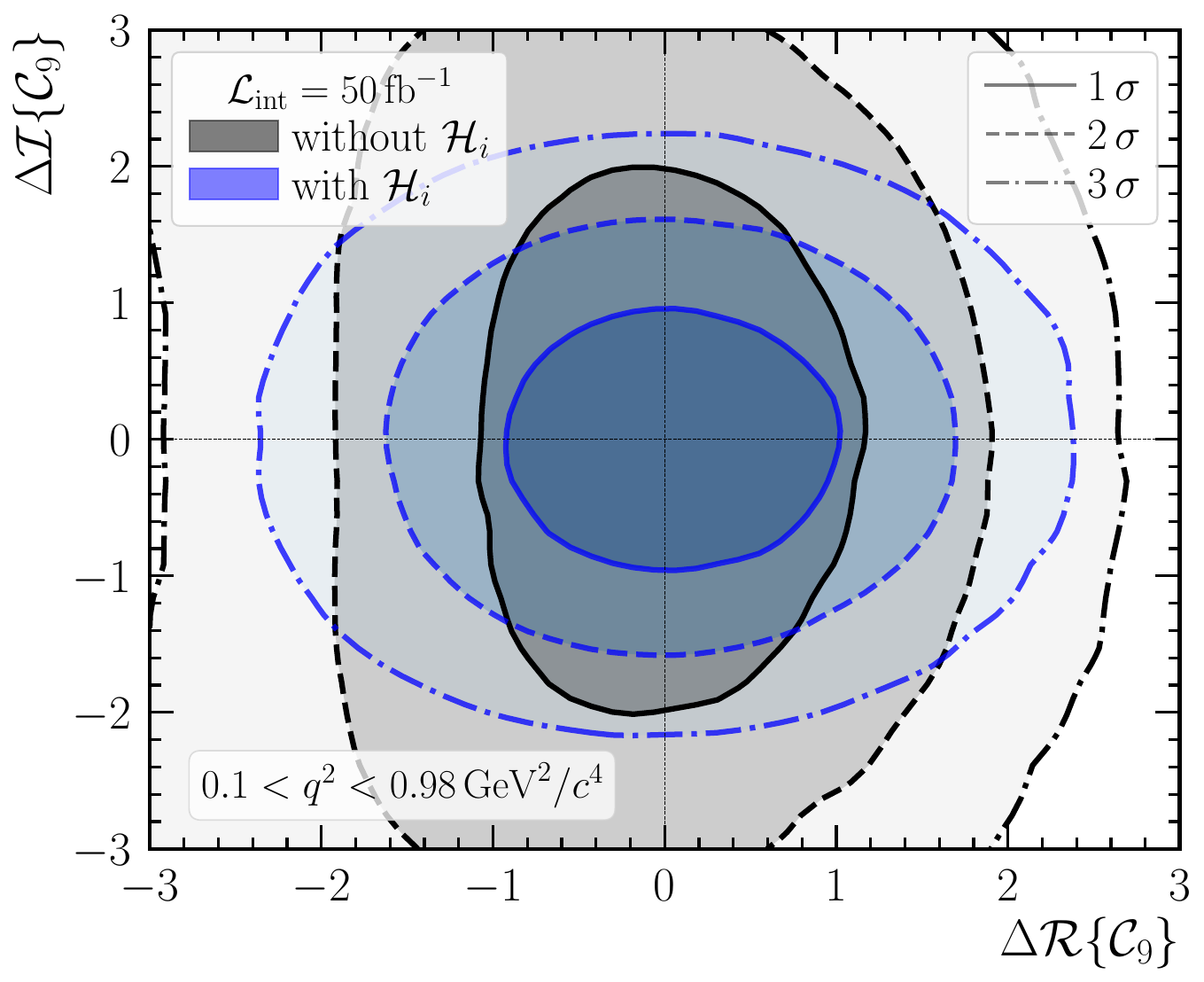}
    \end{minipage}\hspace{0.019\textwidth}%
    \begin{minipage}{0.49\textwidth}
    \includegraphics[width=\textwidth]{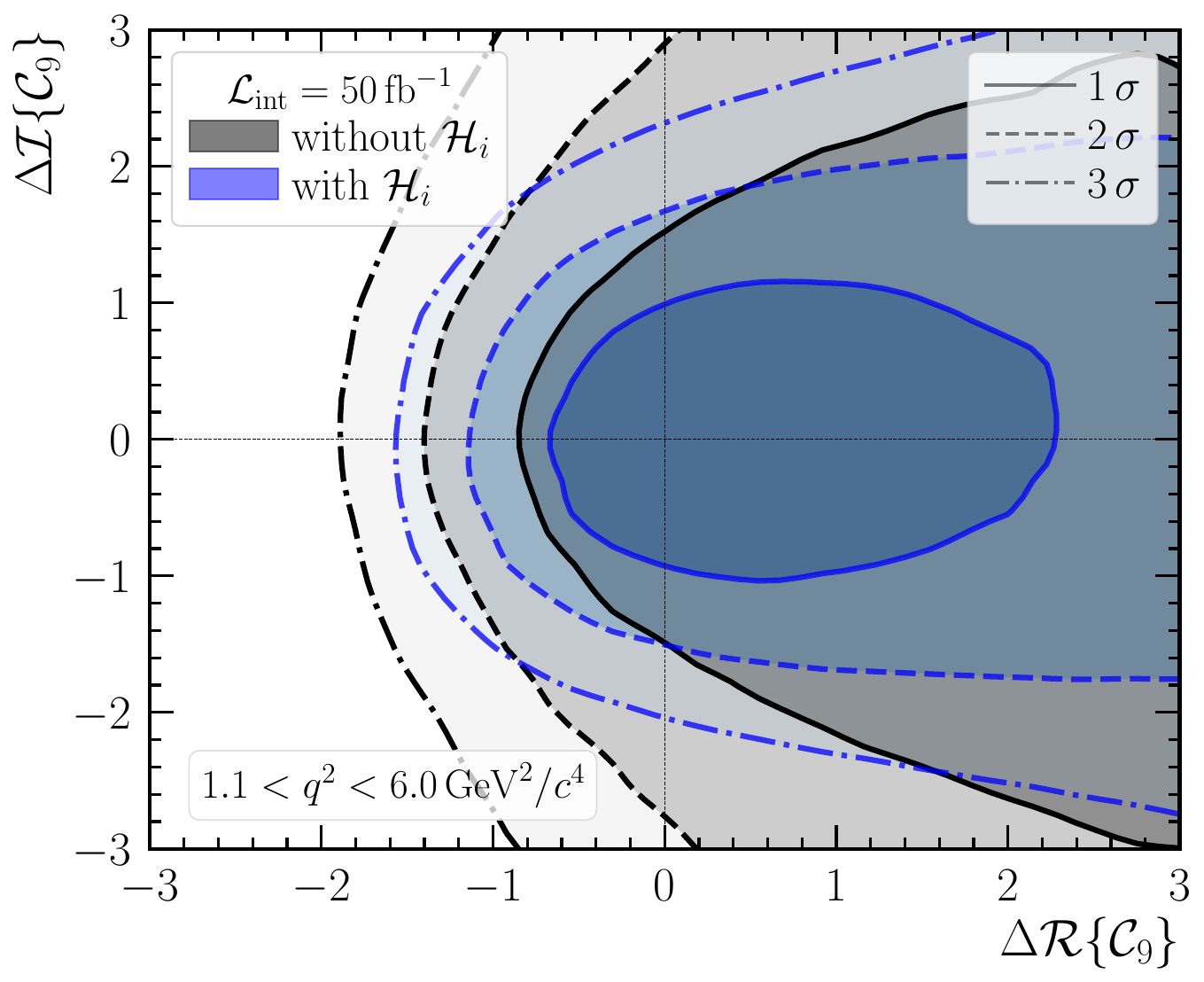}
    \end{minipage}
    
    \caption{Comparison of the constraints obtained for the $\mathcal{R}\{\mathcal{C}_9\}$ and $\mathcal{I}\{\mathcal{C}_9\}$ when fitting the data in the \qsq-region between $0.1$ and $0.98\gevgevcccc$ (left) and $1.1$ and $6.0\gevgevcccc$ (right) when including the $\mathcal{H}_i$. }
    \label{fig:wcfit-time-comp-cqsq}
\end{figure}

A comparison of the constraints obtained for the different fit-setups is shown in Fig.~\ref{fig:np-fits-2d-obs-comp} for an integrated luminosity of $50\invfb$ and a flavour-tagging power of $6\,\%$. 
The constraints on the imaginary part of the Wilson coefficients are improved when including the time-dependent observables and almost doubled when we include the full set of observables accessible if flavour-tagging is performed. 
For the comparison of the time-dependent and time-independent setups, the $\mathcal{H}_i$ are removed from the list of observables that are fitted. 
The improvement in sensitivity from measuring the $\mathcal{H}_i$ in a single \qsq-region is greater than that obtained when combining multiple \qsq-regions. 
Figure~\ref{fig:wcfit-time-comp-cqsq} shows the constraints obtained when including the $\mathcal{H}_i$ and fitting only the data in the \qsq-intervals from $0.1$ to $0.98\gevgevcccc$ and $1.1$ to $6.0\gevgevcccc$, respectively. 
An improved per-\qsq sensitivity to the Wilson coefficients is valuable because it enables more precise fits in the different \qsq-regions, where the relative impact of nonlocal and nonfactorisable SM contributions differ. 
The treatment of these nonfactorisable contributions is under active scrutiny from the theory community, see for example Ref.~\cite{Gubernari:2022hxn}, making it particularly valuable to study how the best fit-value of the Wilson coefficients vary over \qsq. 

\section{Conclusion and Discussion}
\label{sec:conclusion}

The \decay{\Bs}{\phi\mumu} decays provide an excellent opportunity to study the SM and constrain possible NP. 
We present new optimised observables associated with this decay and also illustrate the possibilities to perform a tagged and time-dependent angular analysis at hadronic colliders. 
The resulting sensitivities to the angular observables are evaluated for various milestones of collected \decay{\Bs}{\Pphi \mumu} decays and are compared to those of the SM predictions. 
For the extrapolations, we use the expected performance of the \lhcb upgrade I detector as a benchmark for the potential performance of a dedicated flavour-physics experiment at a hadronic machine.  
The uncertainty on the performance is taken into account by varying the expected properties within reasonable margins. 
The conclusions do not depend on those details. 

We present the tagged time-dependent PDF for \decay{\Bs}{\Pphi \mumu} and, by fitting this to realistic pseudoexperiments, we demonstrate for the first time that the complete set of angular observables associated with \decay{\Bs}{\Pphi \mumu} can be extracted with $\mathcal{O}(10^3)$ signal events under realistic assumptions for background levels and decay-time resolution expected at a hadronic machine. 
We show that this extraction can significantly improve constraints on the underlying Wilson coefficients, and that the sensitivity to their imaginary parts can be 
further enhanced through the optimized observables proposed in this work. 
Among these new observables, $Q_4^\prime$ and $M_8^\prime$ demonstrate a marked gain in sensitivity to short distance new physics effects, compared to their unoptimised counterparts. 
These observables should be measurable with a precision of around $0.3$ in the central-\qsq region by the end of the LHC era, assuming an effective tagging power between $4\text{--}6\,\%$. 
Of the tagging-accessible observables, $P_5^{t\prime}$ is particularly interesting, as its analogue in \decay{\Bz}{\Kstarz\mumu} has exhibited significant tensions with the SM. 
We project that $P_5^{t\prime}$ can be measured with a precision of up to $0.1$ in the central \qsq region, again assuming an effective tagging power between $4\text{--}6\,\%$, by the end of the LHC era. 
Moreover, while the greatest sensitivity is achieved with a tagged analysis, we demonstrate that even an untagged time-dependent measurement, which is experimentally not as challenging as a tagged measurement, yields substantial improvements compared to the time-integrated analyses performed to date. 
This is especially true when examining the variation of the extracted Wilson coefficients across different \qsq-regions, which can provide crucial insight into the origin of the shift in $\mathcal{R}\{C_9\}$ associated with the flavour anomalies.
 \pagebreak 
\clearpage

\section*{Acknowledgements}

\noindent We express our gratitude to Martin Novoa-Brunet, Keri Vos, Peter Stangl, and Méril Reboud for the discussions about the implementation of the angular coefficients into flavio and eos. 
This project received support from the National Science Foundation Grants No. \texttt{CSSI-2411204} and No. \texttt{PHY-2310073}. 

\section*{Data Availability}

The data that support the findings in this article are openly available~\cite{code-repository-phimumu}, embargo periods may apply. 

 \pagebreak 
\clearpage

\section*{Appendices}

\appendix

\section{Angular amplitudes}
\label{app:angular-amplitudes}

\begin{figure}[tb!]
    \centering
    \includegraphics[width=.8\textwidth]{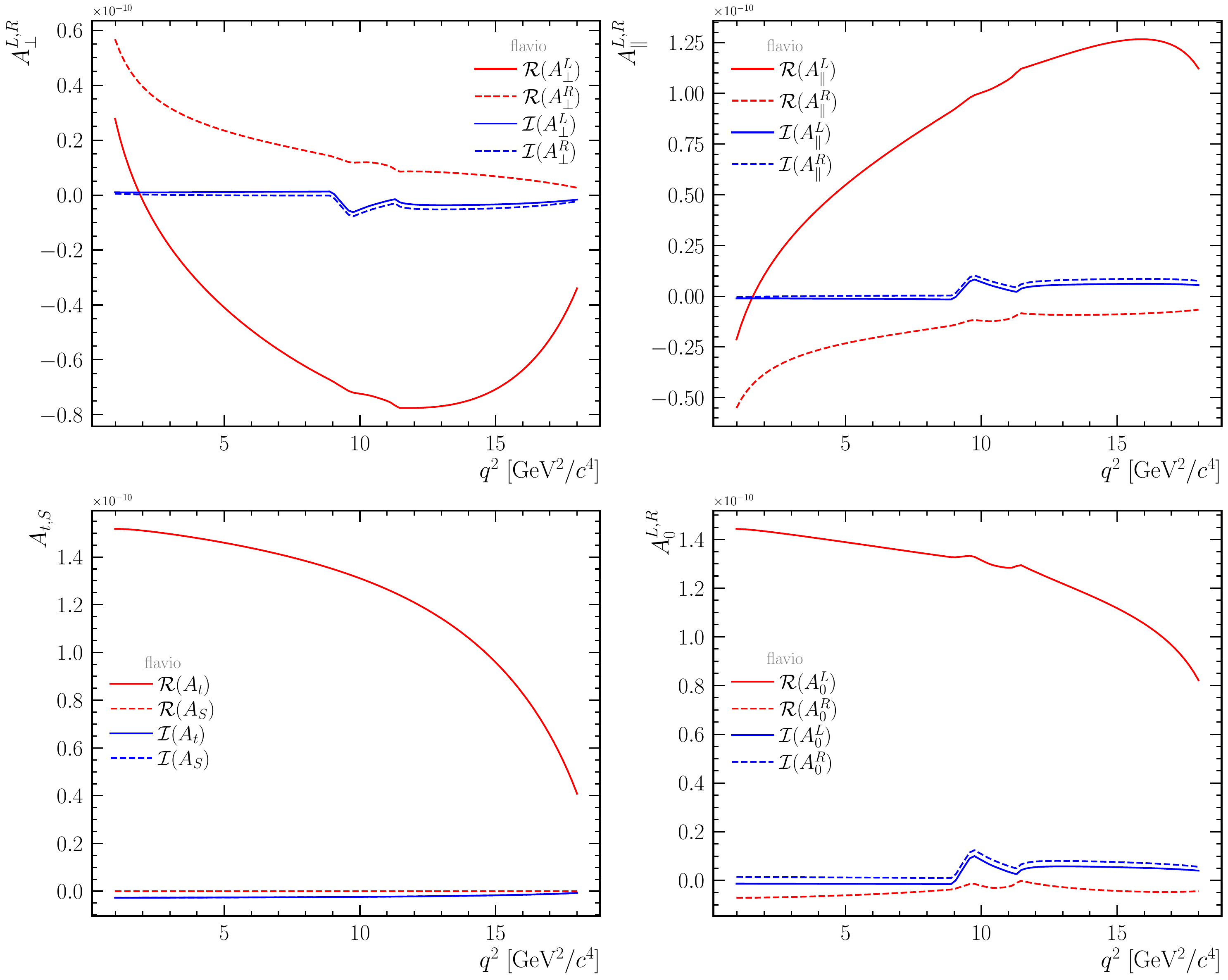}
    \caption{Real and imaginary parts of the transversity amplitudes depending on \qsq implemented in the flavio package~\cite{straub:2018flavio}. }
    \label{fig:app:transversity-amps}
\end{figure}

Following Ref.~\cite{Altmannshofer:2008dz}, the eight transversity amplitudes $A_X^{(H)}$ can be expressed using Wilson coefficients $\mathcal{C}_i$ and seven form-factors $A_i$, $V$, and $T_i$. 
We use effective Wilson coefficients where contributions of the quark-current operators are absorbed into the photon, vector, and axial-vector couplings, $\mathcal{C}_{7,9,10}$. 
The transversity amplitudes are then given by
\begin{align}
    A_{\perp}^{L(R)} & = N\sqrt{2\lambda} \left[ (\mathcal{C}_9^{+} \varmp \mathcal{C}_{10}^{+}) \frac{V}{m_B + m_V} + \frac{2 m_B T_1}{\qsq} \mathcal{C}_7^{+} \right], \\
    A_{\parallel}^{L(R)} & = -N \sqrt{2}(m_B^2 - m_V^2)\left[ (\mathcal{C}_9^{-} \varmp \mathcal{C}_{10}^{-}) \frac{A_1}{m_B - m_V} + \frac{2 m_b T_2}{\qsq} \mathcal{C}_7^{-} \right], \\
    A_{0}^{L(R)} & = - \frac{N}{2 m_V\sqrt{\qsq}} \Biggl [ (\mathcal{C}_9^{-} \varmp \mathcal{C}_{10}^{-}) \\ \notag
    & \hspace{1.3em} \hspace{5.25em} \left( \{ m_B^2 - m_V^2 - \qsq \} \{ m_B + m_V \} A_1 - \frac{\lambda A_2}{m_B + m_V} \right) \\
    & \hspace{1.3em} \hspace{5.25em} \left. + 2m_b \mathcal{C}_7^{-} \left( \{ m_B^2 + 3 m_V^2 - \qsq \} T_2 - \frac{\lambda T_3}{m_B^2 - m_V^2} \right) \right], \notag \\
    A_t & = N \sqrt{\frac{\lambda}{\qsq}} A_0 \left[ 2 \mathcal{C}_{10}^{-} + \frac{\qsq}{m_\ell} \mathcal{C}_P^{-} \right], \\
    A_S & = -2 N \sqrt{\lambda} \mathcal{C}_{S}^- A_0,
\end{align}
where the shorthand notation $\mathcal{C}^\pm_i = \mathcal{C}_i \pm \mathcal{C}_i^\prime$ is used and 
\begin{equation}
    N = V_{tb}V_{ts}^{\ast} \sqrt{ \frac{ G_F^2 \alpha_{\rm em.}^2 \qsq \beta_\ell}{ 3 \cdot 2^{10} \pi^5 m_B^3 } } \lambda^{\frac{1}{4}},
\end{equation}
with $\lambda$ being the Källén function
\begin{equation}
    \lambda = m_B^4 + m_V^4 + q^4 - 2(m_B^2m_V^2 + m_V^2\qsq + m_B^2\qsq),
\end{equation}
and $\beta_\ell = \sqrt{1 - 4 m_\ell^2/\qsq}$. 
The real and imaginary parts of the transversity amplitudes are visualised in Fig.~\ref{fig:app:transversity-amps}. 

Using the transversity amplitudes, the angular coefficients are written following Refs.~\cite{Descotes-Genon:2015hea,Altmannshofer:2008dz} as
\begin{align}
    J_{1s} &= \frac{2 + \beta_\ell^2}{4} \left[ |A_\perp^L|^2 + |A_\parallel^L|^2 + (L \rightarrow R) + \frac{4 m_\ell^2}{\qsq} \mathcal{R}\{A_\perp^L A_\perp^{R\ast} + (L \rightarrow R)\} \right], \\
    J_{1c} &= |A_0^L|^2 + |A_0^R|^2 + \frac{4m_\ell^2}{\qsq}\left[ |A_t|^2 + 2\mathcal{R}\{A_0^L A_0^{R\ast}\} \right] + \beta_\ell^2 |A_S|^2, \\
    J_{2s} &= \frac{\beta_\ell^2}{4} \left[ |A_\perp^L|^2 + |A_\parallel^L|^2 + (L \rightarrow R) \right], \hspace{1cm} J_{2c} = -\beta_\ell^2 \left[ |A_0^L|^2 + |A_0^R|^2 \right], \\
    J_3 &= \frac{\beta_\ell^2}{2} \left[ |A_\perp^L|^2 - |A_\parallel^L|^2 + (L \rightarrow R) \right], 
    \hspace{1cm} J_4 = \frac{\beta_\ell^2}{\sqrt{2}} \mathcal{R}\{A_0^L A_\parallel^{L\ast} + (L \rightarrow R)\}, 
\end{align}
\begin{align}
    J_5 &= \sqrt{2}\beta_\ell \left[ \mathcal{R}\{ A_0^L A_\perp^{L\ast} - (L \rightarrow R) \} - \frac{m_\ell}{\sqrt{\qsq}} \mathcal{R}\{A_\parallel^L A_S^\ast + (L \rightarrow R)\} \right], \\
    J_{6s} &= 2 \beta_\ell \mathcal{R}\{ A_\parallel^L A_\perp^{L\ast} - (L \rightarrow R) \}, 
    \hspace{1cm} J_{6c} = 4\beta_\ell \frac{m_\ell}{\sqrt{\qsq}} \mathcal{R}\{ A_0^L A_S^\ast + (L \rightarrow R) \}, \\
    J_7 &= \sqrt{2}\beta_\ell \left[ \mathcal{I}\{ A_0^L A_\parallel^{L\ast} - (L \rightarrow R) \} + \frac{m_\ell}{\sqrt{\qsq}} \mathcal{I}\{ A_\perp^L A_S^\ast - (L \rightarrow R) \}\right], \\
    J_8 &= \frac{\beta_\ell^2}{\sqrt{2}} \mathcal{I}\{ A_0^L A_\perp^{L\ast} + (L \rightarrow R) \}, 
    \hspace{1.25cm} J_9 = \beta_\ell^2 \mathcal{I}\{ A_\parallel^{L\ast} A_\perp^{L} + (L \rightarrow R) \}
\end{align}
From these expressions the simplifications of the PDF for the massless lepton case that are mentioned in the main body become apparent\footnote{For the second relation also $|A_S| = 0$ in the SM is used.}, as 
\begin{align}\label{app:eq:massless-approx}
    J_{1s} = 3 J_{2s}\text{ and } J_{1c} = -J_{2c}. 
\end{align}
The coefficients $\bar{J}_i$ are obtained by substituting $A_X$ with $\bar{A}_X$, where all weak phases are complex-conjugated. 

\begin{figure}
    \centering
    \begin{minipage}{.49\textwidth}
    \includegraphics[width=\textwidth]{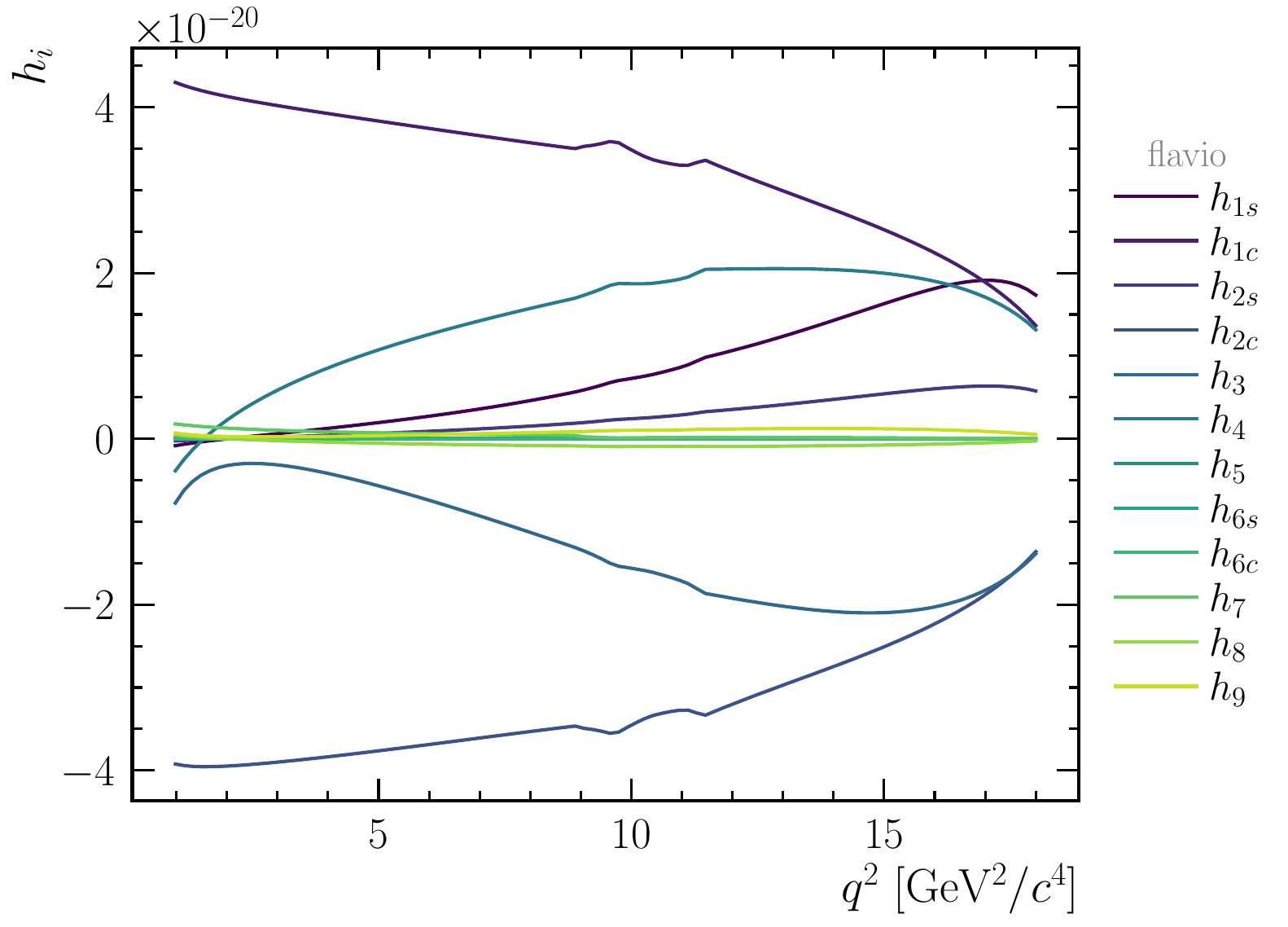}
    \end{minipage}\hspace{.019\textwidth}%
    \begin{minipage}{.49\textwidth}
    \includegraphics[width=\textwidth]{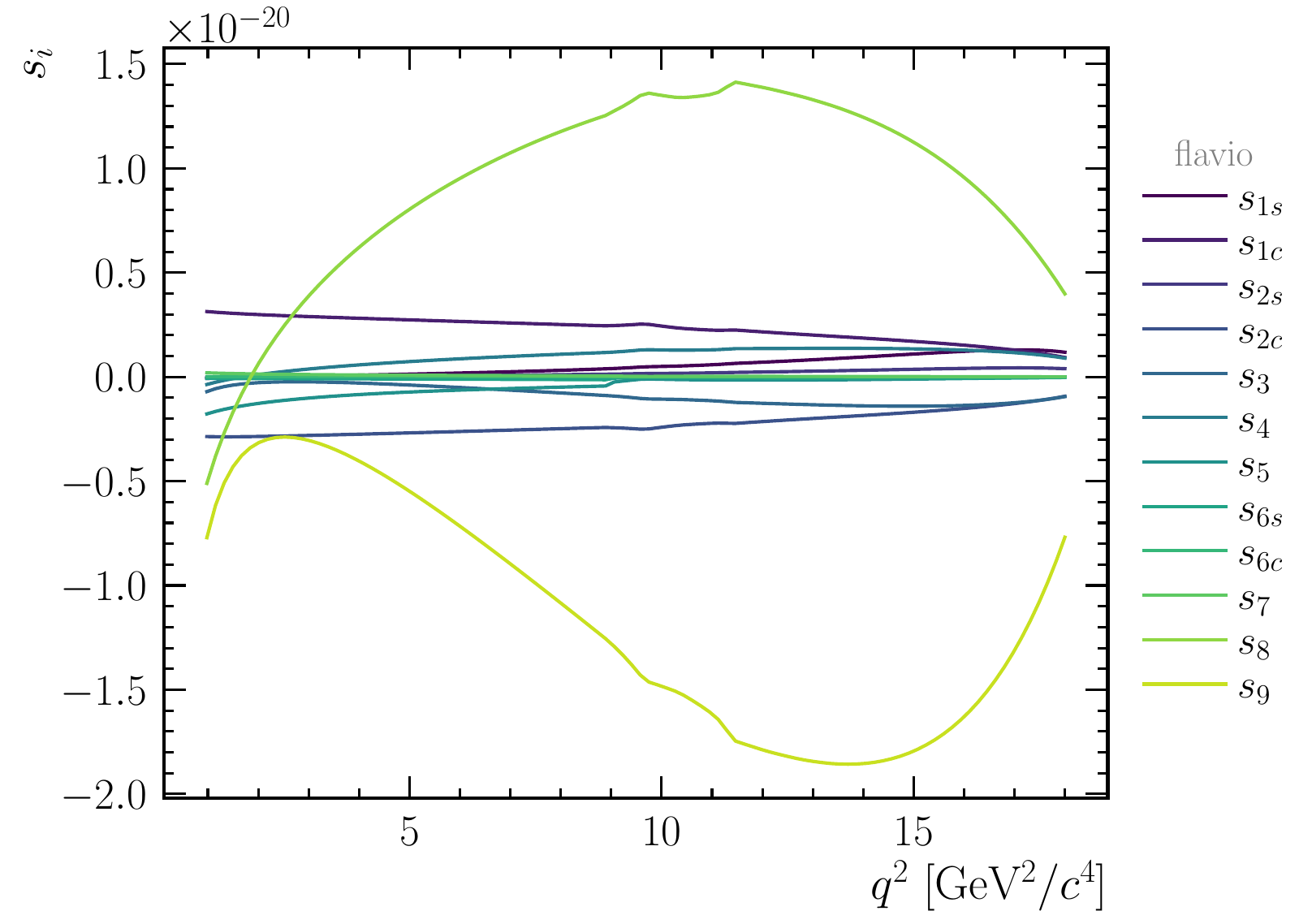}
    \end{minipage}%
    \caption{Angular coefficients $h_i$ (left) and $s_i$ (right) implemented within flavio~\cite{straub:2018flavio}, the $s_i$ have been implemented for this work. At large \qsq, the relations obtained for the massless-lepton case are visible for the $h_i$ and $s_i$. }
    \label{app:fig:s_i-and-h_i}
\end{figure}

The angular coefficients $h_i$ and $s_i$ that enter the time-dependent decay rates for \decay{\Bs}{\Pphi\mumu} can also be expressed using the transversity amplitudes. Following Ref.~\cite{Descotes-Genon:2015hea}, the $h_i$ read
\begin{align}
    h_{1s} &= \frac{2 + \beta_\ell^2}{2} \Re\{ \eip [ \tilde{A}_\perp^L A_\perp^{L\ast} + \tilde{A}_\parallel^L A_\parallel^{L \ast} + (L \rightarrow R) ] \} \\ 
    & \hspace{1.3em} + \frac{4 m_\ell^2}{\qsq} \Re\{ \eip [ \tilde{A}_\perp^L A_\perp^{R\ast} + \tilde{A}_\parallel^L A_\parallel^{R\ast} ] + \emip [ A_\perp^L \tilde{A}_\perp^{R\ast} + A_\parallel^L \tilde{A}_\parallel^{R \ast} ] \} , \notag \\
    h_{1c} &= 2 \Re\{ \eip [ \tilde{A}_0^L A_0^{L\ast} + \tilde{A}_0^R A_0^{R \ast} ] \} + 2 \beta_
    \ell^2 \Re\{ \eip \tilde{A}_S A_S^\ast \}  \\ 
    & \hspace{1.3em} + \frac{8 m_\ell^2}{\qsq} \left[ \Re\{ \tilde{A}_t A_t^\ast \} + \Im\{ \eip \tilde{A}_0^L A_0^{R \ast} + \emip A_0^L \tilde{A}_0^{R\ast} \} \right], \notag \\
    h_{2s} &= \frac{\beta_\ell^2}{2} \Re\{ \eip [ \tilde{A}_\perp^L A_\perp^{L\ast} + \tilde{A}_\parallel^L A_\parallel^{L \ast} + \tilde{A}_\perp^R A_\perp^{R\ast} + \tilde{A}_\parallel^R A_\parallel^{R \ast} ] \}, \\
    h_{2c} &= -2 \beta_\ell^2 \Re\{ \eip [ \tilde{A}_0^L A_0^{L\ast} + \tilde{A}_0^R A_0^{R \ast} ] \}, \\
    h_3 &= \beta_\ell^2 \Re\{ \eip [ \tilde{A}_\perp^L A_\perp^{L\ast} - \tilde{A}_\parallel^L A_\parallel^{L\ast} + (L \rightarrow R) ] \}, \\
    h_4 &= \frac{\beta_\ell^2}{\sqrt{2}} \Re\{ \eip [ \tilde{A}_0^L A_\parallel^{L\ast} + \tilde{A}_0^R A_\parallel^{R\ast} ] + \emip [ A_0^L \tilde{A}_\parallel^{L\ast} + A_0^R \tilde{A}_\parallel^{R\ast} ] \}, \\
    h_5 &= \sqrt{2} \beta_\ell \Bigg [ \Re\{ \eip [ \tilde{A}_0^L A_\perp^{L\ast} - \tilde{A}_0^R A_\perp^{R\ast} ] + \emip [ A_0^L \tilde{A}_\perp^{L\ast} - A_0^R \tilde{A}_\perp^{R\ast} ] \} \\
    & \hspace{1.3em} \hspace{2.65em} - \frac{m_\ell}{\sqrt{\qsq}} \Re\{ \eip [ \tilde{A}_\parallel^L A_S^\ast + \tilde{A}_\parallel^R A_S^\ast ] + \emip [ A_\parallel^L \tilde{A}_S^\ast + A_\parallel^R \tilde{A}_S^\ast ] \} \Bigg ], \notag \\
    h_{6s} &= 2\beta_\ell \Re\{ \eip [ \tilde{A}_\parallel^L A_\perp^{L\ast} - \tilde{A}_\parallel^R A_\perp^{R\ast} ] + \emip [ A_\parallel^L \tilde{A}_\perp^{L\ast} - A_\parallel^R \tilde{A}_\perp^{R\ast} ] \}, \\
    h_{6c} &= \frac{4 \beta_\ell m_\ell}{\sqrt{\qsq}} \Re\{ \eip [ \tilde{A}_0^L A_S^\ast + \tilde{A}_0^R A_S^\ast ] + \emip [ A_0^L \tilde{A}_S^\ast + A_0^R \tilde{A}_S^\ast ] \}, \\
    h_7 &= \sqrt{2}\beta_\ell \Bigg [ \Im\{ \eip [ \tilde{A}_0^L A_\parallel^{L\ast} - \tilde{A}_0^R A_\parallel^{R\ast} ] + \emip [ A_0^L \tilde{A}_\parallel^{L\ast} - A_0^R \tilde{A}_\parallel^{R\ast} ] \}, \\
    & \hspace{1.3em} \hspace{2.65em} + \frac{m_\ell}{\sqrt{\qsq}} \Im\{ \eip [ \tilde{A}_\perp^L A_S^\ast + \tilde{A}_\perp^R A_S^\ast ] + \emip [ A_\perp^L \tilde{A}_S^\ast + A_\perp^R \tilde{A}_S^\ast ] \} \Bigg] , \notag \\
    h_8 &= \frac{\beta_\ell^2}{\sqrt{2}} \Im\{ \eip [ \tilde{A}_0^L A_\perp^{L\ast} + \tilde{A}_0^R A_\perp^{R\ast} ] + \emip [ A_0^L \tilde{A}_\perp^{L\ast} + A_0^R \tilde{A}_\perp^{R\ast} ] \}, \\
    h_9 &= -\beta_\ell^2 \Im\{ \eip [ \tilde{A}_\parallel^L A_\perp^{L\ast} + \tilde{A}_\parallel^R A_\perp^{R\ast} ] + \emip [ A_\parallel^L \tilde{A}_\perp^{L\ast} + A_\parallel^R \tilde{A}_\perp^{R\ast} ] \},
\end{align}
where $\eip = -q/p$, which is in agreement with latest combinations for the \Bs system~\cite{HeavyFlavorAveragingGroupHFLAV:2024ctg}. 
It can be seen that the same relations hold for the massless lepton approximation, namely $h_{1s} = 3 h_{2s}$ and $h_{1c} = -h_{2c}$. 
The $s_i$ have a similar form to the $h_i$, notably exchanging $\mathcal{R}$ for $\mathcal{I}$. 
The $s_i$ coefficients are also given in Ref.~\cite{Descotes-Genon:2015hea} and read
\begin{align}
    s_{1s} &= \frac{2 + \beta_\ell^2}{2} \Im\{ \eip [ \tilde{A}_\perp^L A_\perp^{L\ast} + \tilde{A}_\parallel^L A_\parallel^{L \ast} + (L \rightarrow R) ] \} \\ 
    & \hspace{1.3em} + \frac{4 m_\ell^2}{\qsq} \Im\{ \eip [ \tilde{A}_\perp^L A_\perp^{R\ast} + \tilde{A}_\parallel^L A_\parallel^{R\ast} ] - \emip [ A_\perp^L \tilde{A}_\perp^{R\ast} + A_\parallel^L \tilde{A}_\parallel^{R \ast} ] \} , \notag \\
    s_{1c} &= 2 \Im\{ \eip [ \tilde{A}_0^L A_0^{L\ast} + \tilde{A}_0^R A_0^{R \ast} ] \} + 2 \beta_
    \ell^2 \Im\{ \eip \tilde{A}_S A_S^\ast \}  \\ 
    & \hspace{1.3em} + \frac{8 m_\ell^2}{\qsq} \left[ \Im\{ \tilde{A}_t A_t^\ast \} + \Im\{ \eip \tilde{A}_0^L A_0^{R \ast} - \emip A_0^L \tilde{A}_0^{R\ast} \} \right], \notag 
\end{align}
\begin{align}
    s_{2s} &= \frac{\beta_\ell^2}{2} \Im\{ \eip [ \tilde{A}_\perp^L A_\perp^{L\ast} + \tilde{A}_\parallel^L A_\parallel^{L \ast} + \tilde{A}_\perp^R A_\perp^{R\ast} + \tilde{A}_\parallel^R A_\parallel^{R \ast} ] \}, \\
    s_{2c} &= -2 \beta_\ell^2 \Im\{ \eip [ \tilde{A}_0^L A_0^{L\ast} + \tilde{A}_0^R A_0^{R \ast} ] \}, \\
    s_3 &= \beta_\ell^2 \Im\{ \eip [ \tilde{A}_\perp^L A_\perp^{L\ast} - \tilde{A}_\parallel^L A_\parallel^{L\ast} + (L \rightarrow R) ] \}, \\
    s_4 &= \frac{\beta_\ell^2}{\sqrt{2}} \Im\{ \eip [ \tilde{A}_0^L A_\parallel^{L\ast} + \tilde{A}_0^R A_\parallel^{R\ast} ] - \emip [ A_0^L \tilde{A}_\parallel^{L\ast} + A_0^R \tilde{A}_\parallel^{R\ast} ] \}, \\
    s_5 &= \sqrt{2} \beta_\ell \Bigg [ \Im\{ \eip [ \tilde{A}_0^L A_\perp^{L\ast} - \tilde{A}_0^R A_\perp^{R\ast} ] - \emip [ A_0^L \tilde{A}_\perp^{L\ast} - A_0^R \tilde{A}_\perp^{R\ast} ] \} \\
    & \hspace{1.3em} \hspace{2.65em} - \frac{m_\ell}{\sqrt{\qsq}} \Im\{ \eip [ \tilde{A}_\parallel^L A_S^\ast + \tilde{A}_\parallel^R A_S^\ast ] - \emip [ A_\parallel^L \tilde{A}_S^\ast + A_\parallel^R \tilde{A}_S^\ast ] \} \Bigg ], \notag \\
    s_{6s} &= 2\beta_\ell \Im\{ \eip [ \tilde{A}_\parallel^L A_\perp^{L\ast} - \tilde{A}_\parallel^R A_\perp^{R\ast} ] - \emip [ A_\parallel^L \tilde{A}_\perp^{L\ast} - A_\parallel^R \tilde{A}_\perp^{R\ast} ] \}, \\
    s_{6c} &= \frac{4 \beta_\ell m_\ell}{\sqrt{\qsq}} \Im\{ \eip [ \tilde{A}_0^L A_S^\ast + \tilde{A}_0^R A_S^\ast ] - \emip [ A_0^L \tilde{A}_S^\ast + A_0^R \tilde{A}_S^\ast ] \}, \\
    s_7 &= -\sqrt{2}\beta_\ell \Bigg [ \Re\{ \eip [ \tilde{A}_0^L A_\parallel^{L\ast} - \tilde{A}_0^R A_\parallel^{R\ast} ] - \emip [ A_0^L \tilde{A}_\parallel^{L\ast} - A_0^R \tilde{A}_\parallel^{R\ast} ] \}\\
    & \hspace{1.3em} \hspace{3.2em} + \frac{m_\ell}{\sqrt{\qsq}} \Re\{ \eip [ \tilde{A}_\perp^L A_S^\ast + \tilde{A}_\perp^R A_S^\ast ] - \emip [ A_\perp^L \tilde{A}_S^\ast + A_\perp^R \tilde{A}_S^\ast ] \} \Bigg] , \notag \\
    s_8 &= -\frac{\beta_\ell^2}{\sqrt{2}} \Re\{ \eip [ \tilde{A}_0^L A_\perp^{L\ast} + \tilde{A}_0^R A_\perp^{R\ast} ] - \emip [ A_0^L \tilde{A}_\perp^{L\ast} + A_0^R \tilde{A}_\perp^{R\ast} ] \}, \\
    s_9 &= \beta_\ell^2 \Re\{ \eip [ \tilde{A}_\parallel^L A_\perp^{L\ast} + \tilde{A}_\parallel^R A_\perp^{R\ast} ] - \emip [ A_\parallel^L \tilde{A}_\perp^{L\ast} + A_\parallel^R \tilde{A}_\perp^{R\ast} ] \}.
\end{align}
Also here the same relations for the massless lepton approximation hold. 
Figure~\ref{app:fig:s_i-and-h_i} illustrates the $h_i$ and $s_i$. 
It should be noted, that while $h_{7,8,9} \propto \Im\{A_X A_Y^*\}$ are small the corresponding $s_{7,8,9} \propto \Re\{A_X A_Y^*\}$ are large and vice-versa. 
In the regions where \qsq is large, the relations from Eqs.~\ref{eq:massless-approx} and~\ref{app:eq:massless-approx} can be visually confirmed. 

Alternative expressions for the $J_i$ are given in Ref.~\cite{Gratrex:2015hna} using the helicity amplitudes, $H_X$, in favour of the transversity amplitudes, where the former is used as default in flavio~\cite{straub:2018flavio}. 
The $h_i$ and $s_i$ coefficients can be obtained from the given formulae for the $J_i$ using the substitutions\footnote{thanks also to Martin Novoa-Brunet for discussions on this topic} 
\begin{align}
    J_i \propto |H_X|^2 \Rightarrow \begin{cases} 
    h_i \propto 2 \mathcal{R}\{ e^{i\phi} \tilde{H}_X H_X^\ast \} \\
    s_i \propto 2 \mathcal{I}\{ e^{i\phi} \tilde{H}_X H_X^\ast\}
    \end{cases},
\end{align}
\begin{align}
    J_i \propto H_X H_Y^\ast \Rightarrow \begin{cases} 
    h_i \propto e^{i\phi} \tilde{H}_X H_Y^\ast + e^{-i\phi} H_X \tilde{H}_Y^\ast \\
    s_i \propto e^{i\phi} \tilde{H}_X H_Y^\ast - e^{-i\phi} H_X \tilde{H}_Y^\ast
    \end{cases}\hspace{-1em},
\end{align}
\begin{align}
    J_i \propto \mathcal{R}\{H_X H_Y^\ast\} \Rightarrow \begin{cases} 
    h_i \propto \mathcal{R}\{ e^{i\phi} \tilde{H}_X H_Y^\ast + e^{-i\phi} H_X \tilde{H}_Y^\ast \} \\
    s_i \propto \mathcal{I}\{ e^{i\phi} \tilde{H}_X H_Y^\ast - e^{-i\phi} H_X \tilde{H}_Y^\ast \}
    \end{cases}\hspace{-1em}\text{, and } 
\end{align}
\begin{align}
    J_i \propto \mathcal{I}\{H_X H_Y^\ast\} \Rightarrow \begin{cases} 
    h_i \propto \mathcal{I}\{ e^{i\phi} \tilde{H}_X H_Y^\ast + e^{-i\phi} H_X \tilde{H}_Y^\ast \} \\
    s_i \propto -\mathcal{R}\{ e^{i\phi} \tilde{H}_X H_Y^\ast - e^{-i\phi} H_X \tilde{H}_Y^\ast \}
    \end{cases}\hspace{-1em}.
\end{align} 
We refer the reader to Appendix~C in Ref.~\cite{Altmannshofer:2008dz} for the expressions of the $J_i$ in terms of the helicity amplitudes $H_X$ and further discussion. 
The same substitution rules hold when transforming between $h_i$ and $s_i$ in the transversity amplitude formalism. 

\section{Time acceptance}\label{app:time-acceptance}

\begin{figure}[tb!]
    \centering
    \includegraphics[width=0.45\textwidth]{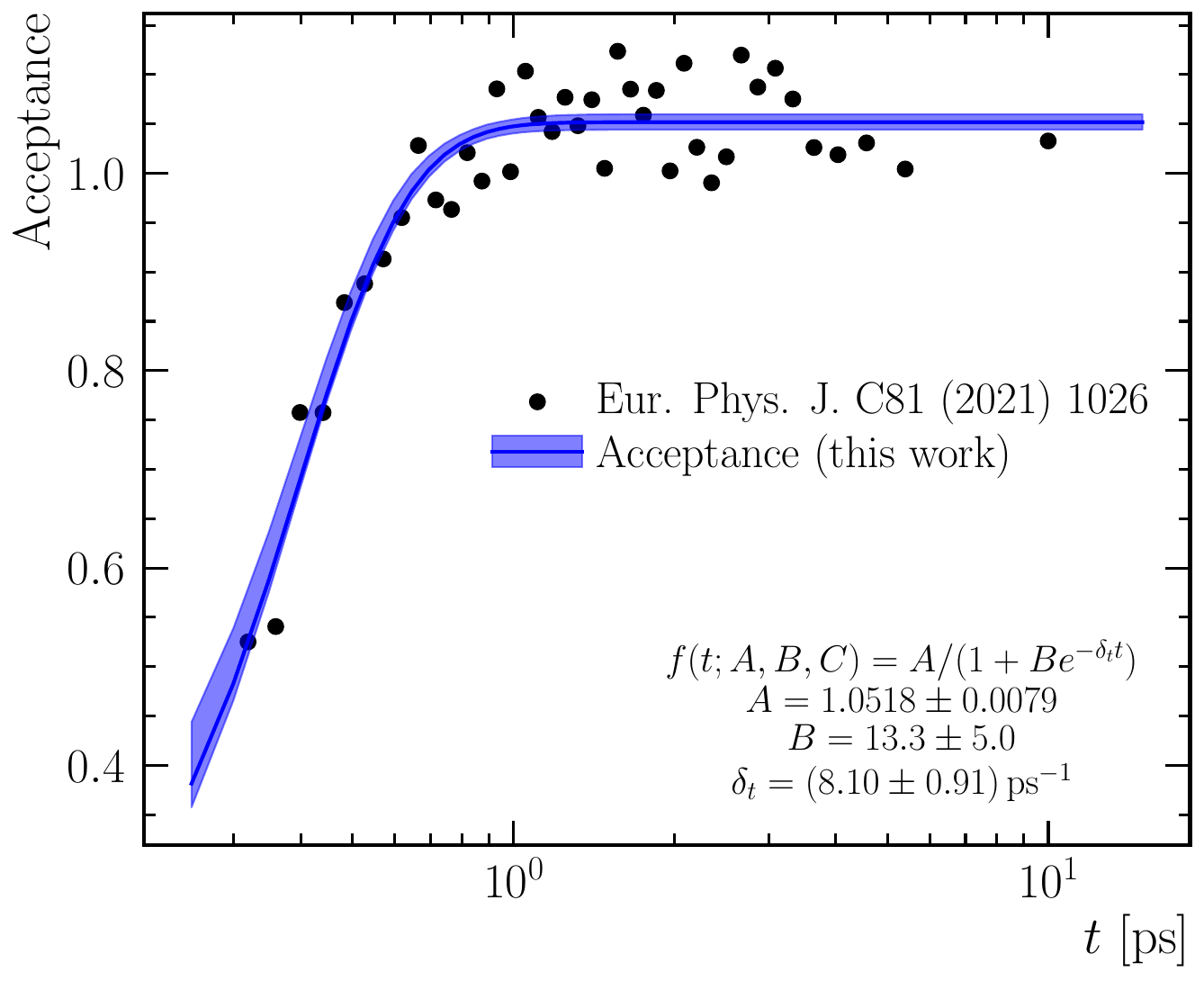}

    \caption{Decay-time acceptance extracted from Ref.~\cite{LHCb-PAPER-2020-042} with our best-fit overlayed. The function captures the behaviour of the acceptance well. The shaded area visualises the $68$th percentiles of the acceptances obtained when randomly sampling the obtained parameters according to the uncertainties. }
    \label{app:fig:time-dependence}    
\end{figure}

In the time-dependent analysis of \decay{\Bs}{\jpsi(\rightarrow e^+e^-)\Pphi(\rightarrow \Kp \Km)} decays, outlined in Ref.~\cite{LHCb-PAPER-2020-042}, the typical \lhcb acceptance depending on the decay-time is shown. 
The points are extracted from Fig.~4 and a simple curve-fit using scipy~\cite{2020SciPy-NMeth} is performed to determine the parameters of the function
\begin{equation}
    \epsilon(t) = \frac{A}{1 + B e^{-\delta_t t}}. 
\end{equation}
The procedure is visualised in Figure~\ref{app:fig:time-dependence}, we find $A = 1.05$, $B = 13.3$, and $\delta_t=8.1\,\mathrm{ps}^{-1}$. 
Note that we neglect the uncertainties on the acceptance, as we are only interested in finding a function that captures the general behaviour of the time-acceptance. 
Our function is positive definite for all decay-times which are analysed, captures the kink in the points well and describes the plateau toward larger decay-times well. 

\FloatBarrier
\section{Fit Projections}\label{app:fit-projections}
\FloatBarrier

\begin{figure}[!tb]
    \centering
    \includegraphics[width=\textwidth]{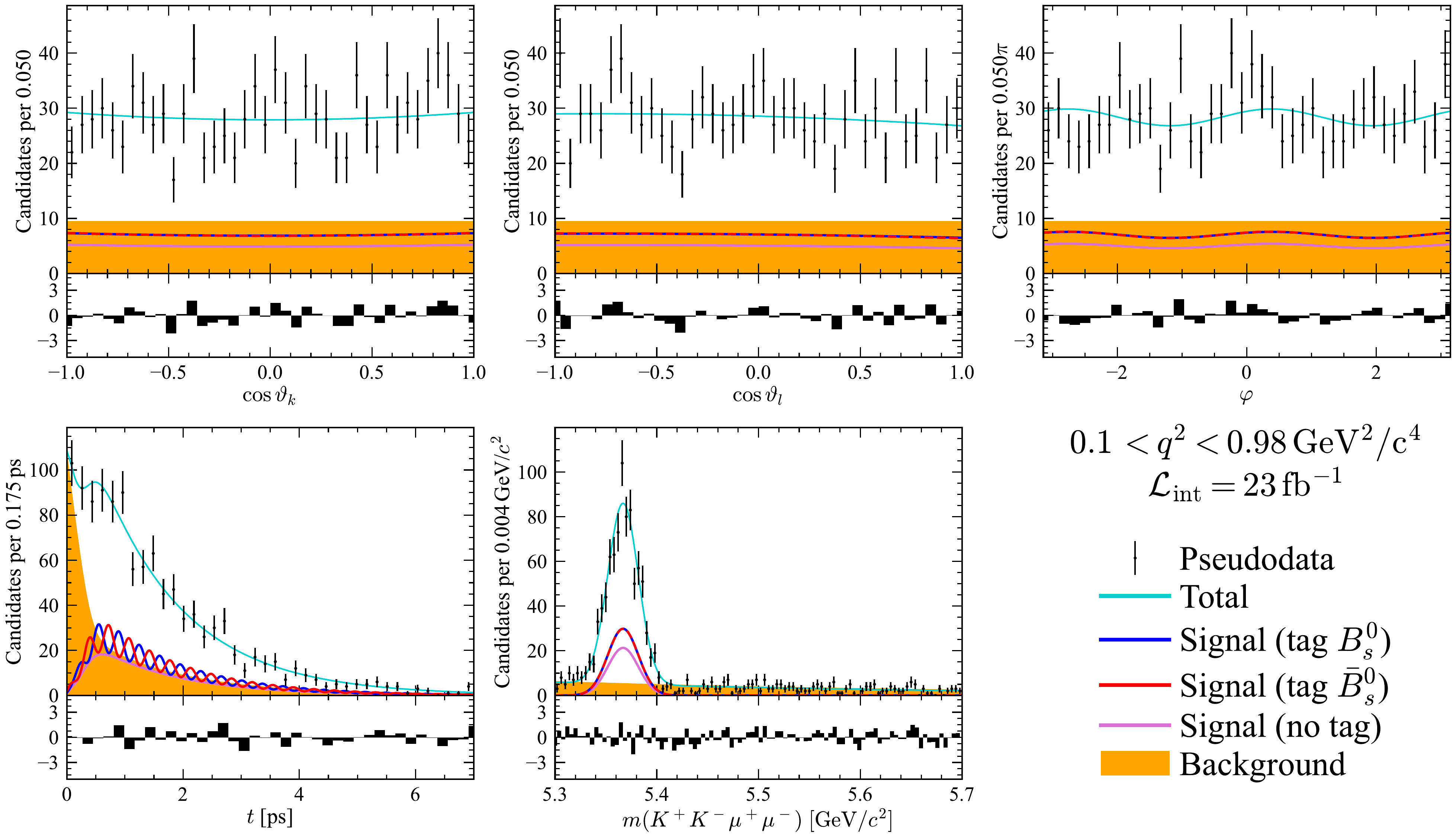}
    \caption{Example of a fit to the generated pseudodata in the region \mbox{$0.1 < \qsq < 0.98\gevgevcccc$} using the model described in the text. The shapes are explained in the text. }
    \label{app:fig:fit-low-23}
\end{figure}

\begin{figure}[!tb]
    \centering
    \includegraphics[width=\textwidth]{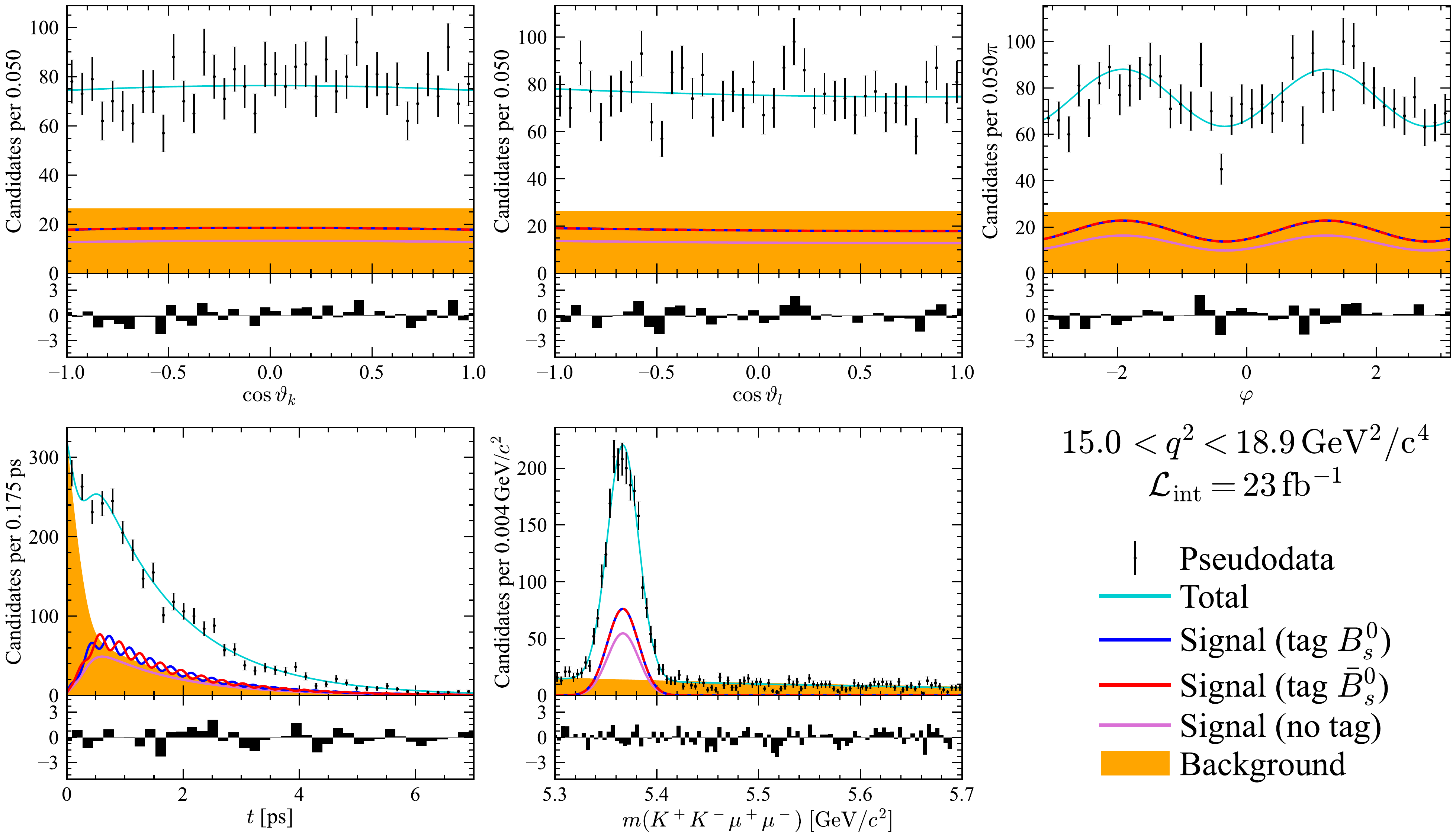}
    \caption{Example of a fit to the generated pseudodata in the region \mbox{$15.0 < \qsq < 18.9\gevgevcccc$} using the model described in the text. The shapes are explained in the text. }
    \label{app:fig:fit-high-23}
\end{figure}

In this section, fit projections for the low \qsq-region from \mbox{$0.1$ to $0.98\gevgevcccc$} and high \qsq-region from $15.0$ to $18.9\gevgevcccc$ are displayed in Figs.~\ref{app:fig:fit-low-23} and~\ref{app:fig:fit-high-23}. 
The signal components are displayed in blue and red for the \Bs and \Bsb components, respectively. 
The untagged component is shown in purple and the combinatorial background in orange. 

\begin{figure}[!tb]
    \centering
    \includegraphics[width=\textwidth]{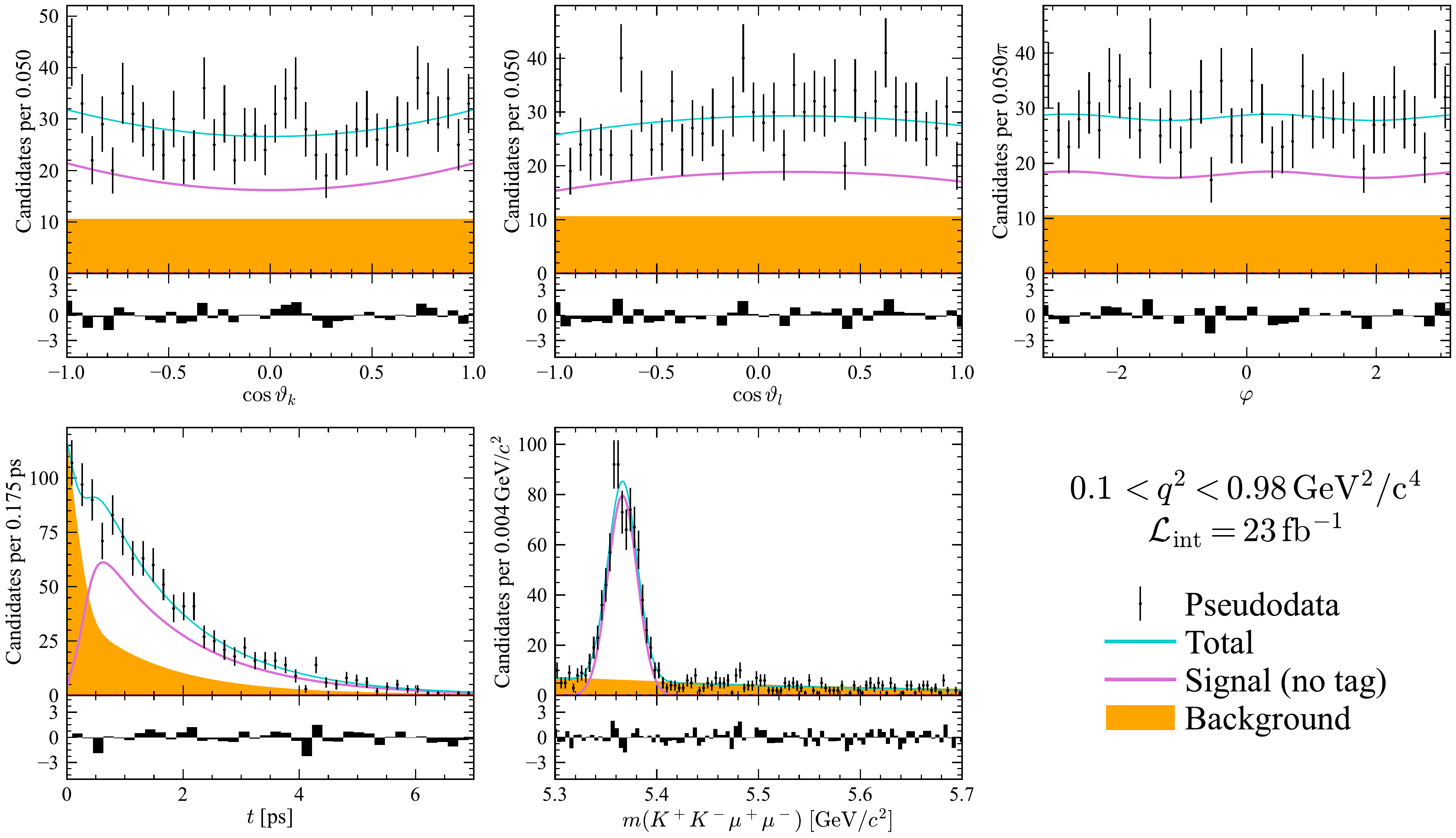}
    \caption{Example of a fit to the generated pseudodata in the region \mbox{$0.1 < \qsq < 0.98\gevgevcccc$} using the model described in the text. The shapes are explained in the text. }
    \label{app:fig:untagged-fit-low-23}
\end{figure}

\begin{figure}[!tb]
    \centering
    \includegraphics[width=\textwidth]{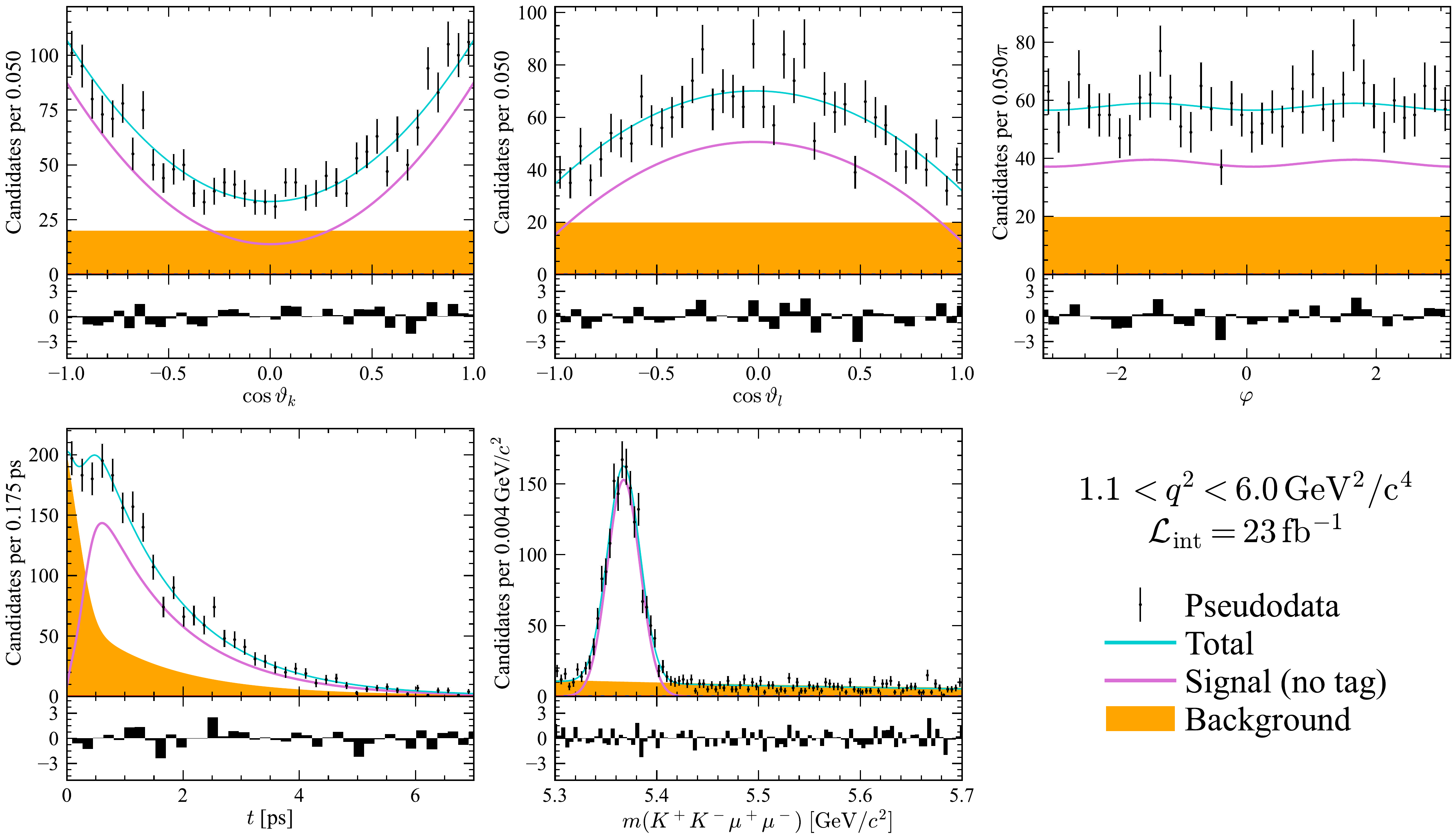}
    \caption{Example of a fit to the generated pseudodata in the region $1.1 < \qsq < 6.0\gevgevcccc$ using the model described in the text. The shapes are explained in the text. }
    \label{app:fig:untagged-fit-central-23}
\end{figure}

\begin{figure}[!tb]
    \centering
    \includegraphics[width=\textwidth]{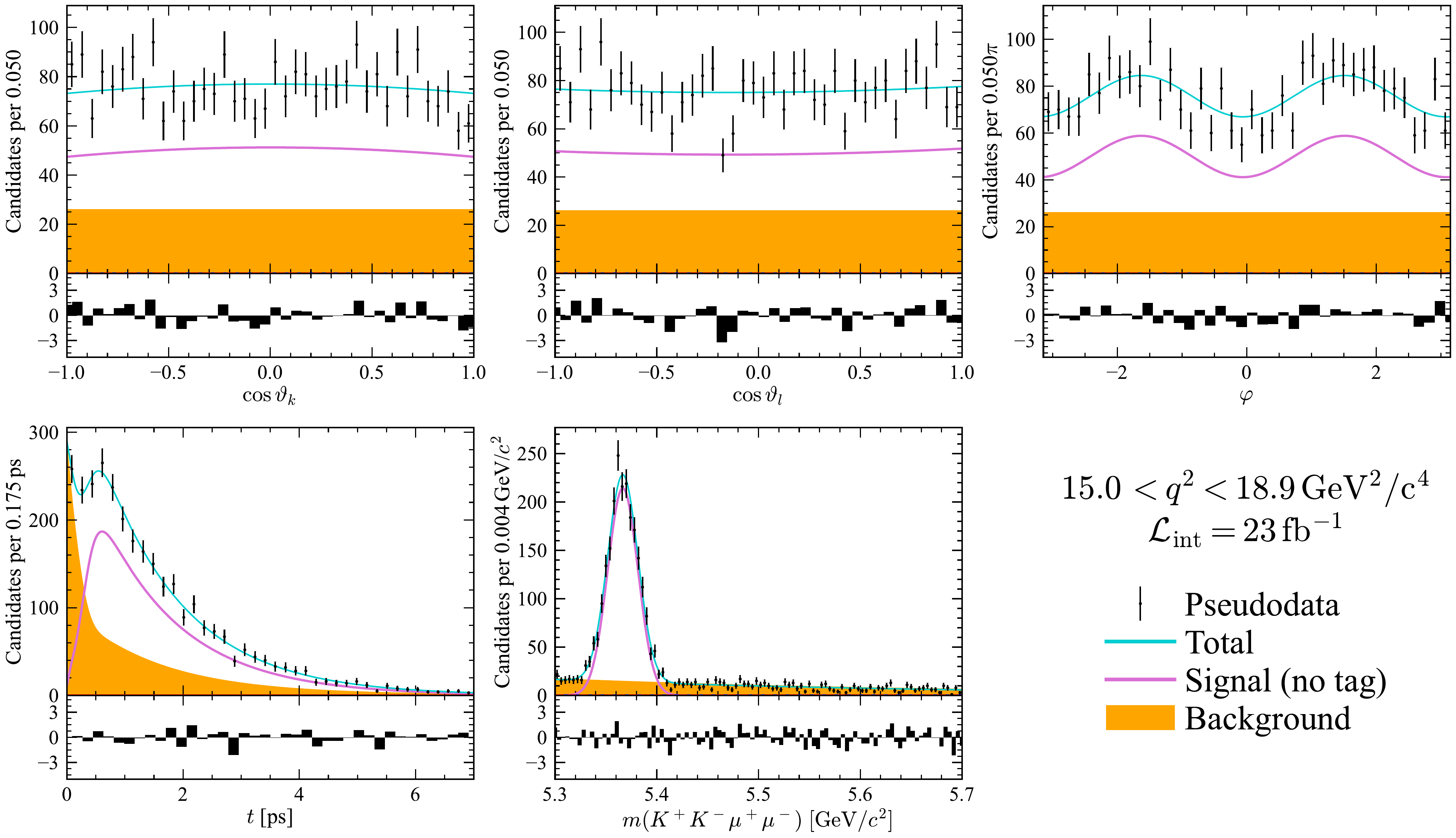}
    \caption{Example of a fit to the generated pseudodata in the region $15.0 < \qsq < 18.9\gevgevcccc$ using the model described in the text. The shapes are explained in the text. }
    \label{app:fig:untagged-fit-high-23}
\end{figure}

The fits in the untagged setup are shown in \cref{app:fig:untagged-fit-low-23,app:fig:untagged-fit-central-23,app:fig:untagged-fit-high-23} for the low-, \mbox{central-,} and high-\qsq regions, respectively. 
The central-\qsq region refers to the region where $1.1 < \qsq < 6.0\gevgevcccc$. 
Here, the signal components for the \Bs and \Bsb are removed and only the untagged sum in purple is shown. 

\FloatBarrier
\section{Fit Validation for the Untagged Fit-Setups}\label{app:pulls-untagged}

\begin{figure}[tb!]
    \centering
    \includegraphics[width=0.95\textwidth]{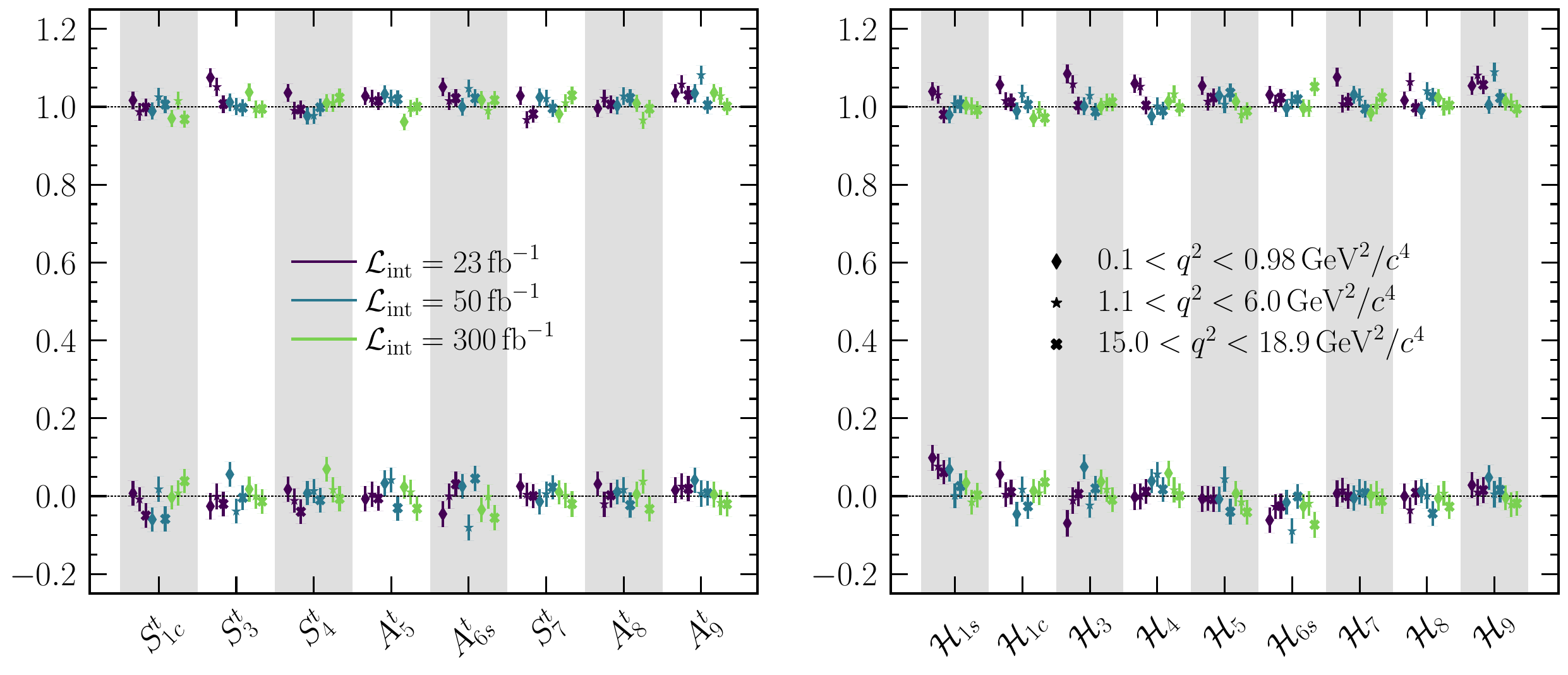}
    \caption{Summary of the bias and coverage for the time-dependent fit without flavour-tagging information. On the left, the $S^t_i$ and $A^t_i$ observables are shown while on the right the $\mathcal{H}_i$ are summarised. }
    \label{app:fig:pull-summary-untagged}
\end{figure}

\begin{figure}[tb!]
    \centering
    \includegraphics[width=0.475\textwidth]{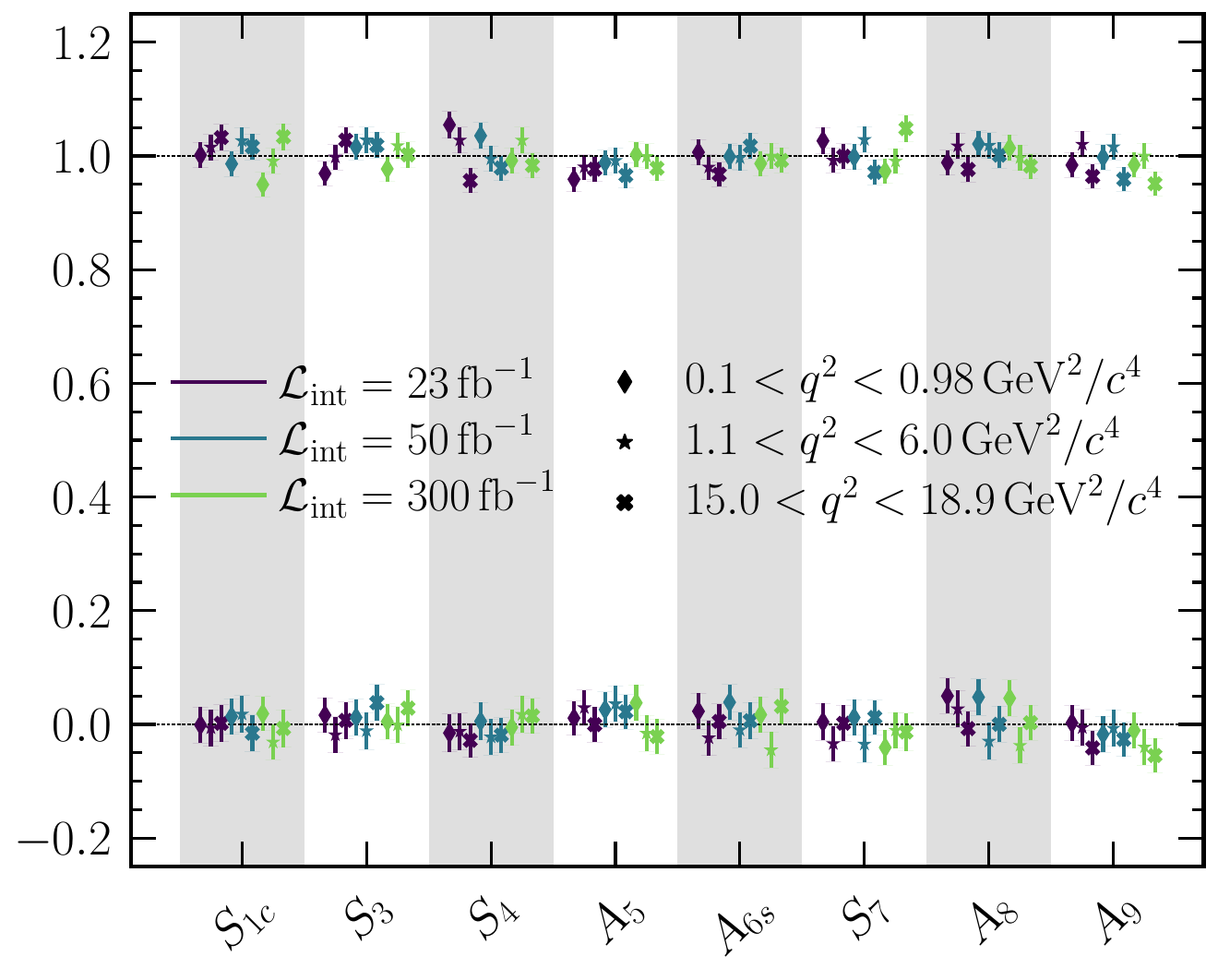}
    \caption{Summary of the bias and coverage for the time-independent fit without flavour-tagging information. }
    \label{app:fig:pull-summary-untagged-timeindependent}
\end{figure}

In this section, the fit validation for the untagged fit-setups are shown. 
Figure~\ref{app:fig:pull-summary-untagged} shows the bias and coverage for the untagged time-dependent fit setup and Fig.~\ref{app:fig:pull-summary-untagged-timeindependent} shows the same for the untagged time-integrated setup. 
Small biases of less than $10\,\%$ of the statistical uncertainty are observed for the $\mathcal{L}_{\rm int}=23\invfb$ setup, which are not present for the larger sample sizes. 
The coverage is generally accurate, as is indicated by the markers being compatible with unity. 
No biases or inaccurate coverages are observed for the time-integrated fit-setup. 

\FloatBarrier 
\section{Sensitivity Extrapolations for the Untagged and Time-Integrated Setup}\label{app:time-idenpendent-results}

\begin{figure}[tb!]
    \centering

    \includegraphics[width=.7\textwidth]{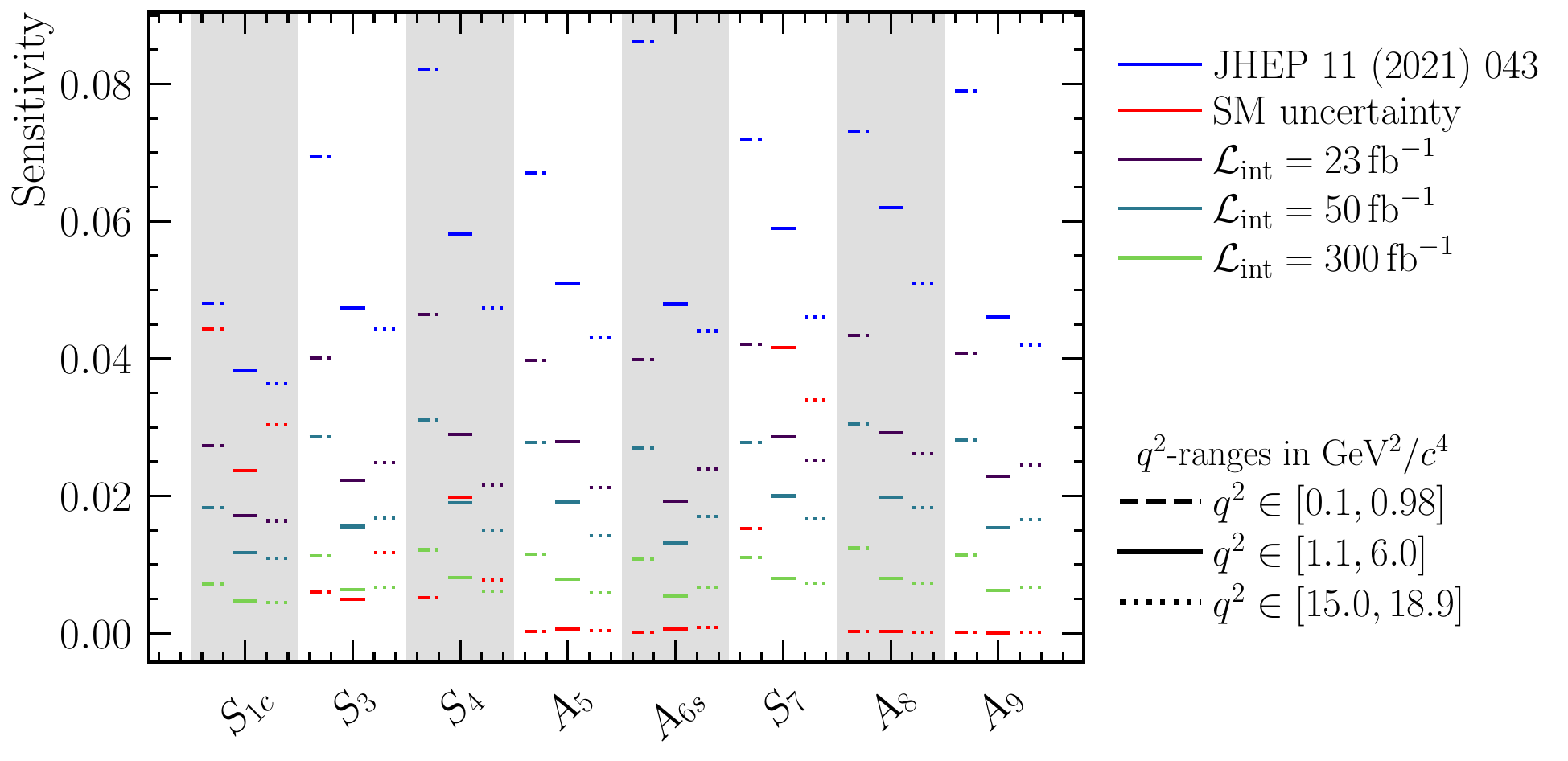}
    \caption{Extrapolated sensitivities to the $S_i$ and $A_i$ observables comparing to both the prediction within the SM evaluated using flavio and the existing measurement by \lhcb. }
    \label{app:fig:time-indep-observables}
    
\end{figure}

In this section, the extrapolations for the time-independent analysis are discussed. 
Effectively, this is equivalent to integrating over the time-dependent $I(\qsq, t)$ and measuring the $S_i$ and $A_i$ observables. 
A measurement in this configuration has been performed using Run~1 and Run~2 data collected by the \lhcb experiment~\cite{LHCb-PAPER-2021-022}. 
Our extrapolations are shown in Fig.~\ref{app:fig:time-indep-observables} and indicate the expected sensitivities which are obtainable for a hypothetical detector with a similar performance compared to the \lhcb upgrade I. 
The larger sample size of \Bs decays is key to improve the sensitivity compared to the existing measurement, leading to improvements over the measurement using data set corresponding to approximately $9\invfb$ of collected luminosity.  

The $S_i$ and $A_i$ are obtained by integrating the full decay-rate over the decay-time and receive contributions from both the $S_i^t$ and $A_i^t$ and the $\mathcal{H}_i$ with the integral of the $\cosh$ and $\sinh$ terms, respectively. 
As derived in Ref.~\cite{Descotes-Genon:2015hea}, for hadronic machines the integral equates to 
\begin{align}
    \langle I_i+\tilde{I}_i \rangle = \frac{1}{\Gamma} \left[ \frac{1}{1-y^2} (J_i + \zeta_i \bar{J}_i) - \frac{y}{1-y^2}h_i \right],
\end{align}
where $I_i + \tilde{I}_i$ relates to $S_i$ and $A_i$, respectively\footnote{This is the case as $\tilde{I}_i = \zeta_i \bar{I}_i$, with $\zeta_i = \pm 1$, depending on the index $i$. }. 
As $y$ is small in the \Bs system, the main contribution to the $S_i$ is formed by the $J_i + \bar{J}_i$ term. 

As can be seen from Fig.~\ref{app:fig:time-indep-observables}, if performing the same time-independent set-up as carried out to date by the LHCb collaboration, one expects at least a five-time increase in sensitivity by the end of the LHC Run~5. 
However, this work does not include systematic effects, which can become increasingly relevant with more data, particularly if performing a time-integrated measurement, as assumptions must be made for the parameterisation of the time-dependent PDF. 

\FloatBarrier
\section{Sensitivity to the Optimised Observables}\label{app:optimised-observable-plots}

\begin{figure}[htb!]
    \centering
    \begin{minipage}{.33\textwidth}
        \includegraphics[width=\textwidth]{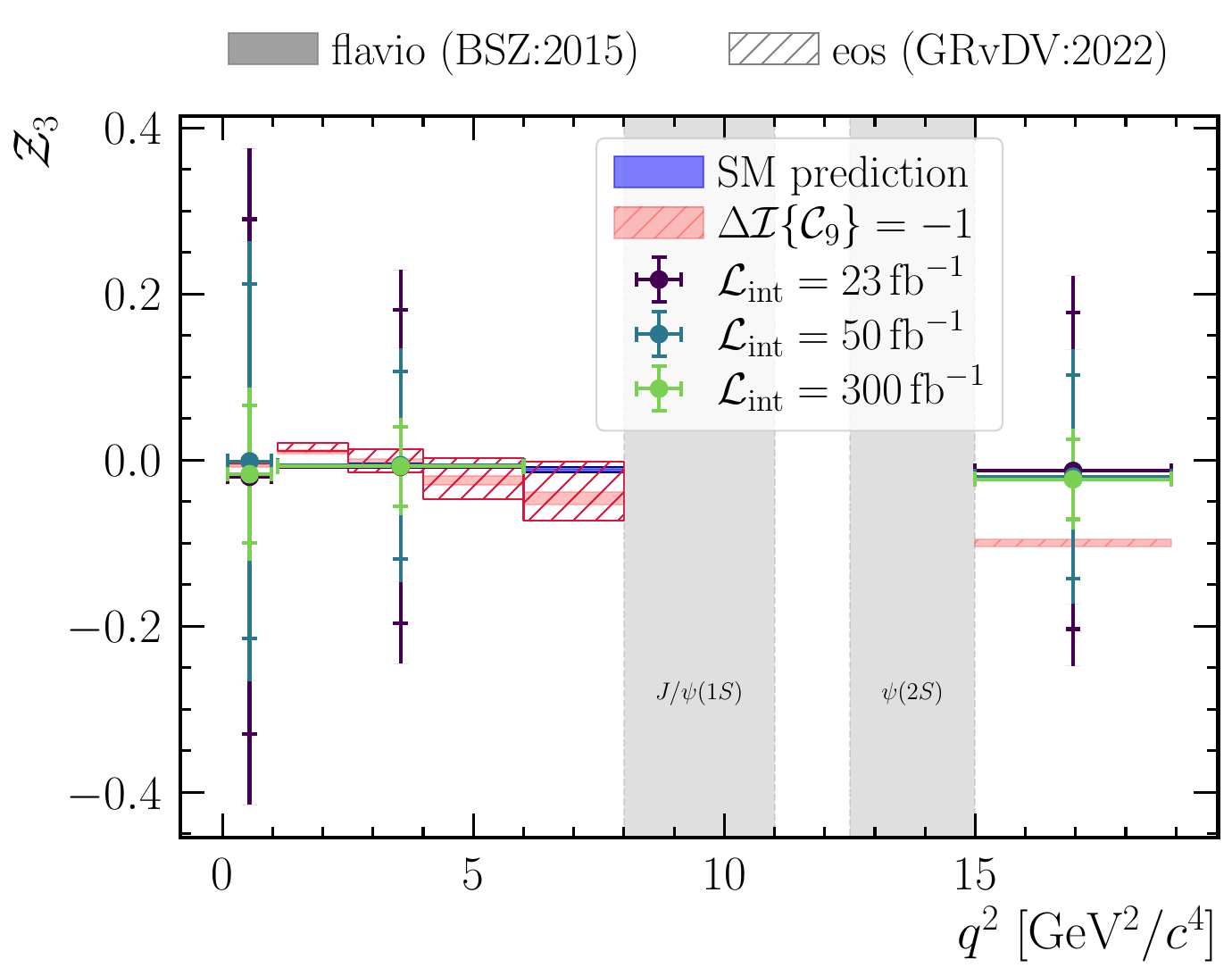}
    \end{minipage}%
    \begin{minipage}{.33\textwidth}
        \includegraphics[width=\textwidth]{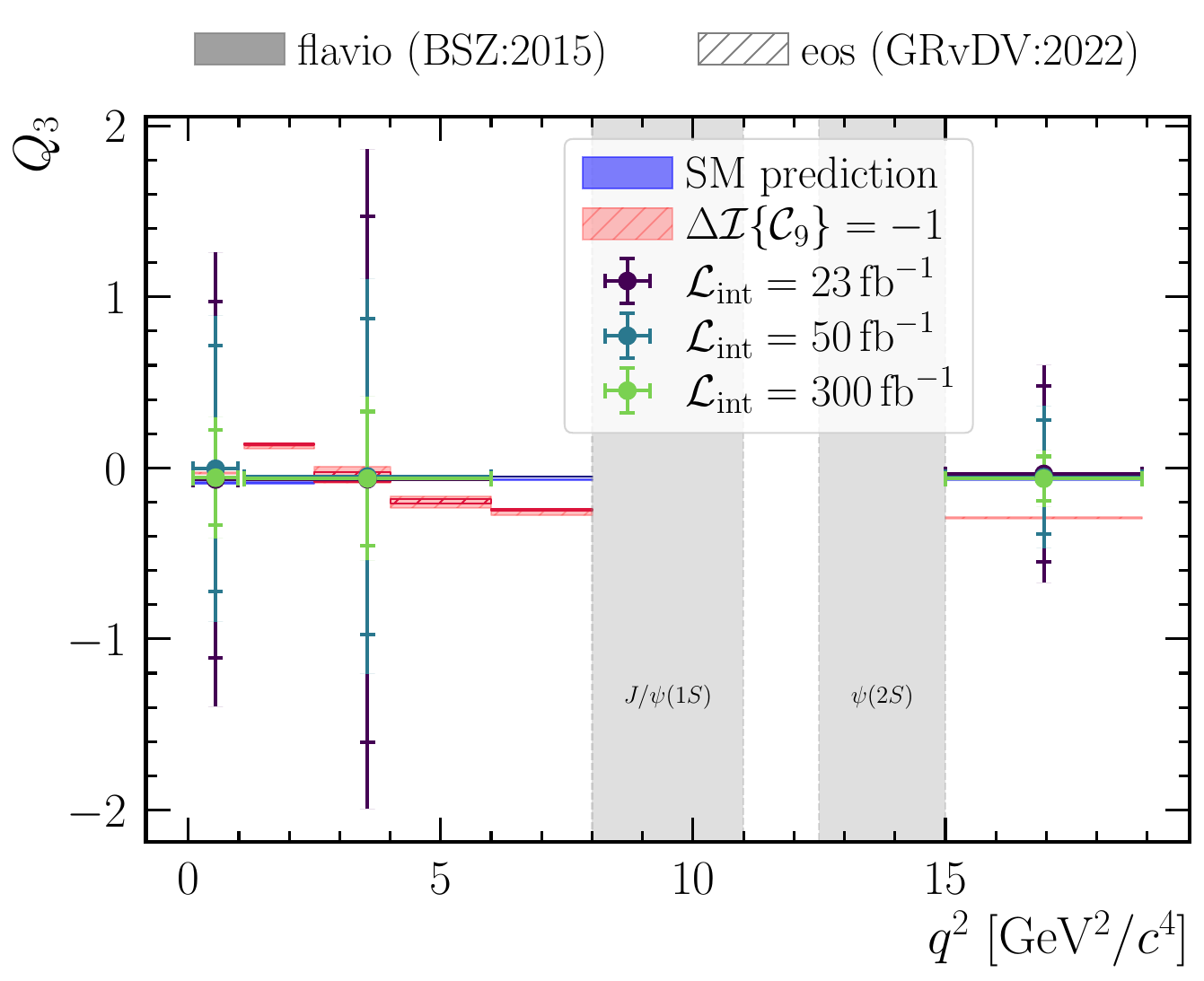}
    \end{minipage}%
    \begin{minipage}{.33\textwidth}
    \hfill
    \end{minipage}
    
    \begin{minipage}{.33\textwidth}
        \includegraphics[width=\textwidth]{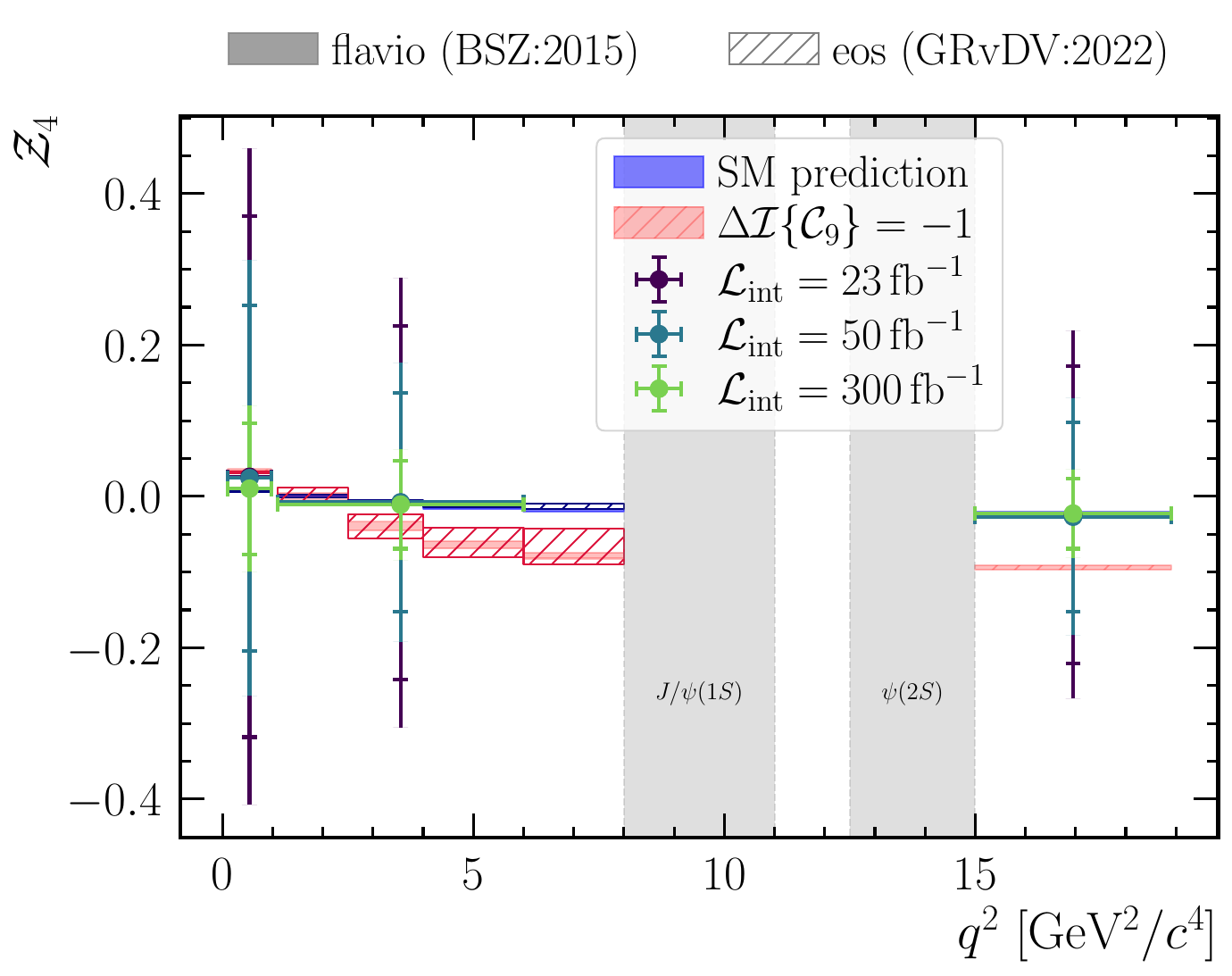}
    \end{minipage}%
    \begin{minipage}{.33\textwidth}
        \includegraphics[width=\textwidth]{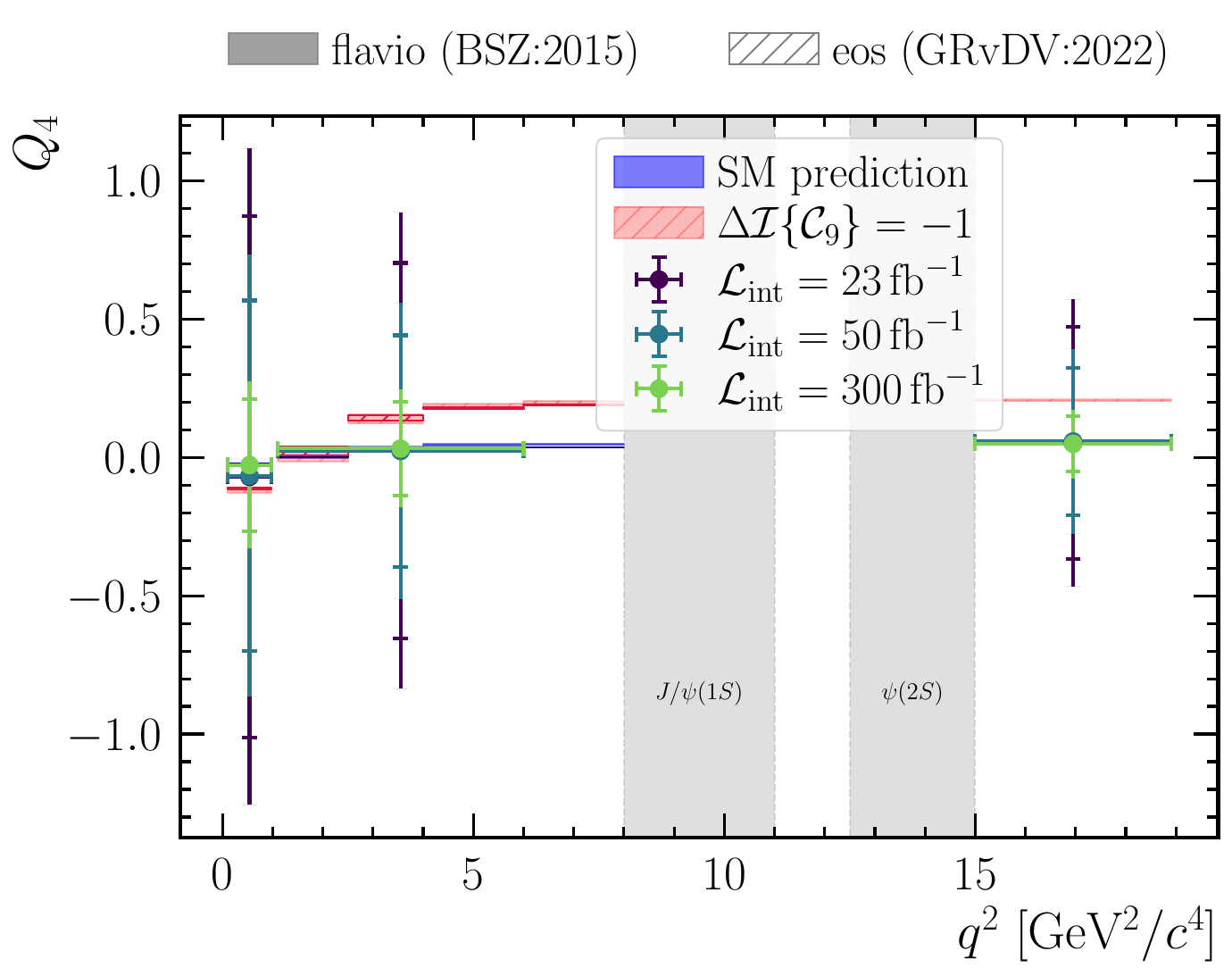}
    \end{minipage}%
    \begin{minipage}{.33\textwidth}
        \includegraphics[width=\textwidth]{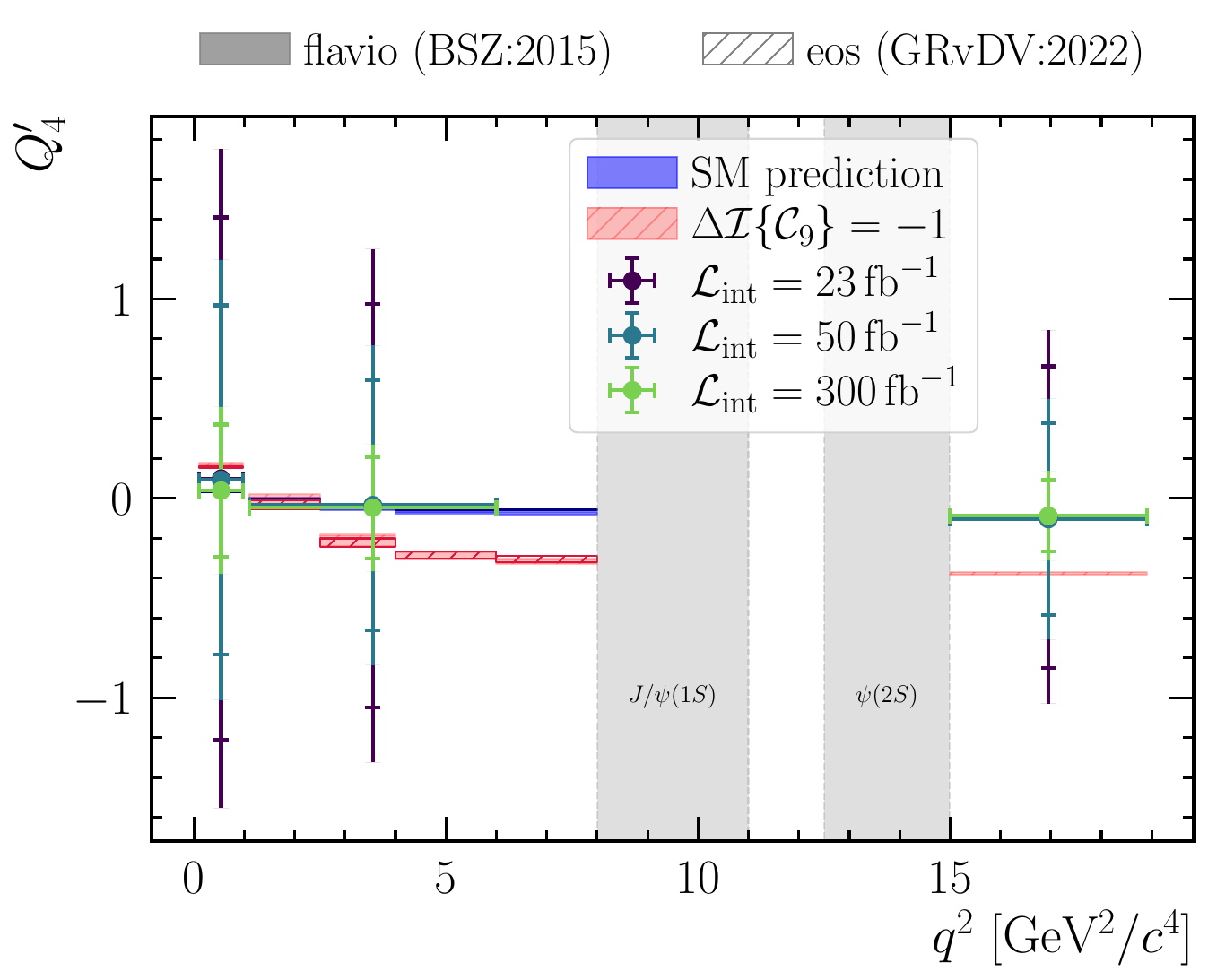}
    \end{minipage}

    \begin{minipage}{.33\textwidth}
        \includegraphics[width=\textwidth]{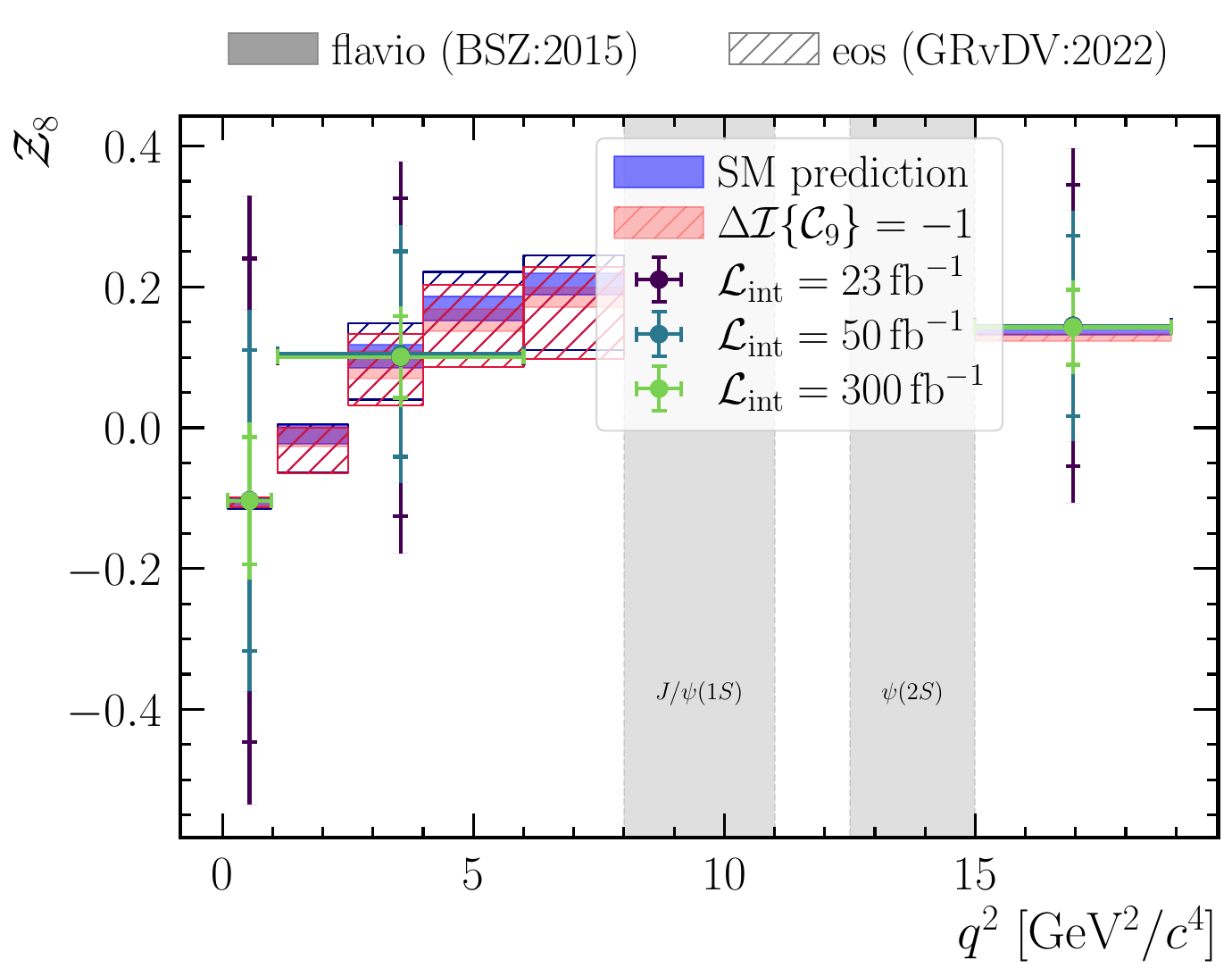}
    \end{minipage}%
    \begin{minipage}{.33\textwidth}
        \includegraphics[width=\textwidth]{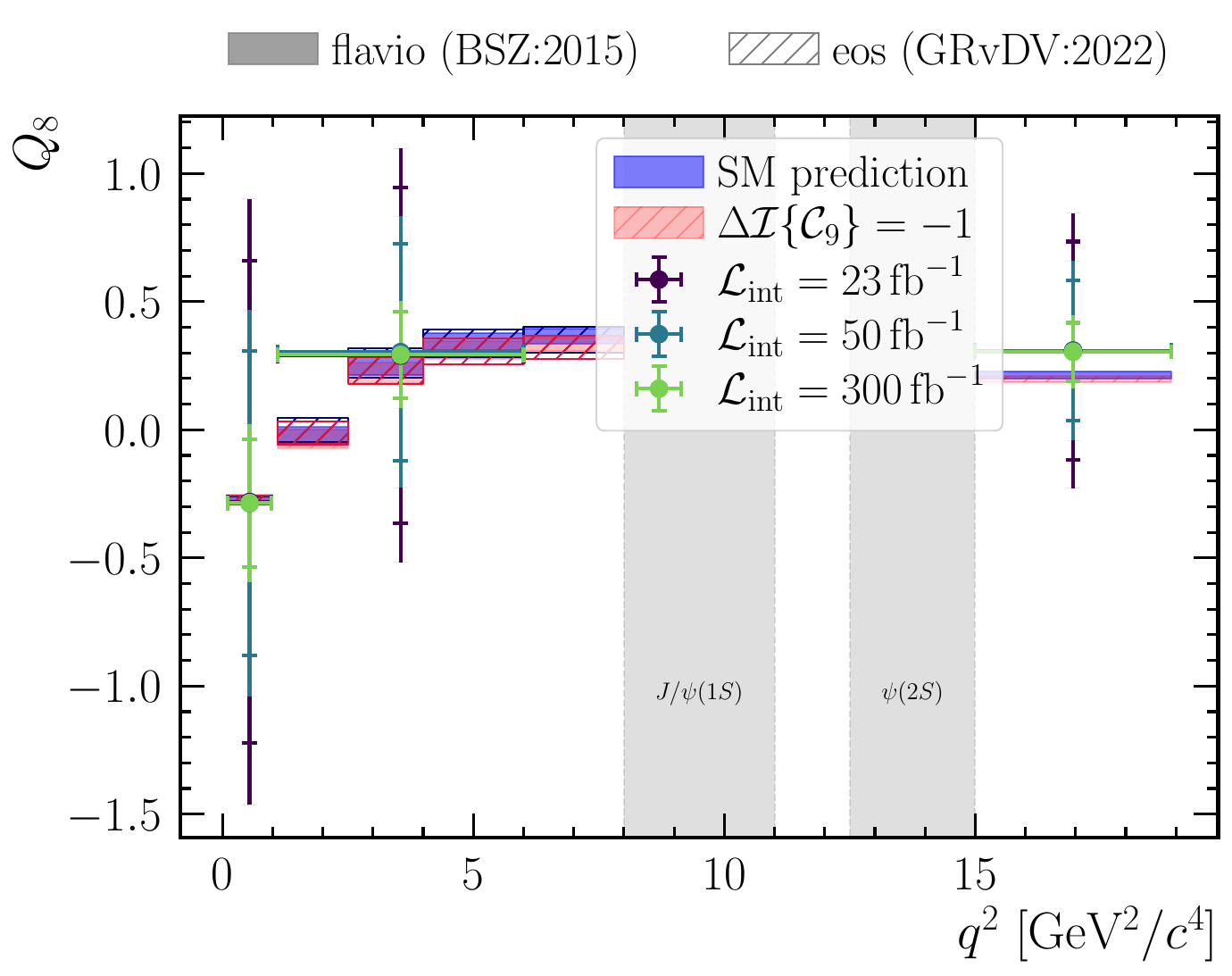}
    \end{minipage}%
    \begin{minipage}{.33\textwidth}
        \includegraphics[width=\textwidth]{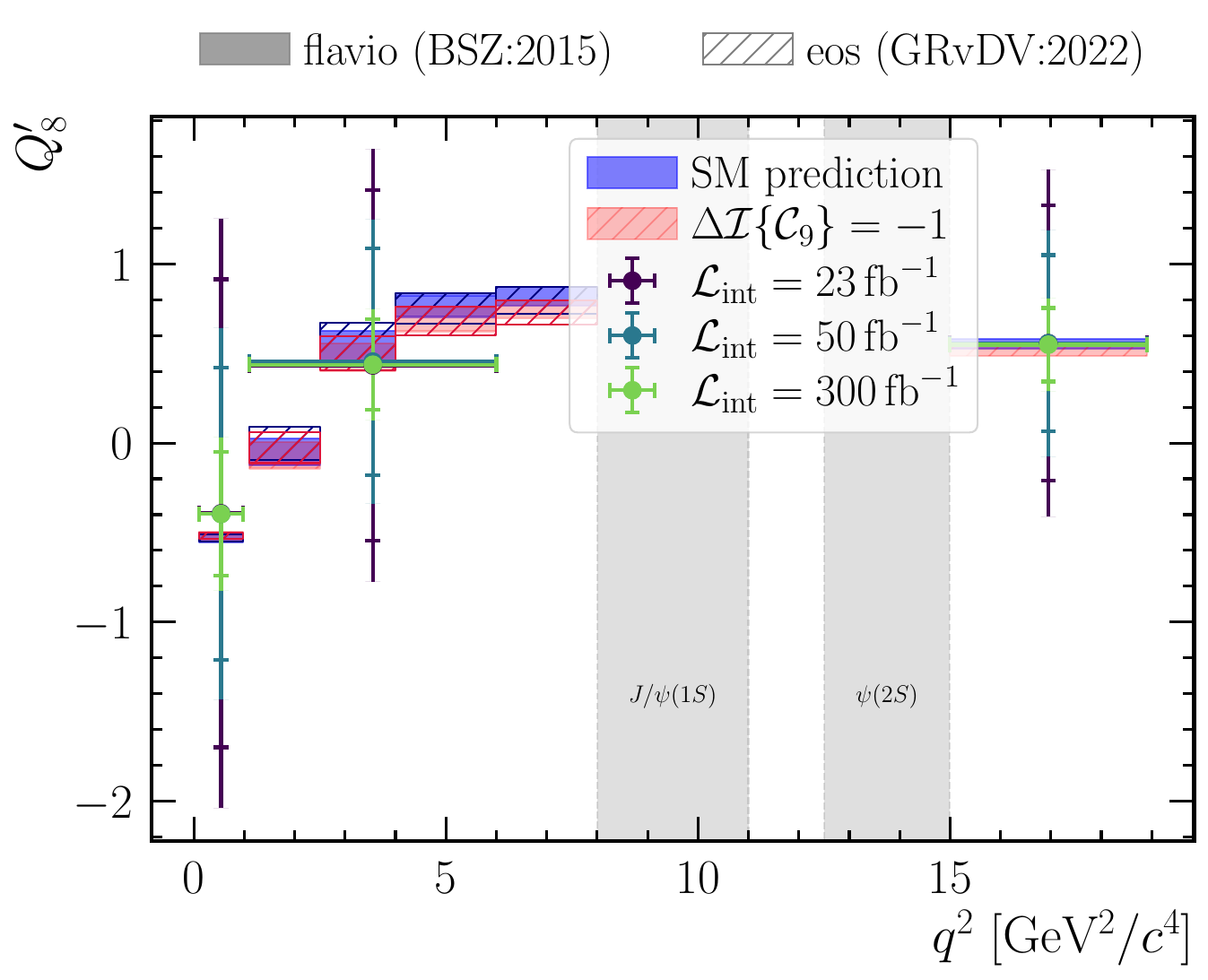}
    \end{minipage}

    \begin{minipage}{.33\textwidth}
        \includegraphics[width=\textwidth]{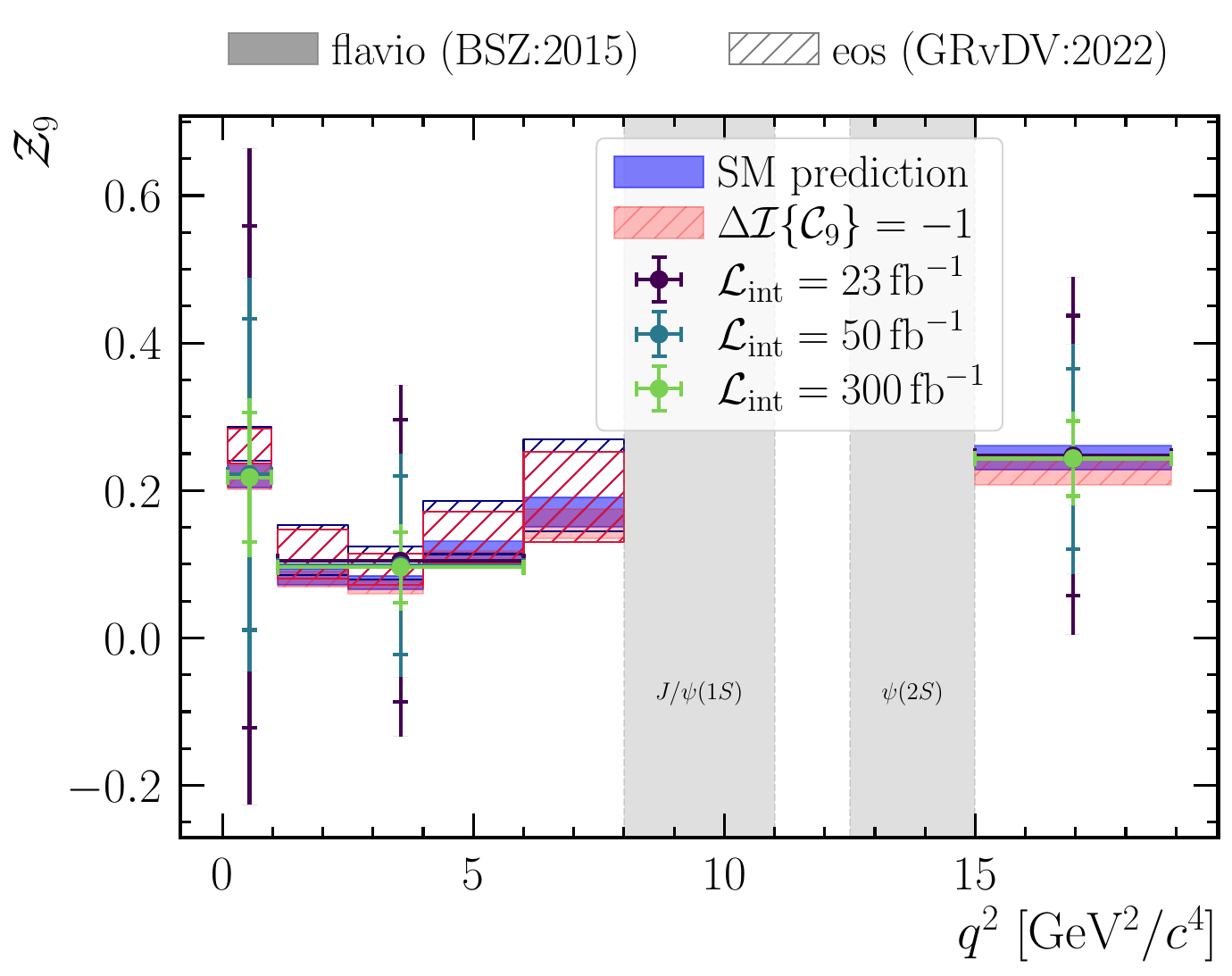}
    \end{minipage}%
    \begin{minipage}{.33\textwidth}
        \includegraphics[width=\textwidth]{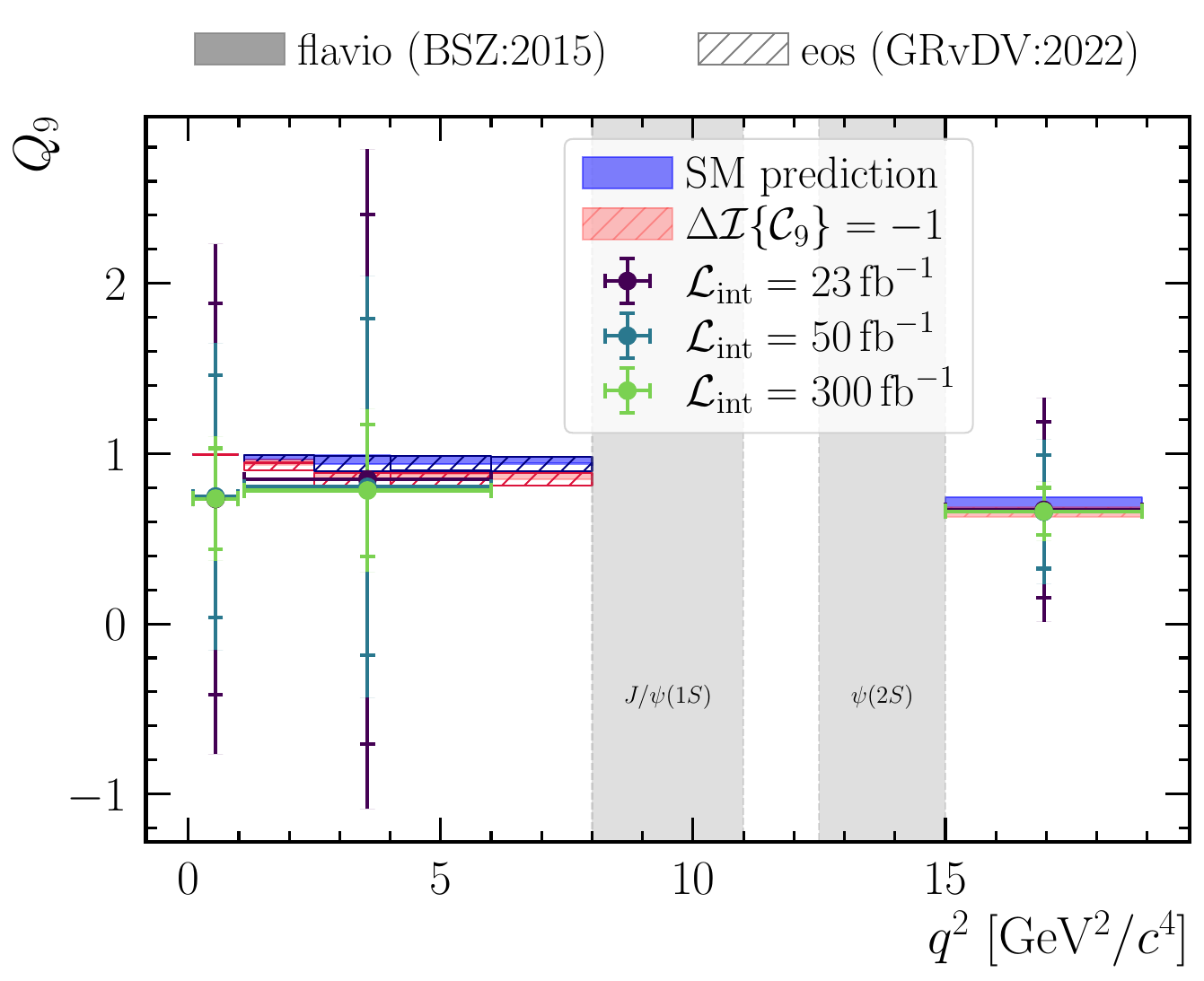}
    \end{minipage}%
    \begin{minipage}{.33\textwidth}
    \hfill
    \end{minipage}

    \caption{Extrapolated sensitivity to $Q_{3}$ (top row), $Q_4^{(\prime)}$ (second row), $Q_8^{(\prime)}$ (third row), and $Q_9$ (bottom row). In addition to the predictions within the SM also an alternative scenario is displayed, where $\Delta\mathcal{C}_9 = -i$ is used. The lower bound on the uncertainty, obtained using the better tagging-power, is given by the end cap of the error bar, whereas the upper bound on the uncertainty is given by the upper end of the coloured error bar.}
    \label{app:fig:opt-obses-qi}
\end{figure}

\begin{figure}[htb!]
    \centering
    \begin{minipage}{.33\textwidth}
        \includegraphics[width=\textwidth]{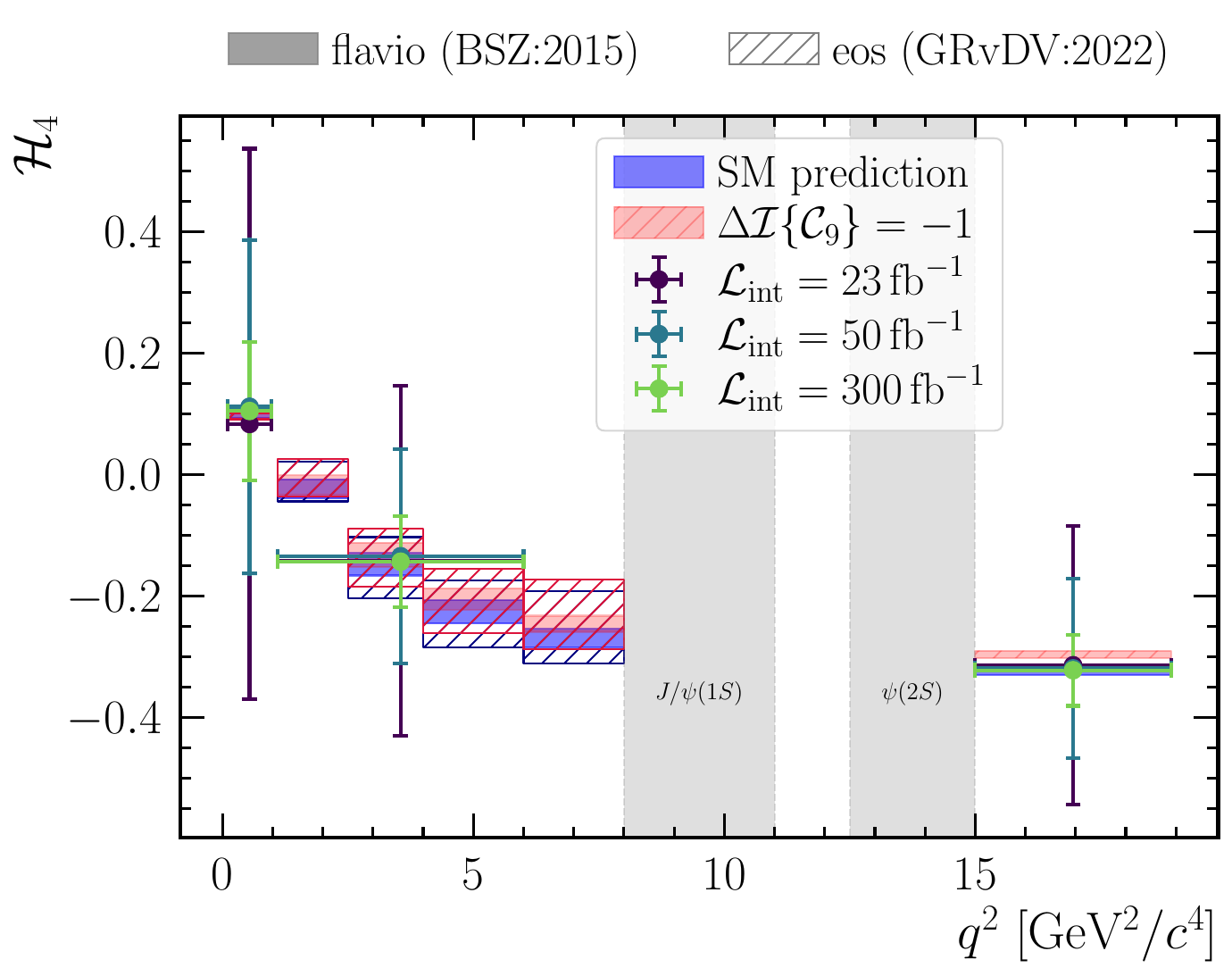}
    \end{minipage}%
    \begin{minipage}{.33\textwidth}
        \includegraphics[width=\textwidth]{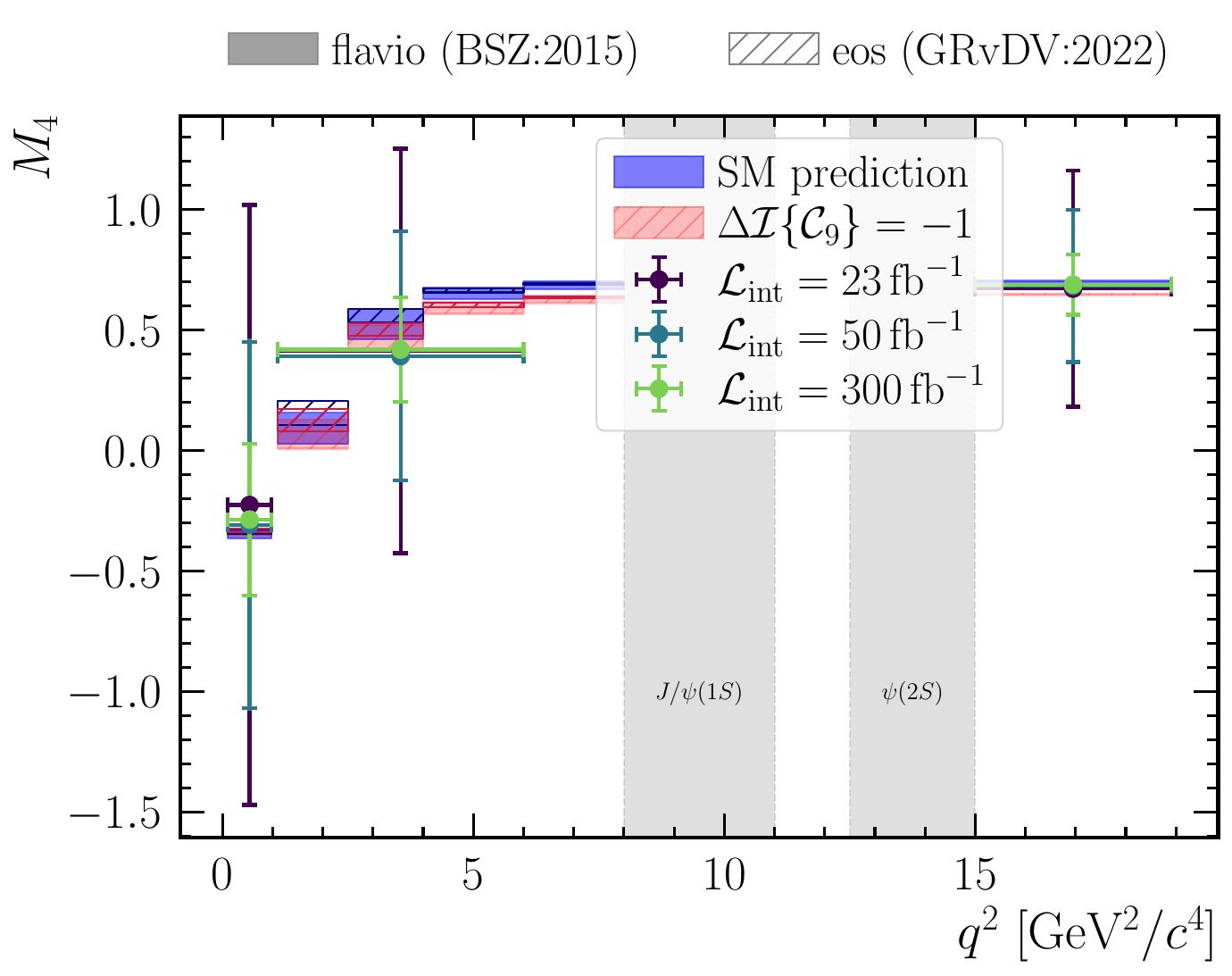}
    \end{minipage}%
    \begin{minipage}{.33\textwidth}
        \includegraphics[width=\textwidth]{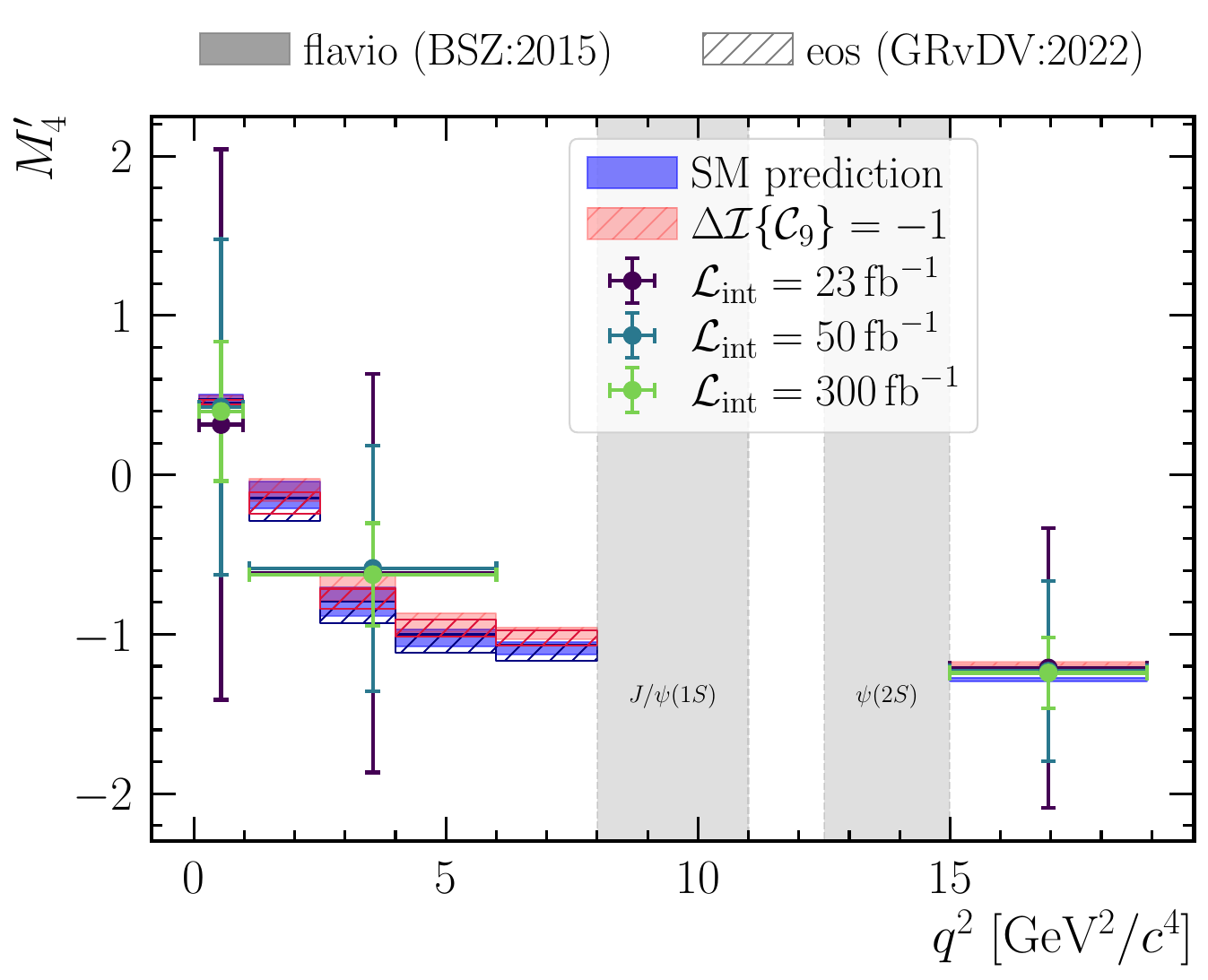}
    \end{minipage}

    \begin{minipage}{.33\textwidth}
        \includegraphics[width=\textwidth]{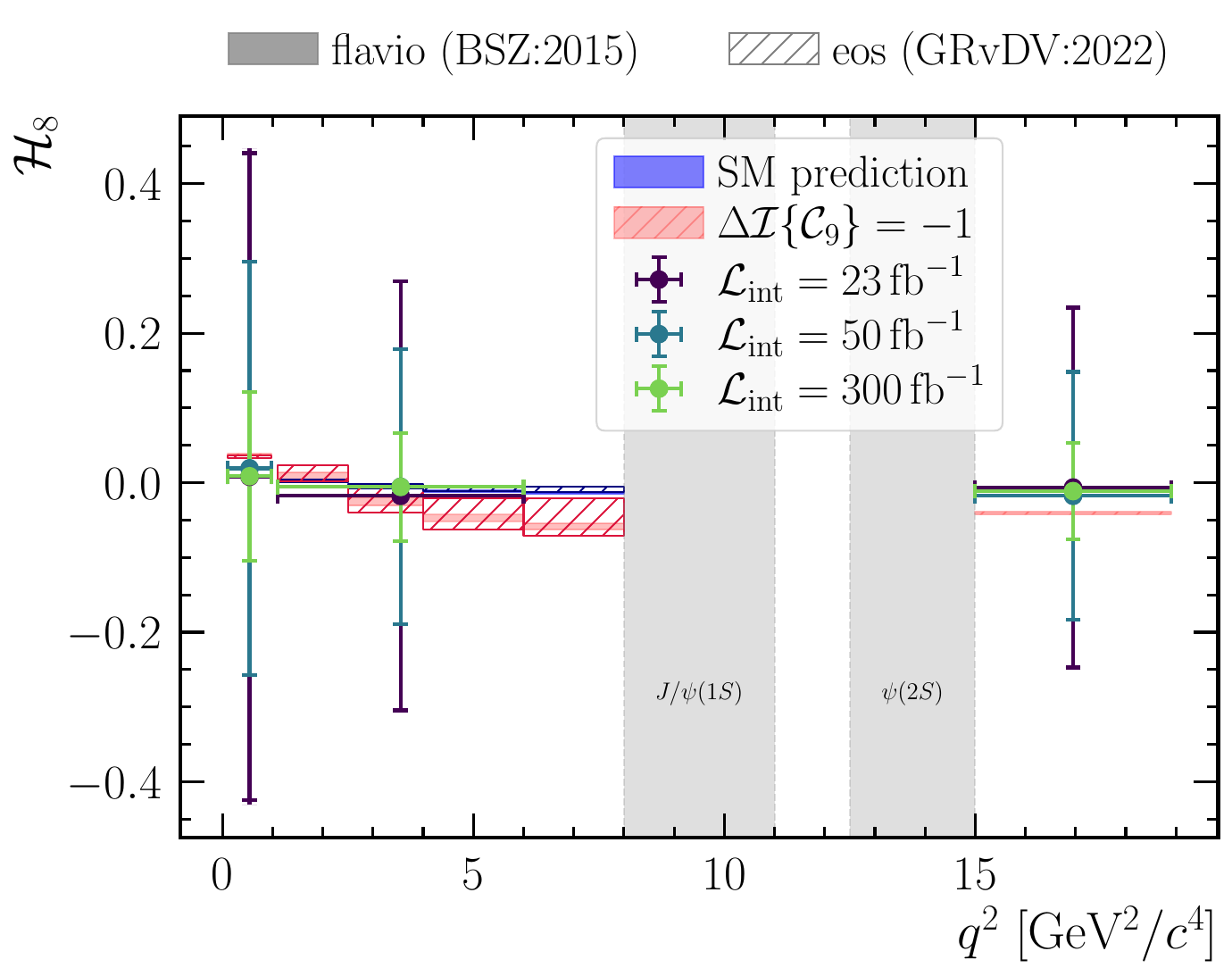}
    \end{minipage}%
    \begin{minipage}{.33\textwidth}
        \includegraphics[width=\textwidth]{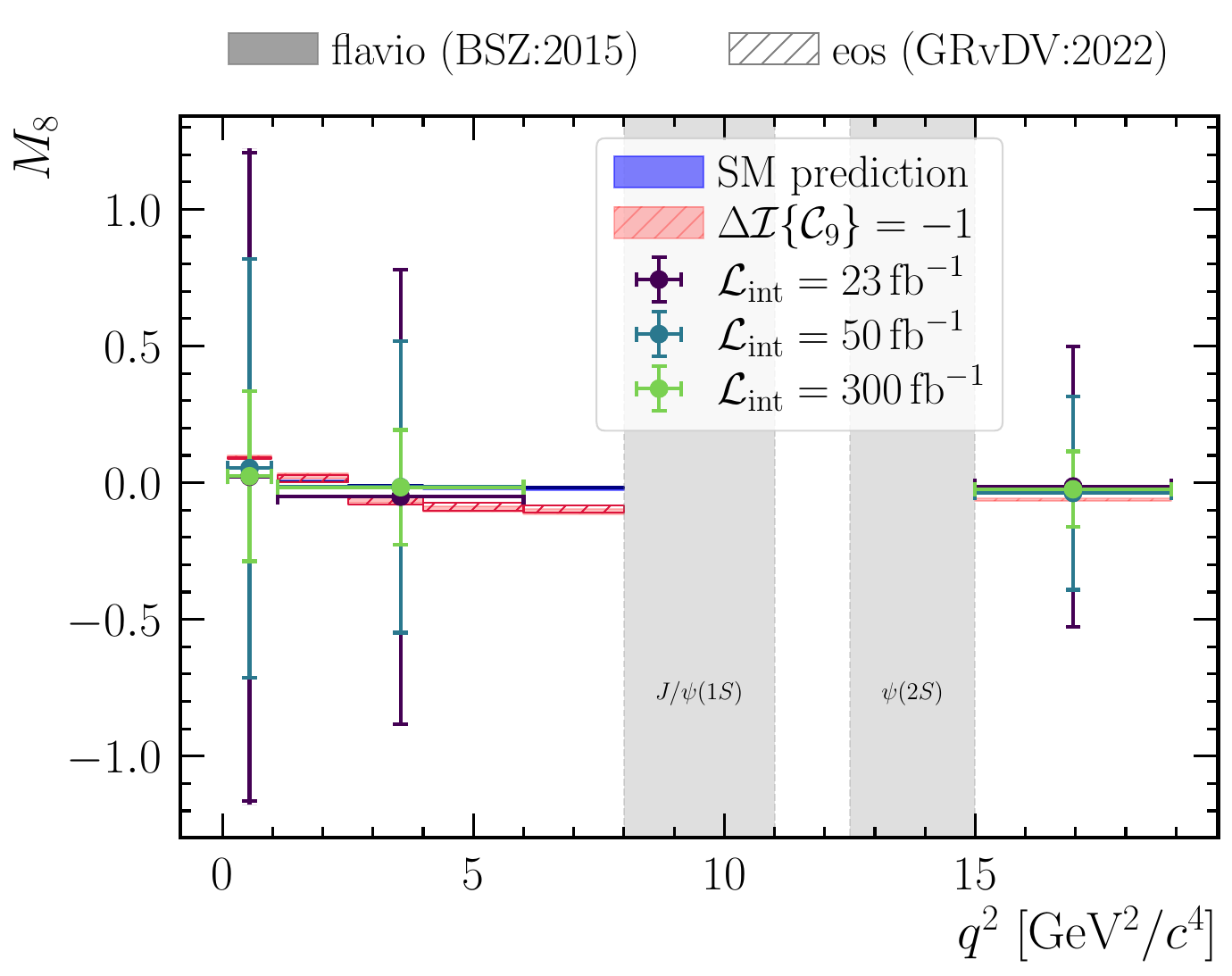}
    \end{minipage}%
    \begin{minipage}{.33\textwidth}
        \includegraphics[width=\textwidth]{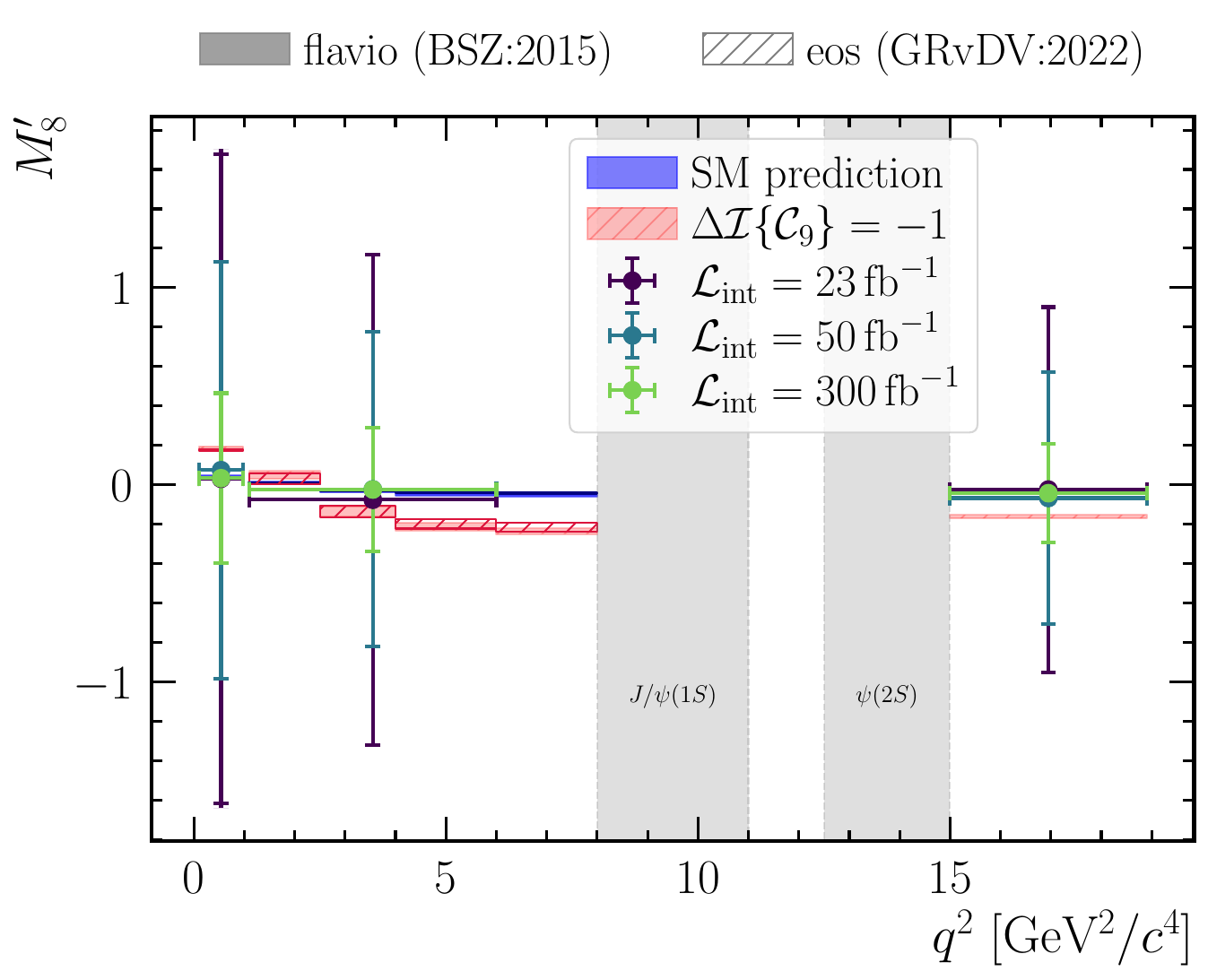}
    \end{minipage}

    \caption{Extrapolated sensitivity to $\mathcal{H}_4$, $M_4$, and $M_4^\prime$ (top row) and $\mathcal{H}_8$, $M_8$, and $M_8^\prime$ (bottom row). In addition to the predictions within the SM also an alternative scenario is displayed, where $\Delta\mathcal{C}_9 = -i$ is used. }
    \label{app:fig:opt-obses-mi}
\end{figure}

\begin{figure}[htb!]
    \centering
    \begin{minipage}{.33\textwidth}
        \includegraphics[width=\textwidth]{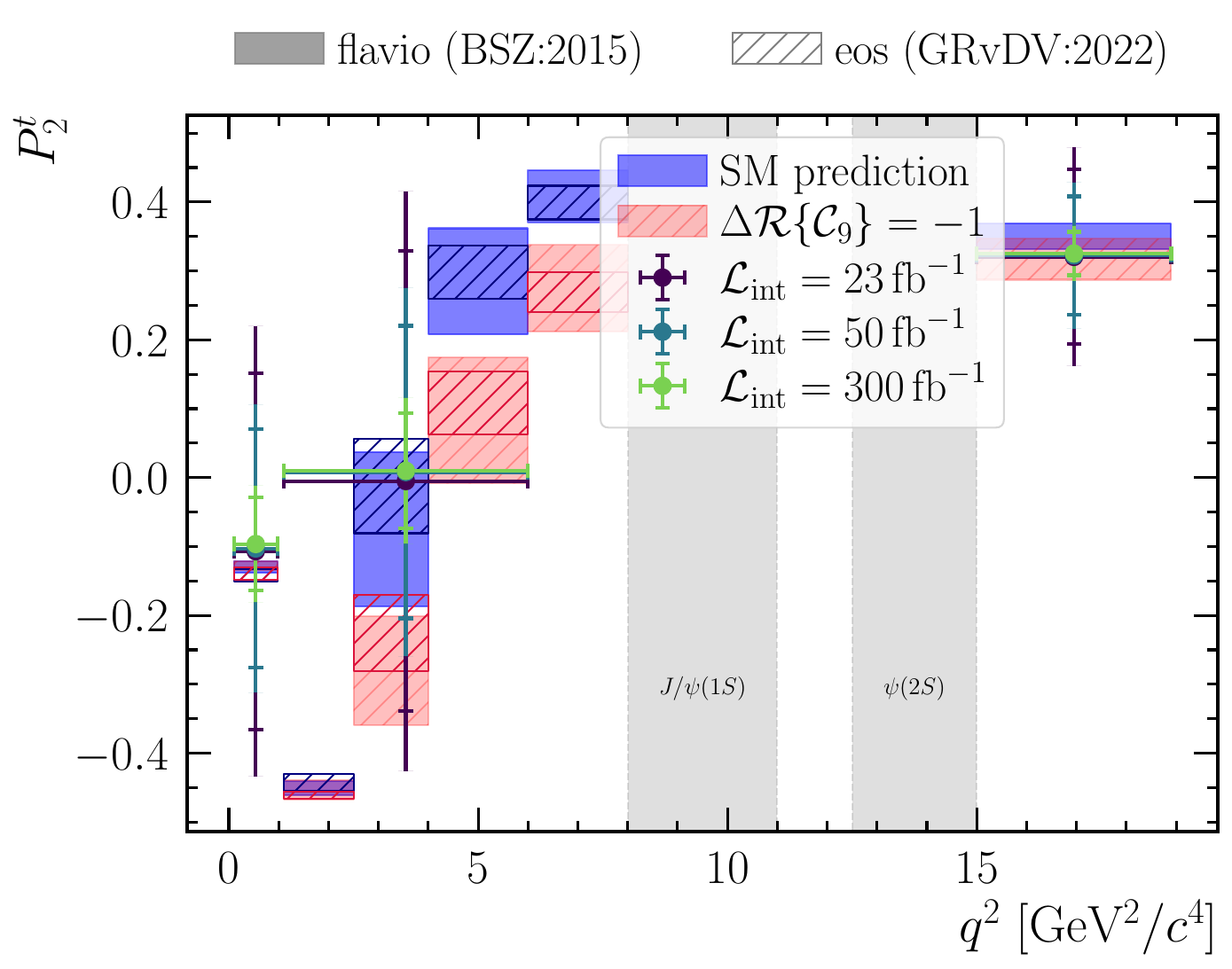}
    \end{minipage}%
    \begin{minipage}{.33\textwidth}
        \includegraphics[width=\textwidth]{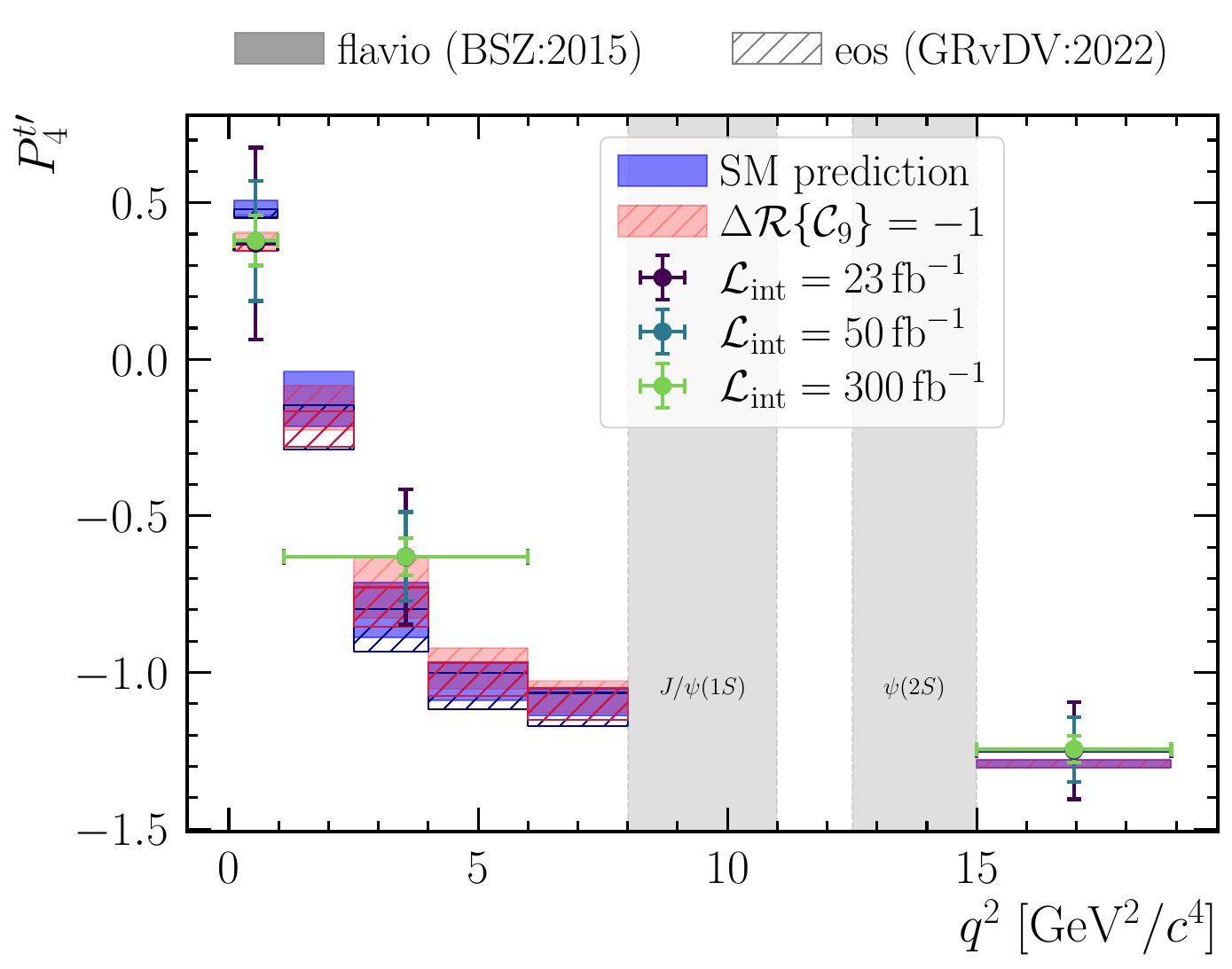}
    \end{minipage}%
    \begin{minipage}{.33\textwidth}
        \includegraphics[width=\textwidth]{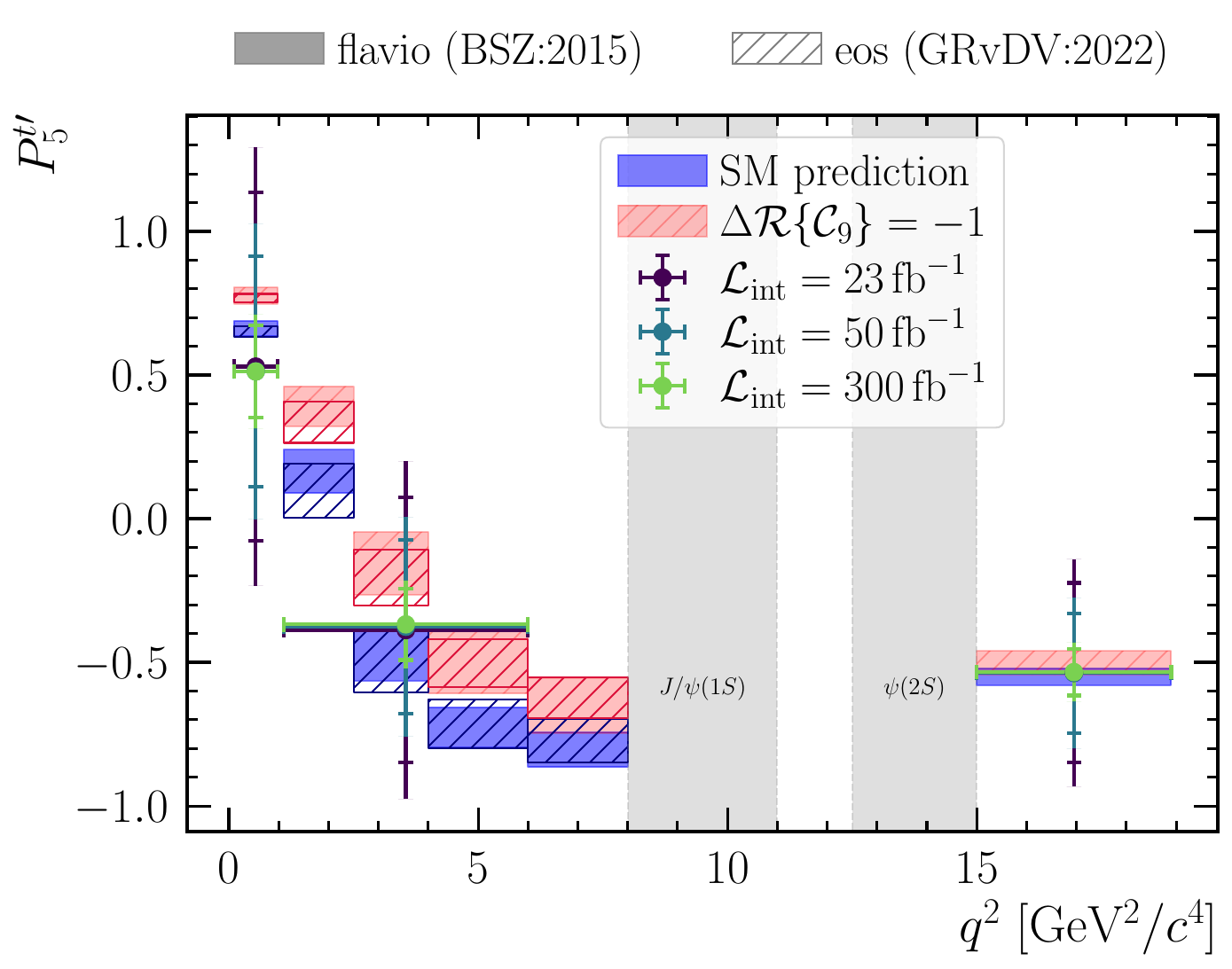}
    \end{minipage}

    \begin{minipage}{.33\textwidth}
        \includegraphics[width=\textwidth]{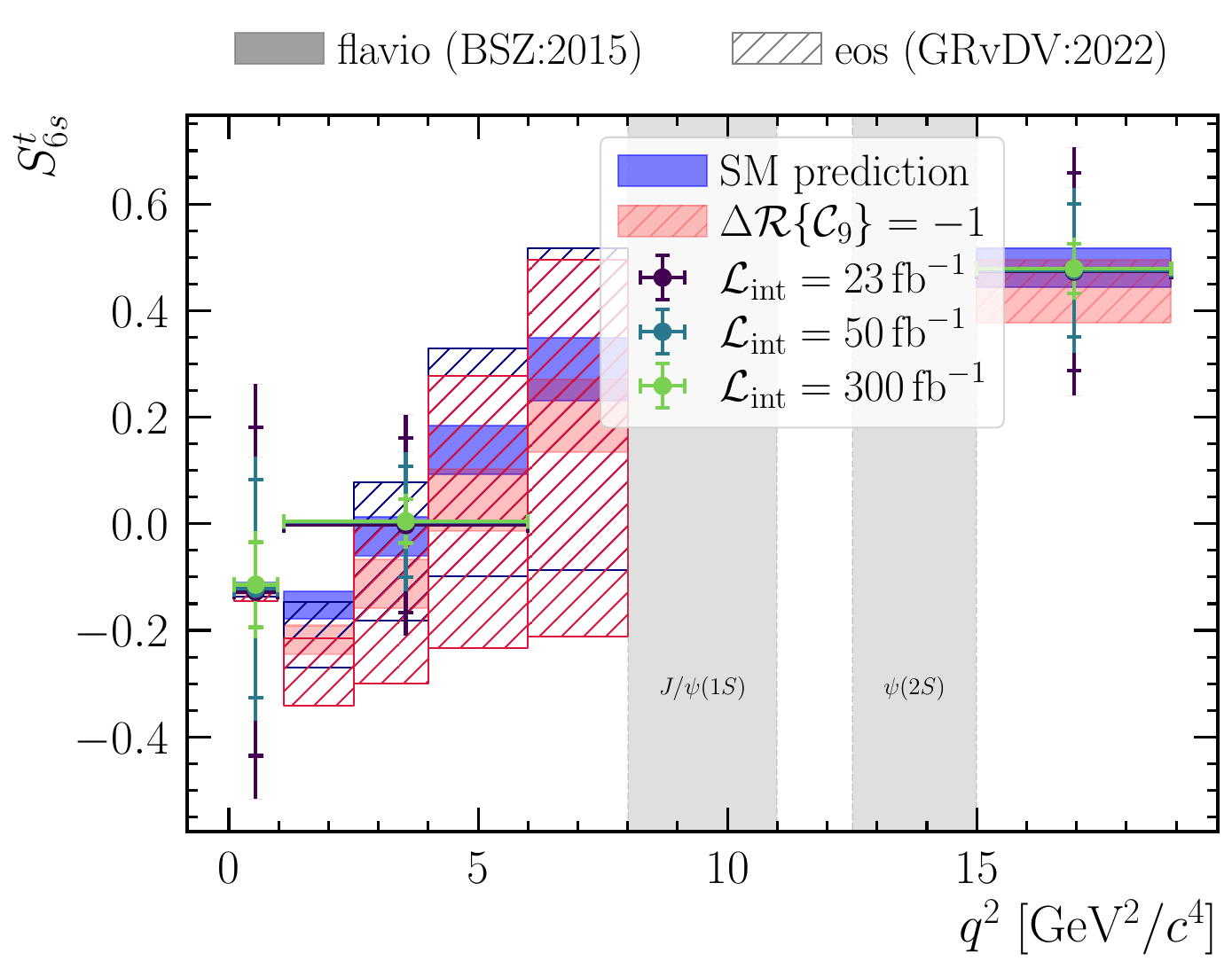}
    \end{minipage}%
    \begin{minipage}{.33\textwidth}
        \includegraphics[width=\textwidth]{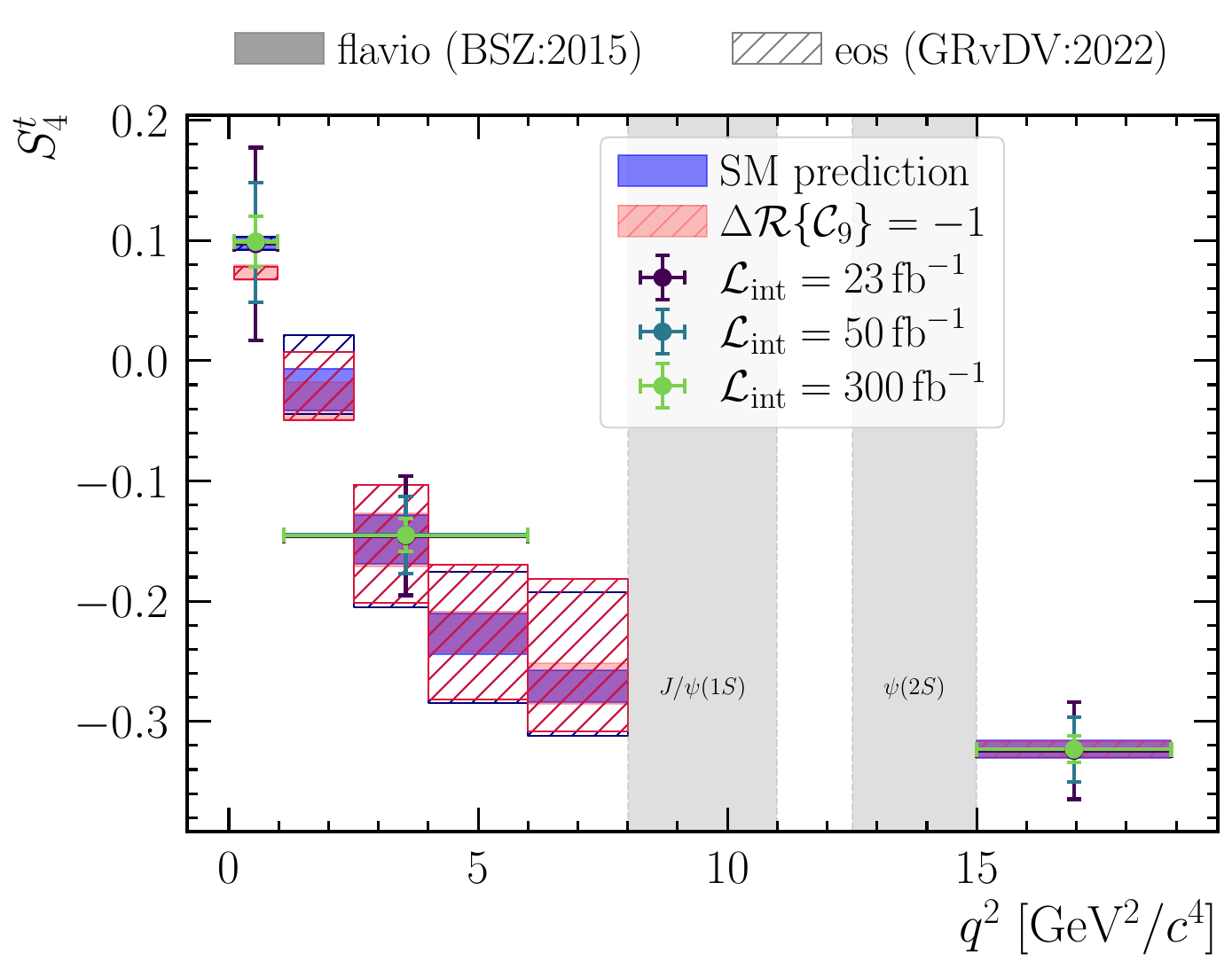}
    \end{minipage}%
    \begin{minipage}{.33\textwidth}
        \includegraphics[width=\textwidth]{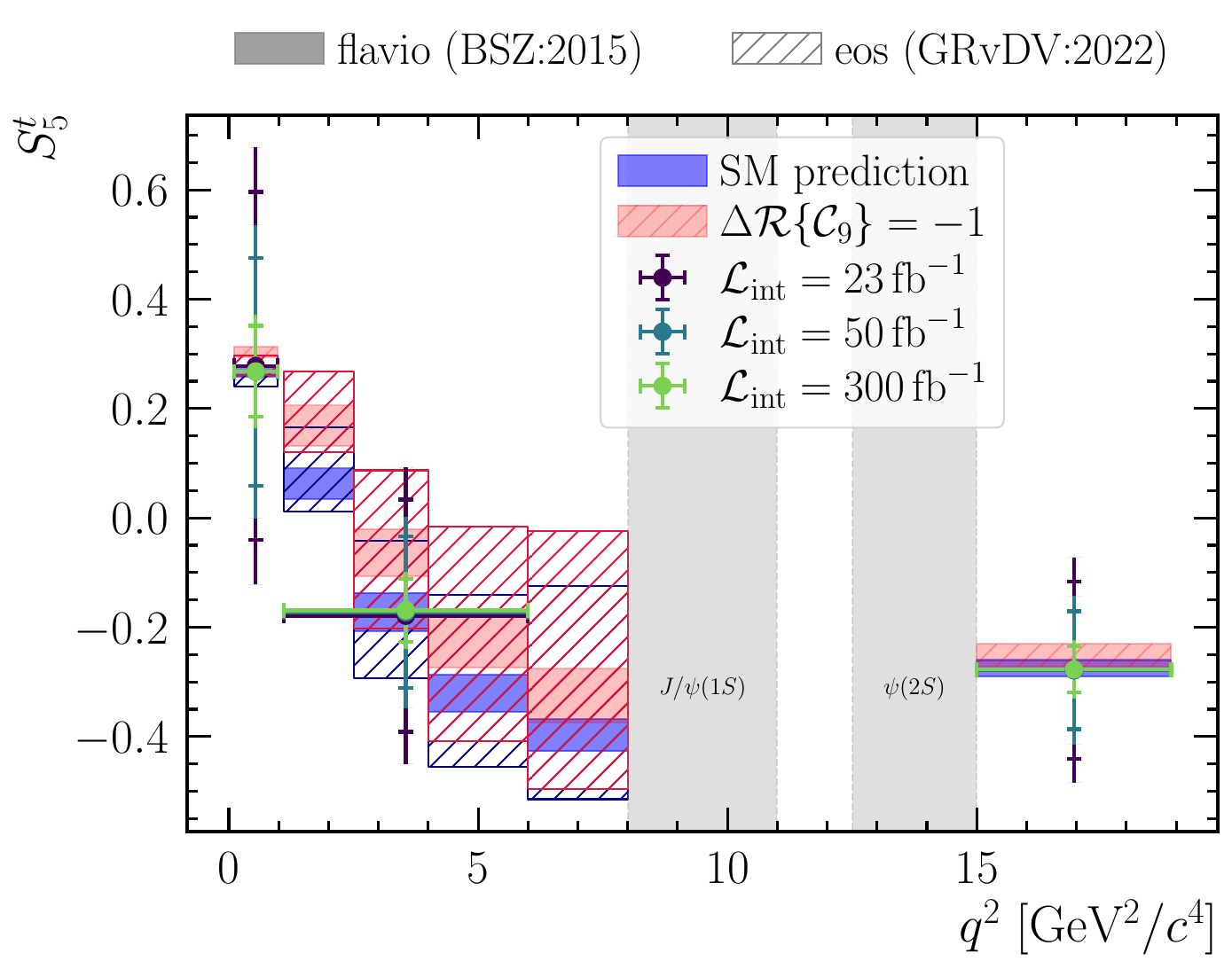}
    \end{minipage}

    \caption{Extrapolated sensitivities to $P_2^t$ (left), $P_4^{t\prime}$ (centre), $P_5^{t\prime}$ (right) and the corresponding unoptimised observables (bottom row). Overlayed are the predictions obtained within the SM and alternative scenarios where the vector coupling $\mathcal{C}_9$ is altered by minus one. Both $P_2^t$ and $P_5^{t\prime}$ are only measurable using flavour-tagging, hence their sensitivities depend on the tagging-power. The lower bound on the uncertainty, obtained using the better tagging-power, is given by the end cap of the error bar, whereas the upper bound on the uncertainty is given by the upper end of the coloured error bar. }
    \label{app:fig:opt-obses-pi}
\end{figure}

This appendix summarises the sensitivity to selected optimised observables and shows the predictions within the SM and when changing the vector-coupling $\mathcal{C}_9$. 
We choose to display observables which show sensitivity to changes in the vector coupling $\mathcal{C}_9$, which comprise $M_{4}^{(\prime)}$, $M_8^{(\prime)}$, $Q_3$, $Q_4^{(\prime)}$, $Q_8^{(\prime)}$, $Q_9$, $P_2^t$, $P_4^{t\prime}$, $P_5^{t\prime}$. 
To visualise the improved sensitivity to NP, also the unoptimised counterparts are shown. 
The $Q_i$ and $M_i$ are displayed in Fig.~\ref{app:fig:opt-obses-qi} and Fig.~\ref{app:fig:opt-obses-mi}, respectively. The $P_i^t$ are shown in Fig.~\ref{app:fig:opt-obses-pi}. 
The different luminosity scenarios are visualised using different colours and the different tagging power is indicated by the two error bar sizes. 
Note that the tagging power influences variables which require tagging, such as the $Q_i$, whereas for example the $M_i$ are measurable without flavour-tagging. 
The definitions of the $M_i$ and $Q_i$ follow Eq.~\ref{eq:opt-obses} and Eq.~\ref{eq:opt-obses-prime} and the $P_i^t$ are given in Eqs.~\ref{eq:opt-obses-pi} and following. 

\FloatBarrier

\pagebreak
\clearpage
 
\addcontentsline{toc}{section}{References}
\bibliographystyle{LHCb}
\ifx\mcitethebibliography\mciteundefinedmacro
\PackageError{LHCb.bst}{mciteplus.sty has not been loaded}
{This bibstyle requires the use of the mciteplus package.}\fi
\providecommand{\href}[2]{#2}

\newpage

\end{document}